\documentclass[12pt, twoside]{article}

\usepackage[a4paper, inner=2.25cm, outer=1.75cm, top=2.5cm, bottom=3cm]{geometry}
\usepackage{natbib}
\usepackage{graphicx}
\usepackage{newtxtext}
\usepackage{newtxmath}
\usepackage{hyperref}
\usepackage{setspace}

\usepackage[figurename=Fig., tablename=Tab.]{caption}
\usepackage{subcaption}

\usepackage{multirow}
\usepackage{tabularx}
\usepackage{makecell}

\usepackage[ruled,vlined,english]{algorithm2e}

\usepackage{siunitx}
\hypersetup{
    colorlinks = true,
    urlcolor   = blue,
    citecolor  = black,
}

\newcommand{\RomanNumeralCaps}[1]
\linenumbers

\DeclareCaptionLabelFormat{andtable}{#1~#2  \&  \tablename~\thetable}

\usepackage{fancyhdr}

\title{Enhancement of Large Eddy Simulation for the prediction of an intake flow rig using sequential Data Assimilation}

\author{Lucas Villanueva$^{1*}$, Karine Truffin$^2$, Jacques Borée$^1$, Marcello Meldi$^3$ \\\vspace{-0.5em}
\small \textit{$^1$Institut Pprime, CNRS - ISAE-ENSMA - Universit\'{e} de Poitiers,} \\
\small \textit{11 Bd. Marie et Pierre Curie, Site du Futuroscope, TSA 41123, 86073 Poitiers Cedex 9, France} \\\vspace{-0.5em}
\small \textit{$^2$Institut Carnot IFPEN Transports Energie, IFP Energies nouvelles,} \\
\small \textit{1-4 avenue de Bois-Préau, 92852 Rueil-Malmaison, France} \\\vspace{-0.5em}
\small \textit{$^3$Univ. Lille, CNRS, ONERA, Arts et Métiers ParisTech, Centrale Lille, UMR 9014-} \\
\small \textit{LMFL- Laboratoire de Mécanique des fluides de Lille - Kampé de Feriet, F-59000 Lille, France} \\
\small \texttt{* Corresponding author: villanueva@cerfacs.fr, lucas.villanueva.pro@gmail.com}}

\begin{document}
\maketitle
\thispagestyle{fancy}

\begin{abstract}
 A Data Assimilation (DA) strategy based on an ensemble Kalman filter (EnKF) is used to enhance the predictive capabilities of scale resolving numerical tools for the analysis of flows exhibiting cyclic behaviour. More precisely, an ensemble of numerical runs using Large Eddy Simulation (LES) for the compressible steady-state flow rig is augmented via the integration of high-fidelity data. This observation is in the form of instantaneous velocity measurements, which are sampled at localized sensors in the physical domain. Two objectives are targeted. The first one is the calibration of an unsteady inlet condition suitable to capture the cyclic flow investigated. The second one is the analysis of the synchronization of velocity field predicted by the LES with the available observation. In order to reduce the computational costs required for this analysis, a hyper-localization procedure (HLEnKF) is proposed and it is integrated in the library CONES, tailored to perform fast online DA. The proposed strategy performs a satisfactory calibration of the inlet conditions, and its robustness is assessed using two different prior distributions for the free parameters optimized in this task. DA state estimation is efficient in obtaining accurate local synchronization of the inferred velocity fields with the observed data. The modal analysis of the kinetic energy of the flow field provides additional information on the quality of the reconstruction of the velocity field, which shows improvements. Thus, the HLEnKF shows promising features for the calibration and the synchronization of scale-resolved turbulent flows, opening perspectives of applications for complex phenomena using advanced tools such as digital twins.
\end{abstract}

\section{Introduction}
\label{sec:intro}
Data Assimilation (DA) refers to a class of methods designed to obtain optimized predictions of physical processes via the combination of results from multiple sources of measurements. Recent advances of this discipline in fluid mechanics \citep{Rochoux2014_nhess,Mons2016_jcp,Meldi2017_jcp,Chandramouli2020_jcp,Zhang2020_cf,LeProvost2021_prf,Zhang2024_jfm,Plogmann2024_cmame,Zheng2024_asme,Ephrati2025_prf,Valero2025_jcp} have provided efficient guidelines of application to the analysis of complex flows, in particular for high Reynolds regimes. While numerous studies focus principally on averaged approaches such as RANS simulation, an increasing number of works deal with scale-resolved simulation such as Large Eddy Simulation (LES) aiming for different key objectives such as model optimization \citep{Mons2021_prf,Moussie2024_ftc,Zhang2024_jfm} or synchronization of instantaneous flow fields \citep{Labahn2020_ftc,Wang2022_jfm,Villanueva2024_ijhff}. The emergence of new approaches in this field is promising and opens future perspectives for flow analysis and control using advanced tools such as digital twins \citep{Rasheed2020_IEEE,Semeraro2021_ci}. In this framework, one of the most important open challenges is the ability to predict and potentially prevent the emergence of extreme events associated with the variability of turbulent flows. These events, which can lead to critical issues, are observed in several kinds of flows dealing with urban environments, transport engineering, and energy harvesting and production. For the latter, one example of applications severely impacted by extreme events are flow configurations for internal combustion engines (ICE). This class of flows can exhibit strong variability between cycles, which are not accurately described even by scale-resolved simulation. The possibility to infuse realistic information in the simulation process, which is one of the promising aspects of DA techniques, is therefore a key element for technological advancement in this field. 

In terms of simulation of ICE flows, the steady-state flow rig is a configuration extensively studied~\citep{thobois2005large,Afailal2019_ogst}. In particular, this geometry is used by engine manufacturers in the early stages of development to optimize intake ducts and valve geometry. It is characterized by a single valve inserted in an intake pipe ejecting the fluid into a cylindrical chamber. The term \textit{steady-state} here refers to the fact that the valve lift distance is fixed, so the geometry is motionless. This flow configuration is characterized by highly anisotropic and inhomogeneous turbulence, compressibility effects, wall effects and potentially swirl \citep{Afailal2021_phd}. The flow moving through the intake pipe and around the valve forms a jet which is affected by high shear stresses. The flow impacts the cylinder wall on either side of the valve, generating large scale structures. As previously discussed, recent investigations studied cycle-to-cycle variations in the internal combustion engine \citep{ding2024part1, ding2024part2}. The cyclic variability of combustion is due to several factors. One of the most complex to identify is the variability of the intake jet, due to the intrinsically non-linear and complex nature of unsteady turbulence combined with interactions with the walls and external fluctuations. The precise study of these events and how they originate remains difficult. Accurate representation of the behaviour of boundary conditions, which contribute to the emergence of these phenomena, could permit the study of these events relying on scale-resolved simulation. To this purpose, DA approaches show promising features to infuse experimental information into the numerical process. However, due to the complexity of the optimization process and the large number of degrees of freedom to be considered, few studies in the literature dealing with data-informed procedures target the calibration of boundary conditions via the assimilation of local available information. 

Methods used for the prediction of inlet conditions include i) the minimization of a cost function with respect to experimental data for the simulation of free-surface flows with an LBM method \citep{Nishi2013_cf}, a gradient-based optimization for a synthetic turbulence inlet condition of an LES simulation for urban flows \citep{Lamberti2018_jweia} or the use of machine learning to reconstruct a hydraulic turbine suction tube inlet condition from downstream flow information for RANS and LES simulations \citep{Veras2023_cf}. In the field of data assimilation, two studies focus on the inference of an inlet condition in the recent years. \cite{Sousa2019_be} use an ensemble Kalman filter (EnKF) to infer the behaviour of an inlet condition for an urban flow configuration from real measurements of anemometry data for RANS simulations. \cite{Moussie2024_ftc} target the reduction of  discrepancies between scale-resolved numerical simulation and experimental measurements using the EnKF via the calibration of a synthetic turbulence inlet condition. The investigation is performed for the flow around a bluff body. These studies are developed for statistically steady test cases. To the knowledge of the Authors, similar techniques have not been comprehensively explored for flows showing cyclic behaviour, which are arguably more prone to exhibit high variability of the solution and extreme events. This point is a key challenge for future applications of the digital twin paradigm for complex flows, in order to provide accurate prediction and safe control of the physical twin.

In this work, a data assimilation strategy based on the EnKF is used to investigate a steady-state flow rig, combining high-fidelity localized data from a reference simulation with LES numerical predictions. In order to reduce the computational costs required to perform this task, a hyper-localization procedure (HLEnKF) inspired by works in the literature is proposed. Two objectives are targeted. The first one is the calibration of unsteady inlet conditions used to predict cyclic flows. The second objective deals with the synchronization of the numerical prediction by LES with the flow configuration from which the observed data are sampled and used in the DA procedure \citep{Villanueva2024_ijhff}. The synchronization is investigated via the analysis of instantaneous features of the flow as well as the modal behaviour of the flow.  The article is structured as follows. In Section \ref{sec:NumTools} the numerical tools used for the investigation are introduced and discussed. In Section \ref{sec:setup} the simulation strategies used to investigate the steady-state flow rig are validated for a simplified inlet condition using available experimental data \citep{Thobois2004_saefl}. In Section \ref{sec:inletCalibration}, the calibration of the parameters describing an unsteady inlet conditions is performed using the HLEnKF. In Section \ref{sec:localSynchronization} the DA strategy is used to investigate the efficacy of synchronization of the velocity fields obtained with the LES runs with the observed data. In Section \ref{sec:conclusion} concluding remarks are provided and future applications are discussed.

\section{Numerical tools}
\label{sec:NumTools}
\subsection{Navier--Stokes equations and numerical discretization \& modelling}
\label{sec:LES_equations}
Numerical simulations in this work are performed using the open-source code OpenFOAM, which relies on finite volume discretization of the Navier-Stokes equations. For compressible flows and Newtonian fluids, they can be formulated as ~\citep{garnier2009large,poinsot2011theoretical}:

\begin{eqnarray}
    \frac{\partial \rho}{\partial t} + \frac{\partial (\rho u_j)}{\partial x_j} &=& 0, \label{eq:mass} \\
    \frac{\partial \rho u_i}{\partial t} + \frac{\partial (\rho u_i u_j)}{\partial x_j} &=& - \frac{\partial p}{\partial x_i} + \frac{\partial \sigma_{ij}}{\partial x_j}, \label{eq:mom} \\
    \frac{\partial \rho E}{\partial t} + \frac{\partial (\rho E +p ) u_j }{\partial x_j} &=&  + \frac{\partial (\sigma_{ij}u_i)}{\partial x_j} - \frac{\partial q_j}{\partial x_j},  \label{eq:temp}
\end{eqnarray}

with $\sigma_{ij} = - \frac{2}{3} \mu \frac{\partial u_k}{\partial x_k}\delta_{ij} + \mu \biggl(\frac{\partial u_j}{\partial x_i} + \frac{\partial u_i}{\partial x_j} \biggl)$ 
and $q_j = -\kappa \frac{\partial T}{\partial x_j}$.

Here, $\mathbf{u}=[u_1, \, u_2, \, u_3] = [u_x, \, u_y, \, u_z]$ is the velocity field, $\rho$ is the density, $p$ is the pressure, $E$ is the total energy, $T$ is the temperature, $\mu$ is the dynamic viscosity, $\kappa$ is the thermal conductivity. Repetition over the index $j$ is employed for the sake of conciseness. Numerical discretization of equations \ref{eq:mass}-\ref{eq:temp} can be used to investigate the flow evolution. However, the complete representation of active flow scales for high Reynolds numbers demands prohibitive computational resources in terms of grid refinement. Among the approaches presented in the literature \citep{Pope2000_cambridge}, LES \citep{Sagaut2006_springer} is a well established methodology to obtain a representation of the instantaneous flow dynamics with reduced computational costs. In the LES formalism, the equations are filtered so that the large scales of motion are directly simulated, while the effects of the small filtered scales are represented by a subgrid-scale model. This operation significantly reduces the number of degrees of freedom to be simulated, therefore reducing the computational burden. The filtered compressible flow variables are typically Favre averaged or density weighted $\widetilde{f} = \overline{\rho f}/\overline{\rho}$. The resulting equations, where the tilde symbol $\; \tilde{} \;$ indicates the filtered variables, are:  

\begin{eqnarray}
    \frac{\partial \overline{\rho}}{\partial t} + \frac{\partial (\overline{\rho} \widetilde{u}_j)}{\partial x_j} &=& 0, \label{eq:LESmass} \\
    \frac{\partial \overline{\rho} \widetilde{u}_i}{\partial t} + \frac{\partial (\overline{\rho} \widetilde{u_i} \widetilde{u}_j)}{\partial x_j} &=& - \frac{\partial \overline{p}}{\partial x_i} + \frac{\partial \overline{\sigma}_{ij}}{\partial x_j} - \frac{\partial \overline{\rho} \tau_{ij}}{\partial x_j}, \label{eq:LESmom}  \\
    \frac{\partial \overline{\rho} \widetilde{E}}{\partial t} + \frac{\partial ( \overline{\rho} \widetilde{E} +\bar p)\widetilde{u}_j}{\partial x_j}  &=& + \frac{\partial \widetilde{u}_i \overline{\sigma}_{ij}}{\partial x_j}  +\biggl(\overline{\kappa}\frac{\partial \widetilde{T}}{\partial x_j}\biggl) - \frac{\partial q_j^{SGS}}{\partial x_j}.\label{eq:LEStemp}
\end{eqnarray}

$\tau_{ij}= (\widetilde{u_i u_j} - \widetilde{u}_i \widetilde{u}_j$) is the subgrid scale stress tensor. In the Smagorinsky model \citep{Smagorinsky1963_mwr}, the deviatoric part of $\tau_{ij}$ is modelled as an eddy viscosity effect, for compressible flows :

\begin{equation}
\label{eq:tauSGS_LES}
    \tau_{ij} - \frac{1}{3} \tau_{kk} \delta_{ij} = -2 \nu_{SGS} \widetilde{S}_{ij},
\end{equation}

where $\nu_{SGS} = (C_S \Delta)^2 \sqrt{2 \widetilde{S}_{ij} \widetilde{S}_{ij}}$ is the subgrid scale viscosity and $\widetilde{S}_{ij} = \frac{1}{2} \left( \frac{\partial \widetilde{u}_i}{\partial x_j} + \frac{\partial \widetilde{u}_j}{\partial x_i} \right)$ is the rate-of-strain tensor of the resolved velocity field, $\Delta$ is the filter width and $C_S$ is a model coefficient that can be selected by the user. This formulation, which is derived from the asymptotic turbulence theory by Kolmogorov, fails to provide an accurate prediction of the interactions between the resolved and filtered physical variables. The reason is that the SGS stress tensor in equation \ref{eq:tauSGS_LES} is inherently dissipative and affects all the simulated scales of the flow \citep{Sagaut2006_springer}. $q_j^{SGS}=\overline{\rho E u_j} - \bar\rho \widetilde{E}\widetilde{u_j}$ is the unresolved flux of the total energy. A gradient approach is used to close this term following: $q_j^{SGS}=-\bar\rho \bar C_p \frac{\nu_{SGS}}{Pr_t} \frac{\partial \widetilde{T}}{\partial x_j}$, with the turbulent Prandtl number $Pr_t$ set to one.

The WALE subgrid scale model for \textit{Wall Adapting Local Eddy-viscosity} was developed to overcome the limitations of the Smagorinsky model, particularly in the proximity of immersed surfaces. ~\cite{nicoud1999subgrid} mention two main limitations of the Smagorinsky model, based on the second invariant of the symmetric part of $\widetilde{S}_{ij}$ (1): its formulation is only related to the rate-of-strain tensor and not to the rotation rate (2): the order of magnitude of the invariant near the wall is $\mathcal{O}(1)$. The goal of the WALE model is to reproduce the expected behaviour of the subgrid viscosity when approaching the wall $\nu_{SGS} = \mathcal{O}(y^3)$ using both the rate-of-strain tensor and the rotation rate. For this, the traceless symmetric part of the velocity gradient square tensor is used:

\begin{equation}
    S_{ij}^d = \frac{1}{2} \biggl(\frac{\partial \widetilde{u}_k}{\partial x_i} \frac{\partial \widetilde{u}_j}{\partial x_k} + \frac{\partial \widetilde{u}_k}{\partial x_j} \frac{\partial \widetilde{u}_i}{\partial x_k}   \biggl) - \frac{1}{3}\delta_{ij} \frac{\partial \widetilde{u}_k}{\partial x_m} \frac{\partial \widetilde{u}_m}{\partial x_k}. 
\end{equation}

The relation given by \citep{nicoud1999subgrid} can be expressed as :

\begin{equation}
    \label{chap2:eq:nusgs_long}
    \nu_{sgs} = \biggl(C_w^2 \Delta\biggl)^2 \frac{(S_{ij}^d S_{ij}^d)^{3/2}}{(\widetilde{S}_{ij}\widetilde{S}_{ij})^{5/2}+(S_{ij}^d S_{ij}^d)^{5/4}}   
\end{equation}

It allows the identification of turbulent structures with high rate-of-strain and/or rotation rates. The subgrid viscosity is also correctly evaluated when approaching the wall thanks to a normalization term used in the expression (details are given in \citep{nicoud1999subgrid}. The implementation of the method in OpenFOAM is based on the subgrid kinetic energy $\mathcal{K}_{SGS}$ :

\begin{equation}
    \label{eq:k_SGS}
    \mathcal{K}_{SGS} = \nu_{SGS}^2/C_k^2\Delta ^2.
\end{equation}

The width of the $\Delta$ filter is also associated with the width of the geometric filter $\Delta_g$, as for the Smagorinsky model. The values of the constants can be modified by the user, by default $C_k = 0.094$ and $C_w = 0.325$.

\subsection{Bayesian data assimilation}
\label{sec:EnKF}
Discussion in section \ref{sec:LES_equations} highlighted the limitations of numerical approaches in terms of accuracy. In fact, it was shown that turbulence closures introduce a bias due to modelling structural limitations. However, even for exact subgrid-scale modelling, the evolution of instantaneous flow fields can rapidly diverge due to uncertainties affecting measurements and calculations. Even minimal variations such as machine error in computations can affect the evolution of the instantaneous field, because of the strong non-linear, multiscale dynamics at high Reynolds regimes \citep{Ge_Rolland_Vassilicos_2023}. While such variations do not usually impact statistical features, they may produce desynchronization of the flow, so that observation of instantaneous features is not reliable.
Bayesian Data Assimilation can be used to resolve such issue. Here, uncertainties are naturally accounted for in the process of state estimation and optimization, providing support to the establishment of connections between the physical and digital twins.     
Among these DA techniques, the Kalman filter (KF) is a well-known DA tool first introduced in 1960 by \cite{Kalman1960_jbe} to estimate an augmented system state by combining model prediction with high-fidelity but sparse external data observation. Such DA tool is able to reconstruct a state estimation combining results from the two sources of information, as well as optimizing free parameters of the model so that its predictive capabilities are augmented. One notable feature of the KF is that both sources of information are considered to be affected by uncertainties, which are modelled as Gaussian random variables. The model uncertainty $\boldsymbol{\eta}$ as well as the observation uncertainty $\boldsymbol{\epsilon}$ are therefore defined as follows:
\begin{eqnarray}
\boldsymbol{\eta}_{k+1} \sim \mathcal{N}(0,\mathbf{Q}_{k+1}), \\
\boldsymbol{\epsilon}_{k+1} \sim \mathcal{N}(0,\mathbf{R}_{k+1}).
\end{eqnarray}

The matrices $\mathbf{Q}_{k+1}$ and $\mathbf{R}_{k+1}$ are the model and the observation covariance matrices, respectively. The Kalman filter operates via  a sequential algorithm which combines model prediction and observation via a minimization of an error covariance matrix for consecutive state updates. The error covariance matrix $\mathbf{P}_{k+1}$ can be defined for a time $k = 0,1,..., k_f$ as:

\begin{equation}
\label{eqn:chap3:P}
    \mathbf{P}_{k+1}^{f/a} = \mathrm{cov}(\mathbf{e}_{k+1}^{f/a}) = E\biggl[\mathbf{e}_{k+1}^{f/a}(\mathbf{e}_{k+1}^{f/a})^\mathrm{T}\biggl],
\end{equation}

with 

\begin{equation}
    \mathbf{e}_{k+1}^{f/a} = \mathbf{u}_{k+1}^{f/a} - \mathbf{u}_{k+1}^{t},
\end{equation}

where the exponent $t$ (\textit{truth}) indicates the true state of the physical system. The superscript $f$ (\textit{forecast}) refers to the forecast of the state matrix obtained via the temporal advancement of the model. It is therefore the predicted state \textit{before} observation is assimilated. The superscript $a$ (\textit{analysis}) represents the final augmented state of the algorithm following the \textit{analysis} phase, therefore \textit{after} DA is performed. The main drawback of the classical Kalman filter consist in the costly time advancement (for the forecast) and manipulations (for the analysis) of the covariance matrix $\mathbf{P}$. For a classical CFD problem for turbulent flows, the state variables calculated on the grid may easily consist of $N \in 10^6-10^{12}$ degrees of freedom. The size of the error covariance matrix is therefore $N\times N$, demanding prohibitive resources for its manipulation.

A strategy extensively analysed in the literature to palliate this issue is the Ensemble Kalman Filter (EnKF) \citep{Evensen2009_IEEE}. This technique has been extensively used in the field of meteorological sciences (see \cite{Asch2016_SIAM}) and more recently in fluid mechanics applications \citep{Mons2016_jcp,LeProvost2021_prf,Zhang2024_jcp,Moldovan2022_cf}. Here, a Monte-Carlo approximation is used to estimate the error covariance matrix $\mathbf{P}$ using a set of pseudo-random realizations. A simplified scheme of the EnKF procedure used in this study is presented in Fig. \ref{fig:EnKF_plot} and a detailed algorithm is provided in Alg. \ref{alg:HLEnKF}. The main elements of the strategy are now discussed. The \textit{forecast} step consists in the time advancement of $N_e$ realizations (ensemble members) using the available numerical model between successive assimilation phases. This step is exemplified by the gray lines in Fig. \ref{fig:EnKF_plot}. If each ensemble member is described by $N$ degrees of freedom, a state matrix $\boldsymbol{\mathcal{U}}$ of size $[N, \, N_e]$ can be assembled at the beginning of each \textit{analysis} phase. Each column of such matrix $i = 1, \cdots, N_e$ corresponds to a physical state $\mathbf{u}_i$ obtained by the $i^{th}$ member of the set. 
The EnKF manipulates the state matrix to provide an ensemble representation of the error covariance matrix $\mathbf{P}$, which is referred to as $\mathbf{P}_e$, via the assumption that the ensemble members are statistically  independent:
\begin{figure}
\centering
\includegraphics[width=0.9\textwidth]{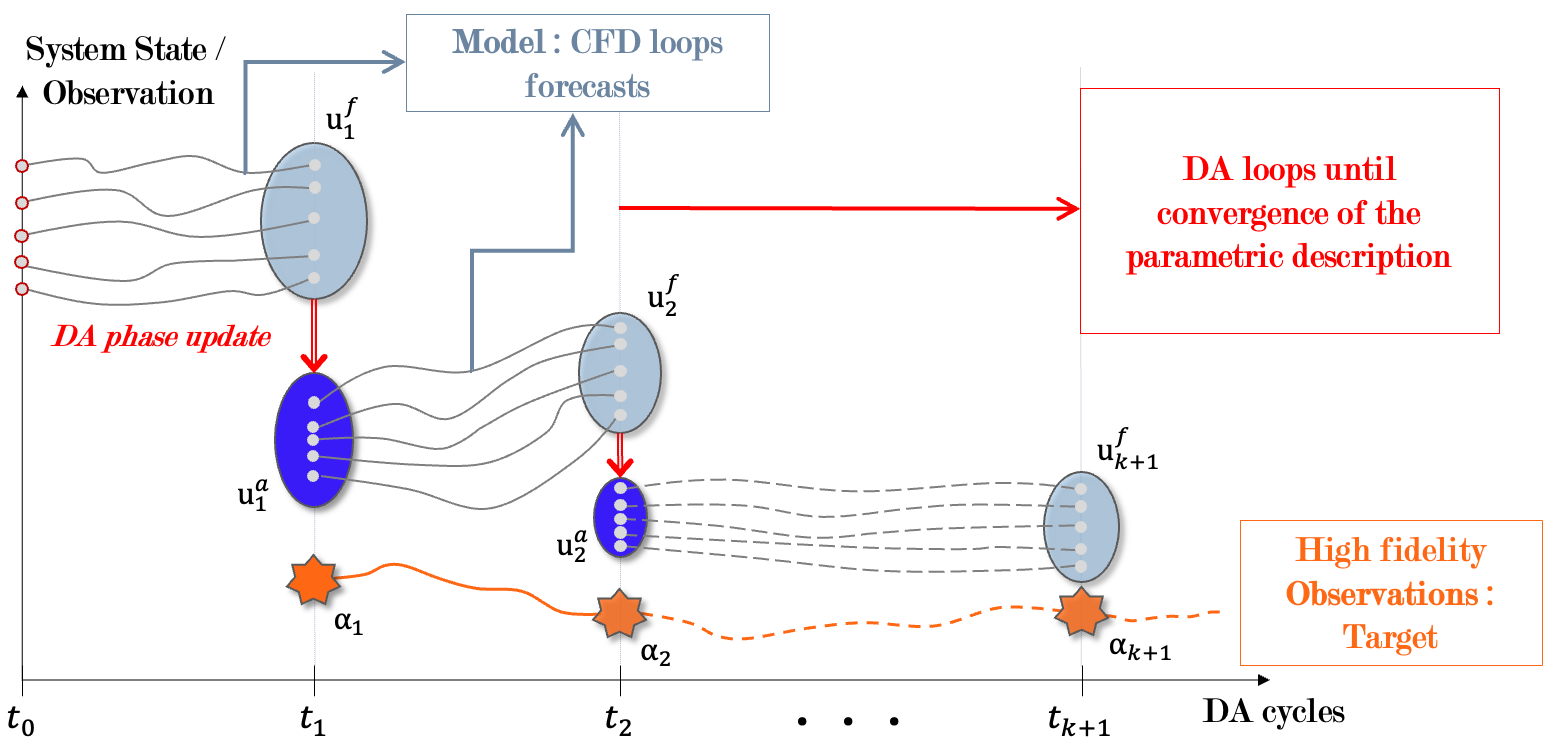}
\caption{Scheme  representing the ensemble Kalman filter. The state of the model $u_k^f$ is updated via the \textit{analysis} phase of the EnKF algorithm by taking into account the observation $\alpha_k$. Each gray line represents the trajectory of a member of the ensemble.} 
\label{fig:EnKF_plot} 
\end{figure} 

\begin{equation} 
\mathbf{P}_e^{f/a} = \frac{1}{N_e-1}\sum_{i=1}^{N_e}(\mathbf{u}_i^{f/a} - \langle \mathbf{u}^{f/a} \rangle)(\mathbf{u}_i^{f/a} - \langle \mathbf{u}^{f/a} \rangle)^\mathrm{T} =\mathbf{\Gamma}\mathbf{\Gamma}^\text{T}, 
\end{equation}

where $\mathbf{\Gamma}$ is the anomaly matrix, which is derived from the state matrix $\boldsymbol{\mathcal{U}}$. It quantifies the deviation of the state vectors from their ensemble mean:

\begin{equation}
[\mathbf{\Gamma}_{k+1}]_i = \frac{\mathbf{u}_{i,k+1}^f-\langle\mathbf{u}\rangle_{k+1}^f}{ \sqrt{N_e-1}} \; , \qquad \langle\mathbf{u}\rangle_{k+1}^f = \frac{1}{N_e}\sum_{i = 1}^{N_e}\mathbf{u}_{i,k+ 1}^f  \;.
\end{equation}

The anomaly matrix plays an essential role in the state estimation. In particular, the Kalman gain $\mathbf{K}$, which governs the state update of the model prediction, is obtained via manipulation of $\mathbf{\Gamma}$ and information obtained at sensors. For the latter, in order to grant an observation matrix with the same rank as the state matrix $\boldsymbol{\mathcal{U}}$, the array including the $N_o$ available observations at time $k$ is artificially perturbed to obtain $N_e$ sets of values. More precisely, a matrix of size $N_o \times N_e$ is obtained adding a bounded Gaussian noise based on the measurement error covariance matrix $\mathbf{R}_{k+1}$ to the observation vector $\boldsymbol{\alpha}_{k+1}$. Each column $i$ of such matrix  is:
\begin{equation}
\label{eqn:obs}
\boldsymbol{\alpha}_{i,k+1} = \boldsymbol{\alpha}_{k+1} + \mathbf{\epsilon}_{i,k+1},\; \text{with} \; \mathbf{\epsilon}_{i,k+1} = \mathcal{N}(0, \mathbf{R}_{k+1}).
\end{equation}

The last element of information needed to construct the Kalman gain $\mathbf{K}$ deals with the projection of the model prediction in the space of the sensors where observation is sampled. This operation is performed via the projection matrix $\mathbf{H}$:
\begin{equation}
\label{eqn:EnKF_Projobs}
\mathbf{s}_{i,k+1} = \mathbf{H} \mathbf{u}_{i,k+1}^f \; .
\end{equation}
In practice, $\mathbf{H}$ connects the variables calculated by the model (velocity field, pressure field...) with the measured quantities which are used as observation. Therefore, $\mathbf{H}$ can for example be an interpolation tool (e.g. to obtain pressure values in the location of the sensors) or an integration operator (e.g. the observed quantity is the drag coefficient of an immersed body).  
The state estimation via EnKF is obtained as a combination of model prediction during the forecast step and the available observation. The augmented state $\mathbf{u}_{i,k+1}^a$ is thus expressed as :

\begin{equation}
\label{eqn:state_estimation}
\mathbf{u}_{i,k+1}^a = \mathbf{u}_{i,k+1}^f + \mathbf{K}_{k+1}(\boldsymbol{\alpha}_{i,k+1}-\mathbf{s}_{i,k+1}).
\end{equation}

The way these two predictions are combined is governed by the Kalman gain $\mathbf{K}_{k+1}$, here expressed at the analysis phase $k+1$~\citep{Asch2016_SIAM,carrassi2018_wcc}:

\begin{equation}
    \label{eqn:EnKF_gain_R}
    \mathbf{K}_{k+1} = \mathbf{\Gamma}_{k+1}(\mathbf{S}_{k+1})^\text{T} \left[\mathbf{S}_{k+1}(\mathbf{S}_{k+1})^\text{T} + \mathbf{R}_{k+1}\right]^{-1},\\
\end{equation}
with
\begin{eqnarray}
\left[\mathbf{S}_{k+1}\right]_i= \frac{\mathbf{s}_{i,k+1}-\langle\mathbf{s}\rangle_{k+1}}{\sqrt{N_e-1}} \; , \qquad \langle\mathbf{s}\rangle_{k+1} = \frac{1}{N_e}\sum_{i = 1}^{N_e}\mathbf{s}_{i,k+1} \; ,
\end{eqnarray}



EnKF-based approaches can also simultaneously optimize the free parameters governing the model to minimize the discrepancy between the model prediction and the obtained state estimation during the \textit{analysis} phase. These free parameters, usually found in boundary conditions and / or turbulence closures, are stored in an array which is referred to as $\boldsymbol{\gamma}$. A simple strategy to perform such an optimization relies on the definition of an \textit{extended state} \citep{Asch2016_SIAM}. Here, the EnKF problem is solved for a state vector $\mathbf{{u}_{ext}}$ defined as:
\begin{equation}
\mathbf{{u}_{ext}} = \begin{bmatrix}
\mathbf{u}. \\
\boldsymbol{\gamma}
\end{bmatrix}
\end{equation}

The size of the extended state is now equal to $N_{ext}=N + N_\gamma$, where $N_\gamma$ is the number of parameters to be optimized. This modification of the DA strategy results in a negligible increase in computational cost if $N_\gamma << N$ and simultaneously provides an updated state estimation and an optimized parametric description for the model at the end of the \textit{analysis} phase.

\subsubsection{Hyperparameters of the EnKF: Inflation}
\label{sec:inflation}
One of the major drawbacks of the Ensemble Kalman Filter is the rapid collapse of the state matrix variability. The consequence of the undesirable reduction of the variance is the convergence of the state matrix towards a localized optimum, strongly related to the provided prior state. If the latter is not well identified, which is usually the case due to lack of information, the accuracy of the optimization via the EnKF can be seriously affected. It is possible to increase the global variability of the system and to decrease the sampling errors by using a higher number of members in the ensemble, thus gaining accuracy in the EnKF prediction. However, this strategy is not conceivable for fluid dynamics applications where computational costs preclude the use of large ensembles. In fact, the number of members typically used for three-dimensional analyses is about $N_e \in [40,100]$ \citep{Mons2021_prf, Moldovan2022_cf}, which is arguably not sufficient to ensure statistical convergence.

\begin{figure}[h]
    \centering
    \includegraphics[width=0.7\textwidth]{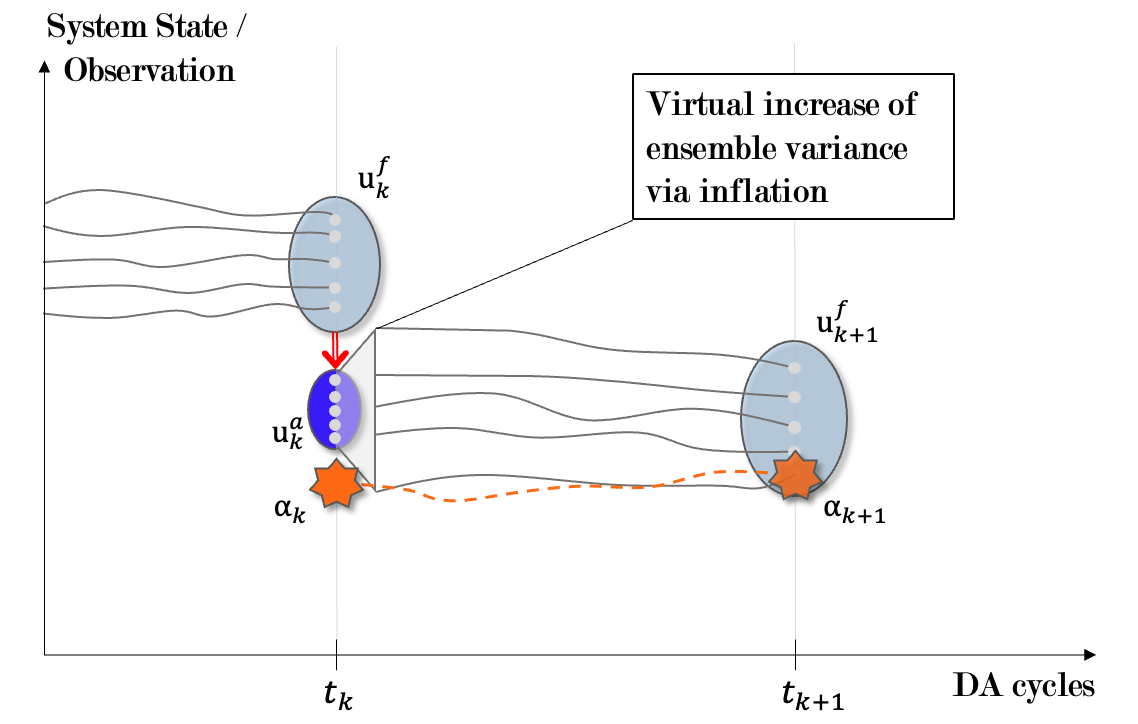}
    \caption{Scheme representing the inflation procedure used for the ensemble Kalman filter. Each gray line represents the trajectory of one member of the ensemble. The increase in the variance of the ensemble via inflation is represented by the gray trapezoid.}
    \label{fig:EnKF_inflation_plot}
\end{figure}

This problem is usually mitigated by inflating the variance of the ensemble \textit{after} the \textit{analysis} phase, as illustrated in Fig. \ref{fig:EnKF_inflation_plot}. This can be achieved by increasing the discrepancy between each state vector $\mathbf{u}_{i,k+1}^a$ and the ensemble mean $\langle\mathbf{u}^a\rangle$ by algebraic operations driven by a coefficient $\lambda$. This procedure is called \textit{multiplicative inflation}. The way this procedure is performed can be \textit{deterministic} or \textit{stochastic}:

\begin{equation}
deterministic \qquad \mathbf{u}_{i}^a \longrightarrow \langle\mathbf{u}^a\rangle + \lambda_i(\mathbf{u}_{i}^a-\langle\mathbf{u}^a\rangle) \qquad \text{with} \; \lambda_i > 1,
\end{equation}

\begin{equation}
\label{eq:stochastic_infl}
stochastic \qquad \mathbf{u}_{i}^a \longrightarrow (1+\lambda_i) \mathbf{u}_{i}^a \qquad \text{with} \; \lambda_i \thicksim \mathcal{N}(0,\sigma).
\end{equation}

The deterministic implementation is usually very efficient in the initial \textit{analysis} phases of the DA run. Considering that it is applied to the discrepancy of each ensemble member from the ensemble mean, this inflation is usually stable and higher values of $\lambda$ can be used. However, it is less efficient when the ensemble exhibits a strong collapse of the variability of the physical solution ($\mathbf{u}_{i}^a-\langle\mathbf{u}^a\rangle \approx 0$). On the other hand, stochastic inflation is efficient to avoid collapse of the variability once the EnKF converges to a stable solution. The Gaussian distribution used to determine $\lambda_i$ in Eq. \ref{eq:stochastic_infl} is usually truncated to avoid the generation of outliers that could lead to divergence of the EnKF.

\subsubsection{Hyperparameters of the EnKF: Localization}
When CFD solvers are used as models for DA procedures, the state matrix is composed of values of the flow variables (e.g. the velocity field) in correspondence of the mesh elements, usually at their centre. As shown in Eq. \ref{eqn:EnKF_gain_R}, the Kalman gain correlates these values with the values of the state matrix projected into the observation space, i.e. over the sensors at which observation is available. Considering that the correlation between flow variables  decreases with distance in continuous systems, the approximations used to determine the Kalman gain in the EnKF framework and in particular sampling errors can lead to non-physical correlations in the analysed state matrix, especially for large domains. These effects can be responsible for critical issues such as unphysical flow fields leading to the divergence of the filter. Again, these problems can be reduced by increasing the number of ensemble members therefore reducing sampling errors, which is not a viable solution for CFD applications. Therefore, different strategies must be envisioned to mitigate the effects of spurious distant correlations. Two approaches are classically adopted in the literature \citep{Asch2016_SIAM}. The \textit{covariance localization} operates on the coefficients of the error covariance matrix $\mathbf{P}_{k+1}^f$, aiming to zero values for model realizations sufficiently far from each sensor. This process is here mathematically performed by a Shur product between a localization matrix $\mathbf{L}$ and the Kalman gain $\mathbf{K}$. This expression can be directly included in the algorithm without any modification:

\begin{equation}
{\mathbf{K}_{k+1}^{loc} = [\mathbf{L}]_{m,n}[\mathbf{K}_{k+1}]_{m,n}}.
\end{equation}

The structure of the matrix $\mathbf{L}$ must be defined by the user. For fluid mechanics problems, the correlation decreases rapidly in space. Therefore, a commonly used structure for the localization matrix is an exponential decay form:

\begin{equation}
\label{eqn:loc_matrix_v2}
\mathbf{L}(m,n) = e^{-\frac{1}{2}(\Delta_{m,n}/l)^2},
\end{equation}

where $\Delta_{m,n}$ is the distance between the given sensor and the centre of the grid element. $l$ is a correlation length scale that can be adjusted according to the local characteristics of the test case.

Another way to localize the Kalman gain is to use the \textit{physical localization}, also called \textit{local analysis} \citep{Evensen2022_book}. Instead of running the EnKF on the entire physical domain, the computation is performed on a \textit{reduced} domain. The reduced space contains all the sensors used for the sampling of observation. This strategy also has the advantage of reducing the number of degrees of freedom operating in the assimilation procedure, which can lead to a significant gain in terms of computational resources required. \textit{Covariance localization} is commonly used with \textit{physical localization} to avoid discontinuities due to the DA state update, in particular at the interface of the reduced domain. This method is very effective in simultaneously speeding up the computation and improving the computational stability and accuracy for a reduced ensemble size, such as those currently usable for CFD-based studies \citep{Villanueva2023_cf, Villanueva2024_ijhff}. The localization methods described are qualitatively illustrated in Fig. \ref{fig:methodes_loc}.
\subsubsection{Sub-setting methods}
\label{sec:LETKF}
\begin{figure}
    \centering
    \includegraphics[width=0.8\textwidth]{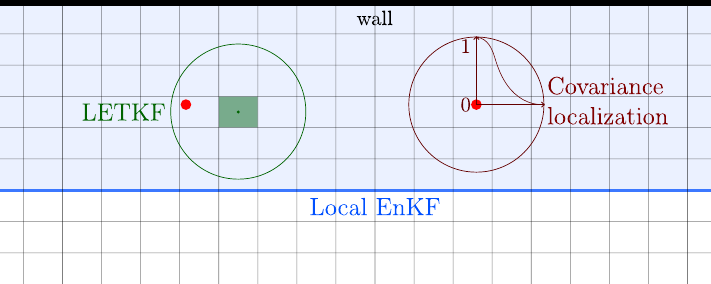}
    \caption{Simplified representation of different DA localization methods on a Cartesian grid. The red dots represent sensors for observation. In blue: the EnKF is executed on a reduced domain. In green: application of the LETKF for one grid element. In dark red: covariance localization is used to control the intensity of the DA state update in a given region. }
    \label{fig:methodes_loc}
\end{figure}

A different way to perform the localization procedure is used in the \textit{Local Ensemble Transform Kalman Filter} (LETKF) \citep{Hunt2007_physD}. The method consists in performing one DA procedure per cell in the domain simulated by the model. 
Let us consider one grid element for which the model prediction is available, generating a subset of the global state matrix $\boldsymbol{\mathcal{U}}_{[g]}$. 
A subset of the observation vector $\boldsymbol{\alpha}_{[g]}$ is also generated including observations which are within a prescribed distance from the grid element. Thus, the LETKF update is a sum of $N_g$ local updates:

\begin{equation}
    (\mathbf{u}_{i,k+1}^a)_{[g]} = \sum_{s=1}^{N_g}  \left(\langle\mathbf{u}\rangle_{k+1}^f + \sqrt{N_e-1}[\mathbf{\Gamma}_{k+1}]_i w_i^{a}\right)_{[s]},
\end{equation}

with $N_g$ the number of grid elements and $w_i^a$ a weight vector belonging to the ensemble space spanned by the forecast ensemble perturbations (see ~\cite{Hunt2007_physD} for more details). In practice, the LETKF performs one  EnKF for each grid element, but the size of the state matrix is significantly reduced. Also, the observations considered for each realization of the LETKF are only the ones included in the sub-volume considered. Despite the increase in the number of DA phases performed, the global computational costs are diminished because each of them is performed over a significantly smaller space and with fewer sensors. However, it has to deal with complex management of the discontinuities for neighbour cells using different sets of observations. The choice of the selected $N_g$ cells for each iteration also requires a more complex set of hyperparameters. The method is qualitatively shown in Fig. \ref{fig:methodes_loc}.\newline

Another method presented in the literature, applied to atmospheric data assimilation, is the \textit{Sequential Ensemble Kalman Filter} (SEnKF)~\citep{houtekamer2001_mwr}. The SEnKF uses a sub-setting method within the observation space. In this procedure, observations are grouped together when sensors are very close and the correlation of their measurements is known. The analysis phase becomes here sequential, performing one update per observation batch. 
The final state is obtained as the sum of state updates provided by each of the $N_b$ observation batches available : 

\begin{equation}
        \mathbf{u}_{i,k+1}^{a_{N_b}} = \mathbf{u}_{i,k+1}^f + \sum_{j=1}^{N_b} \mathbf{K}_{k+1}^{a_{j}}(\boldsymbol{\alpha}_{i,k+1}^{a_{j}}-\mathbf{s}_{i,k+1}^{a_{j}}).
\end{equation}

The SEnKF is conceived to work efficiently with very large number of sensors, avoiding the problem of computationally expensive matrix inversions. The procedure is not acting as a physical localization method, even though it could be coupled to standard localization procedures. However, It can be conceived as a procedure localizing the observations and their correlations. 

\subsection{Hyper-Localized Ensemble Kalman Filter : HLEnKF}
\label{sec:HLEnKF}
\begin{figure}
    \centering
    \includegraphics[width=0.8\textwidth]{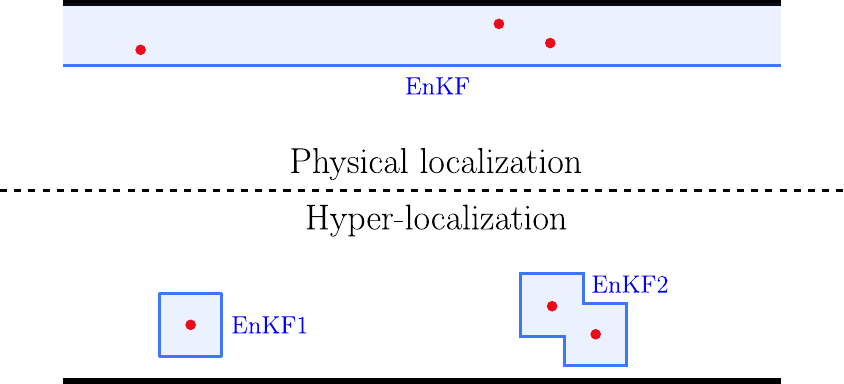}
    \caption{Scheme representing the principle of hyper-localization. The red dots represent sensors for observation. A comparison is shown with classical physical localization.}
    \label{fig:hyper-localization}
\end{figure}

A localization procedure for the EnKF is here proposed, which is inspired by the LETKF and by the SEnKF previously introduced. The \textit{Hyper-localized} Ensemble Kalman Filter (HLEnKF) consists of $N_s$ sequential runs of a classical EnKF, where $N_s$ is tied to the number of sensors used in the DA procedure. More precisely, regions around the sensors are identified and physically localized EnKF are performed. In order to avoid discontinuity in the physical state, covariance localization is also applied to reduce the DA state update moving towards the external surfaces of the physical regions investigated. To this purpose, Eq. \ref{eqn:loc_matrix_v2} is used, tuning the free parameters so that the state update corresponds to zero at the interface of the region. The corresponding state update of the physical field can be written as:

\begin{equation}
    \mathbf{u}_{i,k+1}^a = \mathbf{u}_{i,k+1}^f + \sum_{s=1}^{N_s} (\mathbf{K}_{k+1})_{[s]}((\boldsymbol{\alpha}_{i,k+1})_{[s]}-(\mathbf{s}_{i,k+1})_{[s]}),
\end{equation}
where the subscript $s$ indicates the iterations over the considered regions and $[s]$ indicates the calculation of the ingredients of the EnKF ($\mathbf{K}$, $\boldsymbol{\alpha}$ and $\mathbf{s}$) in each of the $N_s$ selected physical regions. One can see a strong similarity with the LETKF, as a large number of smaller EnKF are here performed. However, the selection for the HLEnKF is performed creating regions around sensors for observation, and not based on each grid element of the model. This latter strategy is reminiscent of the SEnKF, even if for the HLEnKF the state update is performed on the model space and not on the observation space. The number of regions $N_s$ is now discussed. If sensors are physically far one from the other, then $N_s$ is equal to the number of sensors, i.e. each DA physical region is centred around a sensor. This choice precludes the possibility to include the correlation between the observations measured at different sensors. Arguably, this apparent drawback can actually help to stabilize the performance of the filter, as the correlation between variables measured at distant locations can be affected by large errors. In case some sensors are clustered in a physical region, the algorithm is able to combine regions identified around one sensor and to perform one EnKF using a batch of close observations. A qualitative comparison between classical physical localization and hyper-localization is shown in figure \ref{fig:hyper-localization}. One key element governing the efficiency of this approach is clearly the size of the DA region around the sensors. A small region grants reduced computational resources and exhibits smaller sampling errors. This approach becomes a nudging approach if the region considered consists of the location of the sensor. On the other hand, large regions allow the DA state update to influence a greater portion of the domain, thereby accelerating parametric optimization and enhancing flow synchronization. In the present work, the regions are selected to be of spherical form around each sensor. The diameter of such spheres is proportional to the integral length scale of the flow. More details about this important aspect are provided in Sec. \ref{sec:localSynchronization}.

At last, the parametric optimization procedure is discussed. The HLEnKF provides $N_s$ values for each optimized coefficient. In the present work, global values are obtained averaging the $N_s$ optimized coefficient by each local EnKF. This strategy, which was extensively investigated in preliminary studies \citep{villanueva2024_phd}, shows a robust convergence of the parametric optimization at the expense of a slightly slower convergence rate. Another notable advantage of the HLEnKF procedure is the computational efficiency. The usage of localized regions for the DA procedure permits the manipulation of smaller arrays and matrices, reducing the total computational costs while preserving similar levels of accuracy when compared with classical implementation of the EnKF \citep{villanueva2024_phd}. This is particularly true for physical systems described by numerous degrees of freedom, such as the numerical simulations performed in this work.  

\begin{algorithm}
    \caption{Algorithm of the Hyper-localized Ensemble Kalman Filter (HLEnKF)}
    \label{alg:HLEnKF}
    \textbf{Input:} $\mathbf{M}$, $\mathbf{H}$, $\mathbf{R}_{k+1}$, and prior distributions for the extended state vectors $\mathbf{u}_{i,0}^a$, where $\mathbf{u}_{i,0}^a \sim \mathcal{N}(\mu_N, \sigma_N^2)$, final time $k_f$, number of members $N_e$ and number of hyper-localized sub-regions $N_s$. $\mathbf{D}$ and $\mathbf{G}$ are respectively subset selection operators for state and observation. \\
    \For{$k = 0$ to $k_f-1$}{
        \textcolor{white}{.}\\
        \For{$i = 1$ to $N_e$}{
            \nl Time advancement of state vectors:\\
            \qquad $\mathbf{u}_{i,k+1}^f = \mathbf{M}\mathbf{u}_{i,k}^a$ \\
        }
        \For{$j = 1$ to $N_s$}{
            \textcolor{white}{.}\\
            \For{$i = 1$ to $N_e$}{
                \nl Selection of state and observation subset:\\
                \qquad $\mathbf{u}_{i,j,k+1}^f = \mathbf{D}_j\mathbf{u}_{i,k+1}^f$ and $\boldsymbol{\alpha}_{j,k+1} = \mathbf{G}_j\boldsymbol{\alpha}_{k+1}$\\
                \nl Creation of the observation matrix from observation data by introducing errors:\\
                \qquad$\boldsymbol{\alpha}_{i,j,k+1} = \boldsymbol{\alpha}_{j,k+1} + \mathbf{\epsilon}_{i,j, k+1}$, with $\mathbf{\epsilon}_{i,j, k+1} \thicksim \mathcal{N}(0,\mathbf{R}_{j,k+1})$\\
                \nl Projection of the model solution onto the observation space:\\
                \qquad$\mathbf{s}_{i,j,k+1} = \mathbf{H}_j\mathbf{u}_{i,j,k+1}^f$\\
            }
            \textcolor{white}{.}\\
            \nl Calculation of ensemble averages:\\
            \qquad$\langle\mathbf{u}\rangle_{k+1}^f = \frac{1}{N_e}\sum_{i = 1}^{N_e}\mathbf{u}_{i,j,k+1}^f$,\,
            $\langle\mathbf{s}\rangle_{k+1} = \frac{1}{N_e}\sum_{i = 1}^{N_e}\mathbf{s}_{i,j,k+1}$,\\
            \textcolor{white}{.}\\
            \For{$i = 1$ to $N_e$}{
                \nl Calculation of anomaly matrices:\\
                \qquad$[\mathbf{\Gamma}_{k+1}]_{i} = \frac{\mathbf{u}_{i,j,k+1}^f-\langle\mathbf{u}\rangle_{k+1}^f}{\sqrt{N_e-1}}$,\,
                $[\mathbf{S}_{k+1}]_{i} = \frac{\mathbf{s}_{i,j,k+1}-\langle\mathbf{s}\rangle_{k+1}}{\sqrt{N_e-1}}$,\\
            }
            \textcolor{white}{.}\\
            \nl Calculation of Kalman gain:\\
            \qquad$\mathbf{K}_{k+1} = \mathbf{\Gamma}_{k+1}(\mathbf{S}_{k+1})^\text{T} \left[\mathbf{S}_{k+1}(\mathbf{S}_{k+1})^\text{T} + \mathbf{R}_{j,k+1}\right]^{-1}$\\
            \nl Localization of the Kalman gain (Schur product):\\
            \qquad$\mathbf{K}_{k+1}^{loc} = [\mathbf{L}]_{m,n}[\mathbf{K}_{k+1}]_{m,n}$\\
            \textcolor{white}{.}\\
            \For{$i = 1$ to $N_e$}{
                \nl Subset state matrix Update:\\
                \qquad$\mathbf{u}_{i,j,k+1}^a = \mathbf{u}_{i,j,k+1}^f + \mathbf{K}_{k+1}^{loc}(\boldsymbol{\alpha}_{i,j,k+1}- \mathbf{s}_{i,j,k+1})$\\
                \nl Subset state matrix inflation:\\
                \qquad$\mathbf{u}_{i,j,k+1}^a = (1+\lambda_i)\mathbf{u}_{i,j,k+1}^a$\\
                \nl Update of the global state matrix:\\
                \qquad $\mathbf{u}_{i,k+1}^a = \mathbf{D}^{-1}_j\mathbf{u}_{i,j,k+1}^a$\\
            }
            \textcolor{white}{.}\\
        }
        \textcolor{white}{.}\\
        \For{$i = 1$ to $N_e$}{
        \nl Calculation of the average of the parameters update of each sub-region : \\
        \qquad $\overline{\gamma_{i}} = \frac{1}{N_r}\sum_{j=1}^{N_r} \gamma_{i,j}$
        }
        \textcolor{white}{.}\\
        \textcolor{white}{.}\\
    }
\end{algorithm}

\subsection{CONES: \textit{Coupling OpenFOAM with Numerical EnvironmentS}}
\begin{figure}
    \centering
    \includegraphics[width=1\textwidth]{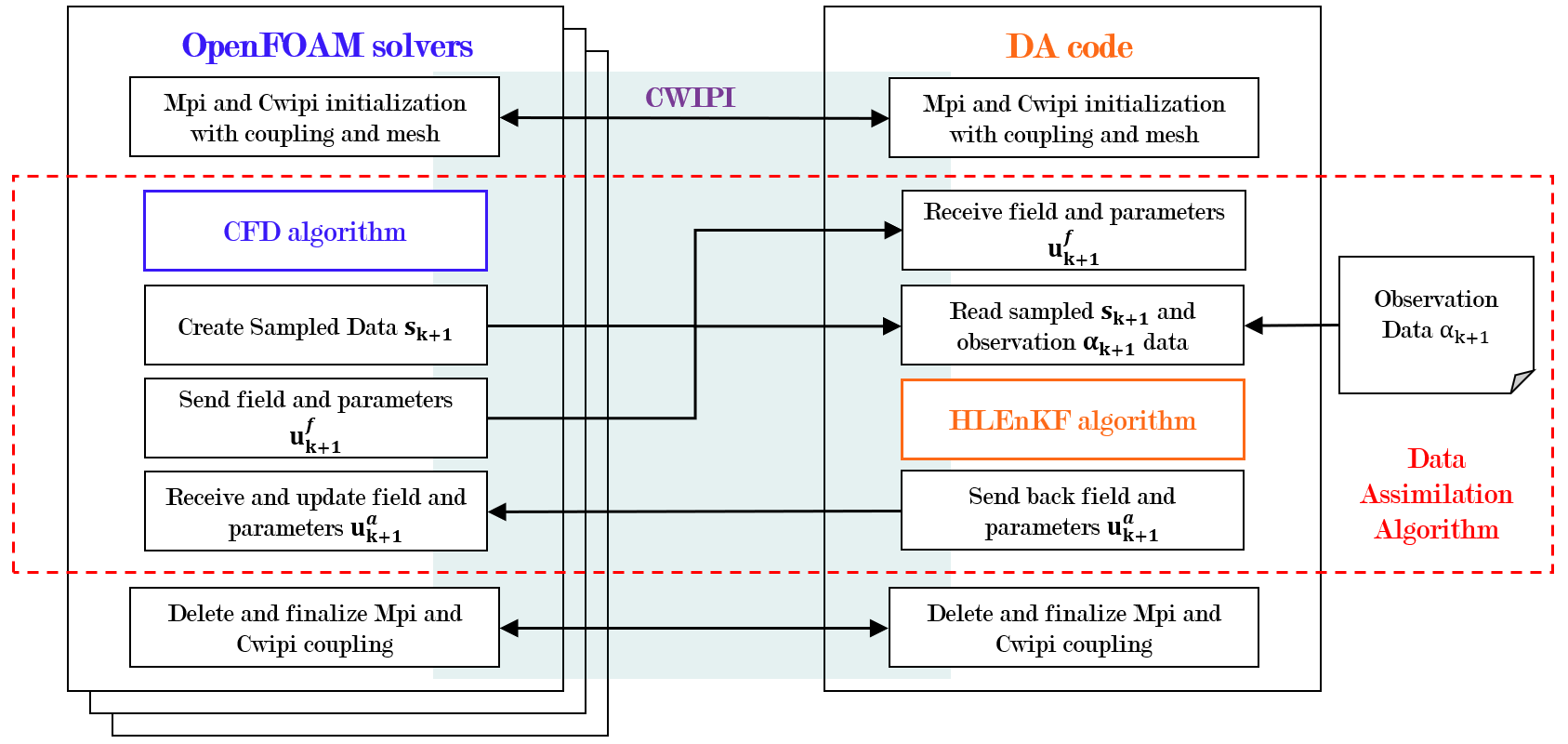}
    \caption{Representation of the different tasks of the library CONES.}
    \label{fig:CONES_scheme}
\end{figure}

CONES (\textit{Coupling OpenFOAM with Numerical EnvironmentS}) is a DA library developed for online applications. It is tailored to perform online coupling between results from numerical calculations performed with the code OpenFOAM and available observation. The latter can be obtained from available repositories as well as from sensors sampling data in real time. The tool relies on an existing coupler, called CWIPI, developed by CERFACS and ONERA French laboratories \citep{Reflox2011_aerlab}. Data Assimilation using an ensemble sequential algorithm requires multiple online simultaneous calculations as well as the processing of their results in the EnKF algorithm. CONES is able to transfer complete physical fields, such as the velocity field, between the OpenFOAM runs and the DA code. The ensemble CFD computations are online and they are paused during data transfer and DA analysis. This implies that stop \& restart of the numerical simulations is avoided, greatly enhancing the efficacy of the procedure in computational terms. Figure \ref{fig:CONES_scheme} summarizes the operations of CONES and in particular the coupling performed by CWIPI. 

In this analysis, the HLEnKF algorithm, which is computationally performed by one core, is coupled with $N_e=35$ CFD realizations (ensemble members). Each simulation is performed using 52 cores, for a total of 1821 cores used for each DA experiment. The steps of the algorithm presented in section \ref{sec:EnKF} are implemented in this way within CONES:
\begin{itemize}
\item First, the \textit{forecast} step is performed by the CFD solver until the time of the first analysis phase. This step includes the time advancement of every simulation of the ensemble.
\item The second step corresponds to the beginning of the DA analysis phase. $N_e$ velocity fields are sent to the DA code by the ensemble members. The information is assembled to form a single large matrix (state matrix using the DA formalism). In addition, the same physical field is interpolated to the coordinates of the sensors used for the observations, in order to obtain the term $s_{k+1}$ in equation \ref{eqn:state_estimation} for each ensemble member. 
\item Similarly, the set of parameters to be optimized, which are included in the extended state matrix, is sent using MPI primitives. Several functions in CONES handle the exchange of different parameters (model constants, boundary conditions...).
\item Observation vectors are prepared. In the present analysis, data is read from the files in the available database.
\item The analysis phase is performed, including the run of the hyper-localized ensemble Kalman filter and the update of the extended state matrix.
\item Lastly, the information stored in the extended state matrix is sent back to the ensemble members and the physical fields are updated. Calculations will resume until the next analysis phase till the end of the DA experiment.
\end{itemize}

When finalizing the computation, the couplings must be completed and the CWIPI and then MPI environments must be finalized.

\section{Test case setup}
\label{sec:setup}

The flow rig test case is now described. The present DA investigation completely relies on numerical tools and extended discussion about the choices performed for the model and the observation will be provided.

\subsection{Validation for a steady flow configuration}
\begin{figure}
    \centering
    \includegraphics[width=1\textwidth]{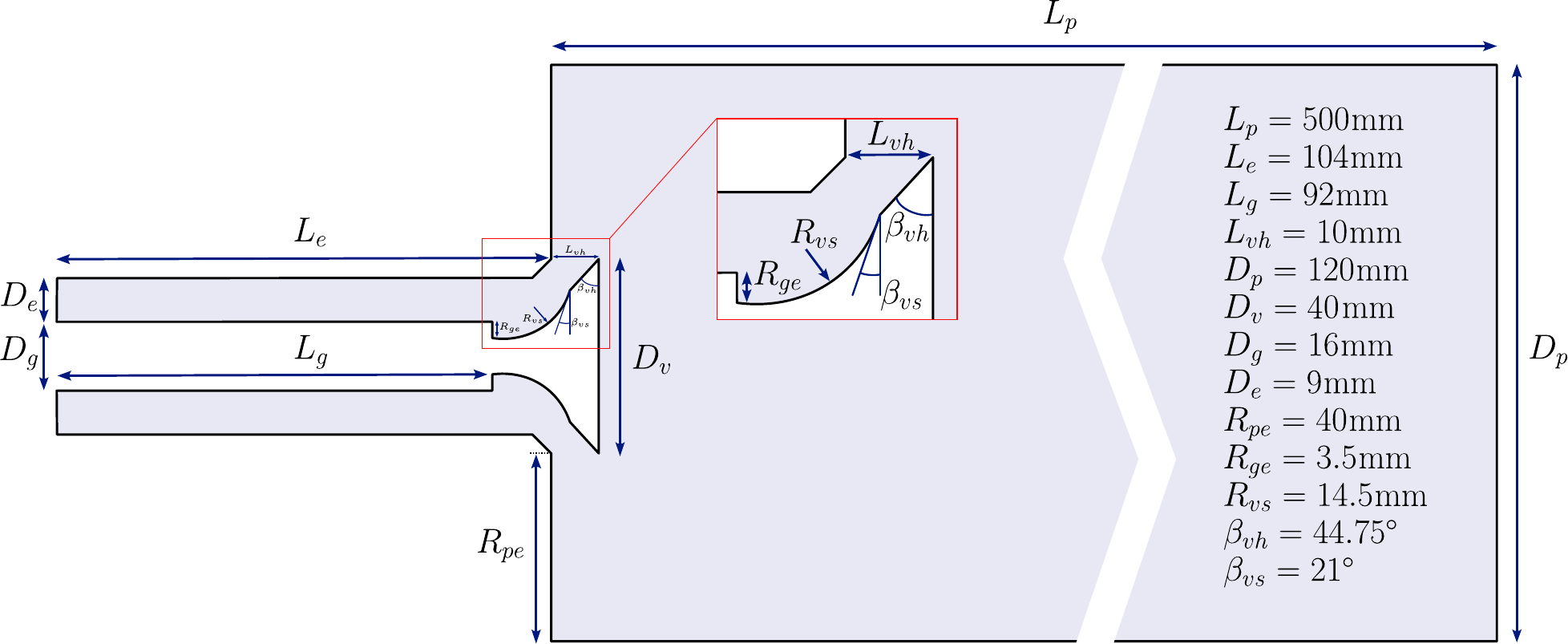}
    \caption{Geometric features of the steady-state flow rig test case.}
    \label{fig:OFR_scheme}
\end{figure}

The physical domain investigated is a simplified engine geometry referred to as the flow rig. This configuration has been previously investigated in several numerical studies, in particular for Large Eddy Simulation \citep{Afailal2019_ogst,Nicoud2018_phd,Thobois2004_saefl,Graftieaux2001_mst} and it is inspired by the asymmetric expansion studied by \cite{Dellenback1988_aiaa} in 1988 for the measurement of a swirled turbulent flow. The geometric features of the test case are shown in Fig. \ref{fig:OFR_scheme}. It consists of an inlet pipe with a guide and an open valve, which lead the flow towards a cylindrical combustion chamber. The geometry does not change over time, meaning the valve stays in a fixed position. This flow configuration is representative of the intake in a spark-ignition engine and it is commonly used in the early development phases by engine manufacturers to optimize pipe dimensions, valve shapes, and even flow rotation levels~\citep{Thobois2004_saefl}. The geometric features are indicated on the scheme. The main characteristics to be summarized are the diameter of the valve $D_v = \SI{40}{\mm}$, the radius of the outlet chamber $R = D_p/2 = \SI{60}{\mm}$, its length $L_p = \SI{500}{\mm}$ as well as the length of the inlet pipe $L_e = \SI{104}{\mm}$. The head of the valve is at a distance $L_{vh} = \SI{10}{\mm}$ from the end of the inlet pipe. The origin of geometry $(x,y,z)=(0,0,0)$ is located on the axis of revolution of the cylinders on the inlet plane at the beginning of the inlet pipe. 

A preliminary simulation for this test case is run with the objective to validate the numerical set-up against available experimental data. The latter have been obtained from the work by \cite{Thobois2006_phd}, who performed velocity measurements by \textit{Laser Doppler Anemometry} (LDA) on the same geometry of the flow rig and with a constant mass flow rate. We refer to this configuration as steady-state inlet. The solver used for the preliminary numerical run is \textit{rhoPimpleFoam} from OpenFOAM, which is tailored for the simulation of compressible flows using the PIMPLE algorithm. A second-order implicit scheme is used for the time advancement of the solution. The time step of the CFD calculation is $\Delta_t = \SI{2e-8}{\second}$, which grants a Courant–Friedrichs–Lewy condition $CFL<0.5$. Second-order centred schemes were used for the discretization of the spatial derivatives, except for the velocity transport term in the momentum equation for which a LUST scheme has been prescribed. This native OpenFOAM scheme combines a second-order centred scheme and a second-order upwind scheme with a ratio $75\%$-$25\%$. Only the air flow is studied and transport of passive-active scalars are not included. The fluid is considered as an ideal gas. Its initial properties, which are selected for $\SI{21.85}{\celsius}$, are imposed in the OpenFOAM configuration files. They are as follows: molar mass $M^{mol} = \SI{28.9}{\g \per \mol}$, specific heat at constant volume $C_v = \SI{718}{\kilo\joule \per \kelvin}$, dynamic viscosity $\mu = \SI{1.82e-5}{\kilo\gram \per \metre \per \second}$ as well as a Prandtl number $Pr = 0.708$. 
The wall condition is a no-slip condition. For the velocity, the outlet condition \textit{pressureInletOutletVelocity} specifies a zero gradient based on the dynamic pressure without recirculation of the flow. The condition \textit{waveTransmissive} is imposed at the outlet for the pressure field with a reference pressure of $p_{atm} = \SI{101325}{\pascal}$. It prevents reflection of waves inside the domain. The pressure gradient is zero at the inlet. A fixed temperature of $\SI{295}{\kelvin}$ is imposed at the inlet, and the temperature gradient is zero at the outlet.
\begin{figure}
    \centering
    \includegraphics[width=0.7\textwidth]{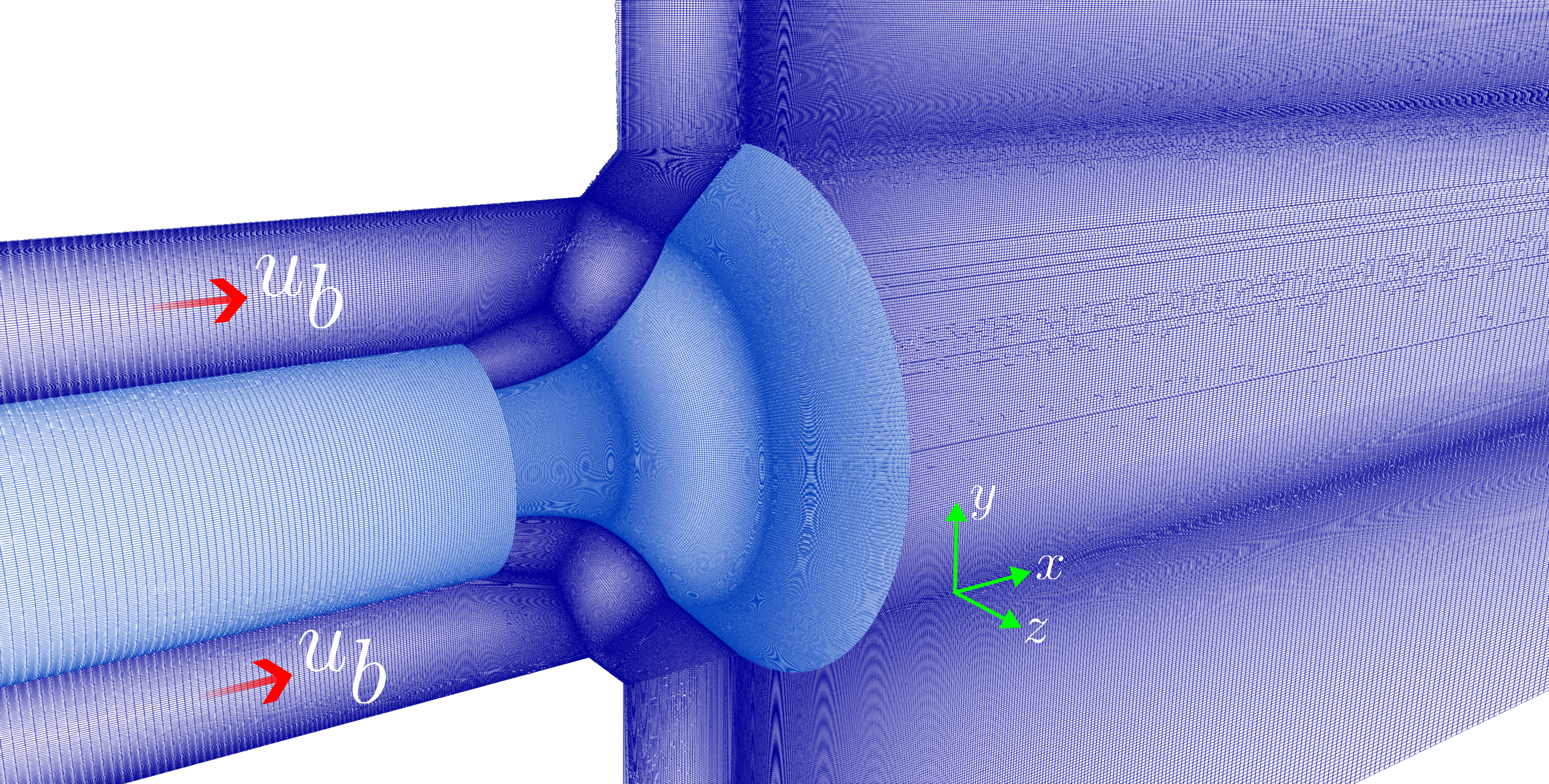}
    \caption{Details of the structured grid used for the high-fidelity reference simulation.}
    \label{fig:LES-HF_mesh}
\end{figure}

\begin{figure}
    \centering
     \begin{subfigure}{0.48\textwidth}
        \centering
        \includegraphics[width=1\textwidth]{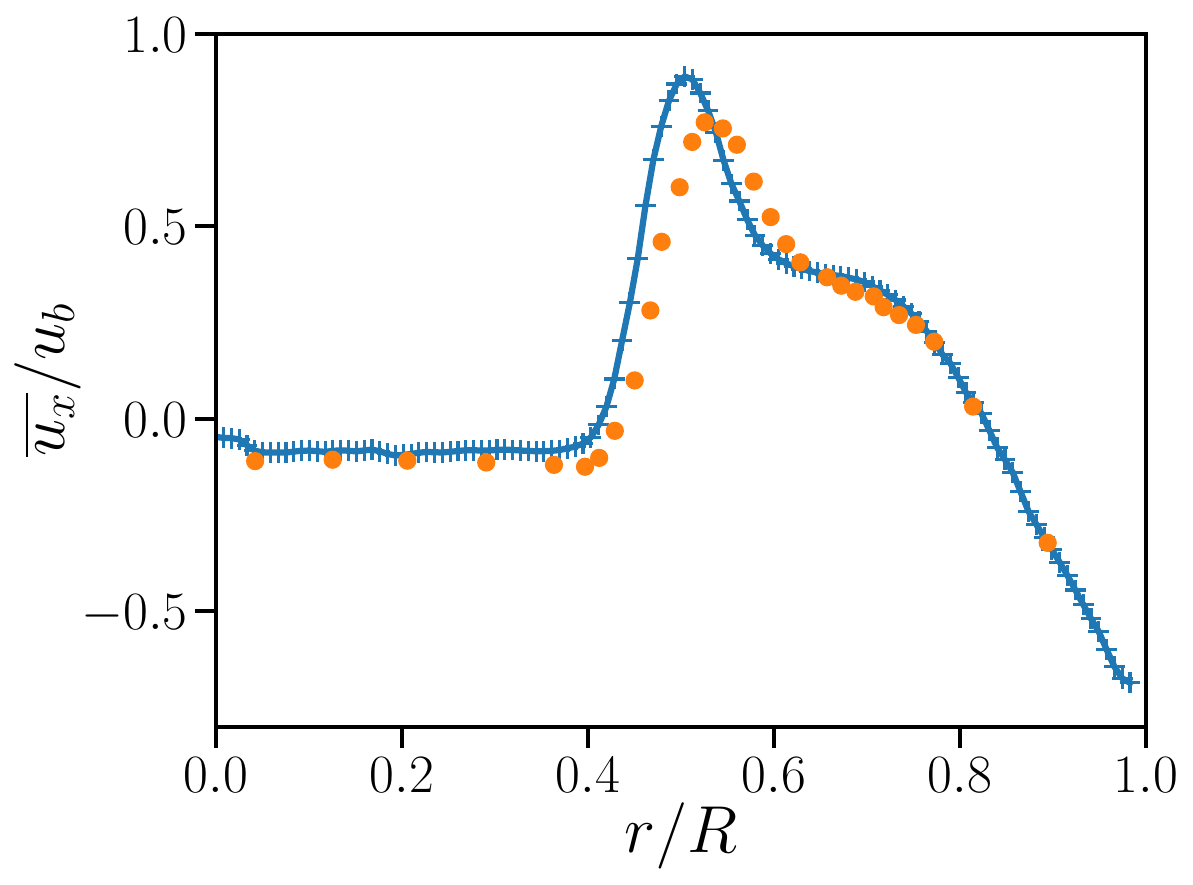}
        \subcaption{}
        \label{fig:validation_profile_124}
     \end{subfigure}
     \begin{subfigure}{0.48\textwidth}
        \centering
        \includegraphics[width=1\textwidth]{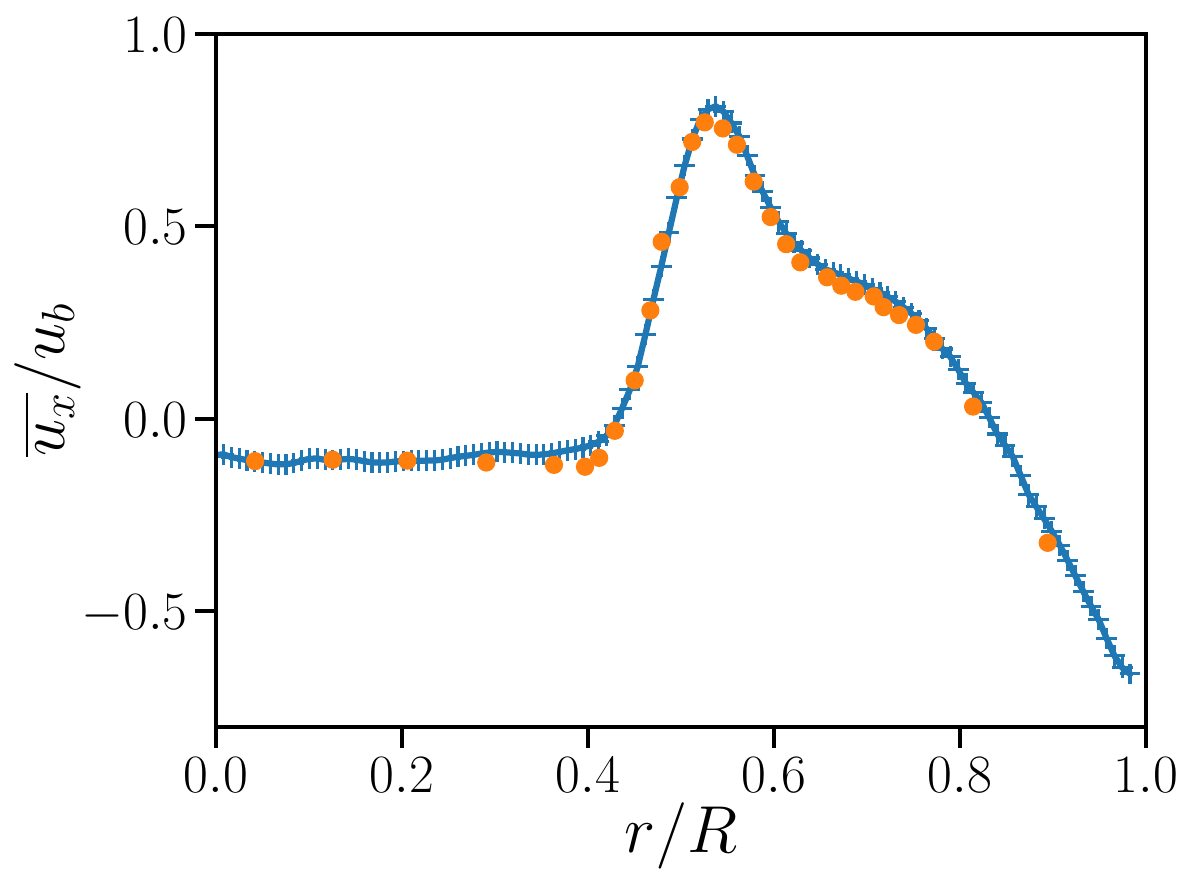}
        \subcaption{}
        \label{fig:validation_profile_126}
     \end{subfigure}
    \caption{Axial velocity profiles for two sections at (a) $x = \SI{0.124}{\m}$ and (b) $x = \SI{0.126}{\m}$. Experimental data from \cite{Thobois2004_saefl}, shown as orange dots, are compared with results from the high-fidelity LES, which are reported in blue.}
    \label{fig:validation_profile}
\end{figure}

The axial bulk velocity imposed at the inlet is selected using the mass flow rate $Q_m = \SI{0.055}{\kilo\gram\per\second}$ as in the experiments by \cite{Thobois2006_phd}. A uniform velocity $U_{inlet}= u_b = \iiint_{V_D} u_x \, dV^{\prime} / V_D = \SI{65}{\metre\per\second}$ is therefore applied at the inlet. The Reynolds number calculated using the diameter of the inlet annular duct is $Re = 37500$. The grid used to perform this simulation, which is shown in Fig. \ref{fig:LES-HF_mesh}, is composed of $131 \, 764 \, 400$ elements. It was generated using the \textit{blockMesh} tool of OpenFOAM. It is structured and composed by hexahedral elements. The size of the grid is a priori calibrated so that the size of the elements at the wall of the intake pipe is $\Delta r_1^+ = r/\delta_{\nu} = r u_{\tau}/\nu \approx 1$, where $u_\tau = \sqrt{\tau_w/\rho}$ is the estimated friction velocity and $\tau_w$ is the shear stress at the wall. The size of the grid elements then progressively increases moving away from the walls. Subgrid scale modelling is performed using a LES with the WALE turbulence model, as presented in Sec. \ref{sec:LES_equations}. For validation in the steady case, the calculation of the velocity averages is performed for a duration of $8$ advective times with $t_A = D_e/u_b$ after the effects of initial conditions are dissipated. The analysis of the rate of convergence of the statistics indicate that this average time grants convergence error lower than $5\%$, which was estimated to be an acceptable precision level for this initial validation, in particular because a steady-state inlet condition are not used in the DA experiment. Fig. \ref{fig:validation_profile} shows the radial profile of mean axial velocity normalized by the bulk velocity $\overline{u_x}/u_b$. The averages $\overline{.}$ are performed in time as well as in the azimuthal direction. Experimental data (orange dots), given at a section for $x=\SI{0.124}{\m}$, are compared with numerical data (blue crosses) taken at a section for $x=\SI{0.124}{\m}$ (Fig. \ref{fig:validation_profile_124}) and at a section for $x=\SI{0.126}{\m}$ (Fig. \ref{fig:validation_profile_126}). One can see that experiments and numerical data exhibit a very good agreement, if a small uncertainty in the positioning of the experimental measurements and in the spatial development of the predicted separated region is considered. This comparison assesses the validity of the numerical set-up used for this case, which will be used both for the model and the production of the observation in the DA procedure.

\subsection{Production of observation: reference high-fidelity simulation (LES-HF)}
\label{sec:LES-HF}
The test case validated via the comparison with experimental data is used to i) produce heterogeneous samples to be used as observation in the DA experiment and ii) to obtain reference statistics to assess the accuracy of the DA algorithm. However, modifications are performed as the DA analysis targets flow reconstruction and parametric optimization for cyclic phenomena such as the ones observed in internal combustion engines. While no modifications of the geometry are performed, a pulsating mass flow rate condition at the inlet is instead considered to induce periodicity of the flow. This condition is obtained via a modification of the velocity at the inlet such that:

\begin{equation}
    U_{inlet} = a_{ref} sin(2\pi f_{ref} \, t + \phi_{ref}) + U_o
    \label{eq:InletBF-HF}
\end{equation}

where $U_o = u_b$ corresponds to the average velocity at the inlet. The phase $\phi_{ref}$ is set to zero. The amplitude is set to consider a maximum variation of $\pm 15\%$ of the inlet velocity, therefore $a_{ref} = \SI{9.75}{\metre\per\second}$. The reference frequency $f_{ref} = \SI{800}{\hertz}$ here used is approximately 6 to 10 times higher than the characteristic frequencies of the pressure and velocity signals in the intake pipes linked to the propagation of acoustic waves \citep{leite2023experimental,Pera2011_sae}. This choice was made in order to perform the DA investigation over an increased number of cycles, which otherwise would have demanded prohibitive computational resources to be performed. The period of a complete oscillation for the velocity at the inlet is $t_p=\SI{1.25e-3}{\second}$ and it is reminded that the time step is equal to $\Delta_t = 2 \times 10^{-8}$ s. The simulation is run for a total time of $10 t_p$ using as initial condition an instantaneous flow field taken from the preliminary simulation. 

\begin{figure}
    \centering
    \includegraphics[width=0.8\textwidth]{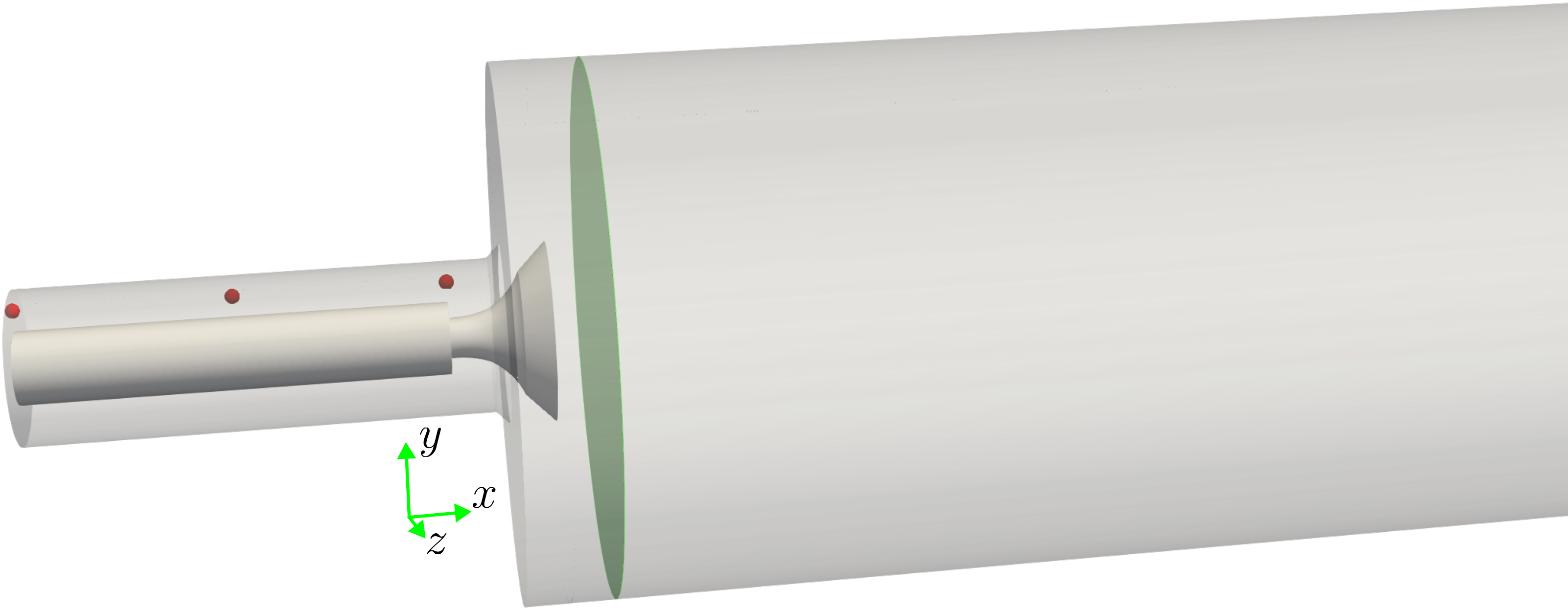}
    \caption{Location of the sensors used to measure the evolution of velocity and pressure in the intake channel. The 2D plane $x=\SI{0.124}{\m}$ used for data sampling is shown in green.}
    \label{fig:SensorsInlet}
\end{figure}

\begin{figure}
    \centering
    \begin{subfigure}{.425\textwidth}
        \centering
        \includegraphics[width=1\textwidth]{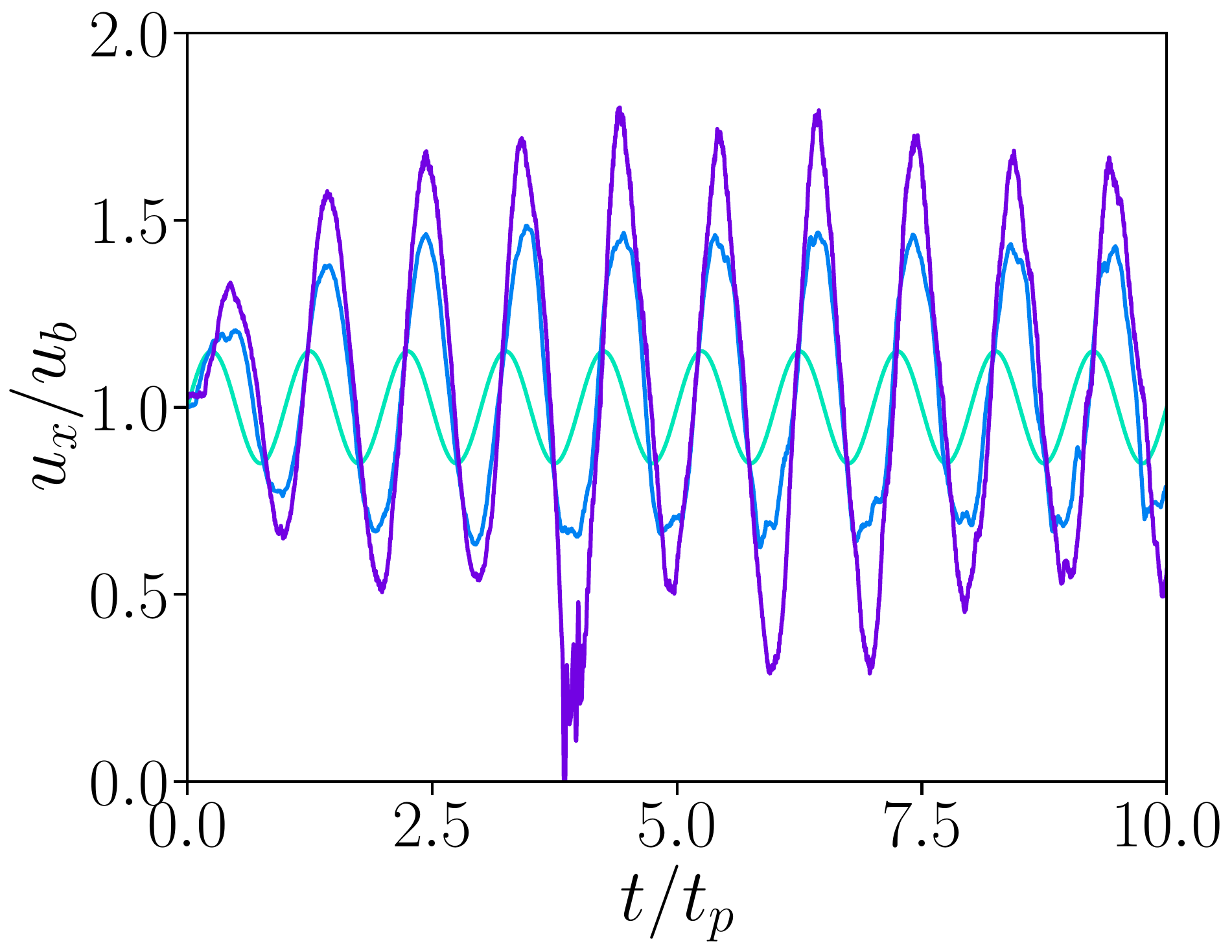}
        \subcaption{}
        \label{fig:portEntry_U_131M}
    \end{subfigure}
    \hspace{2em}
    \begin{subfigure}{.425\textwidth}
        \centering
        \includegraphics[width=1\textwidth]{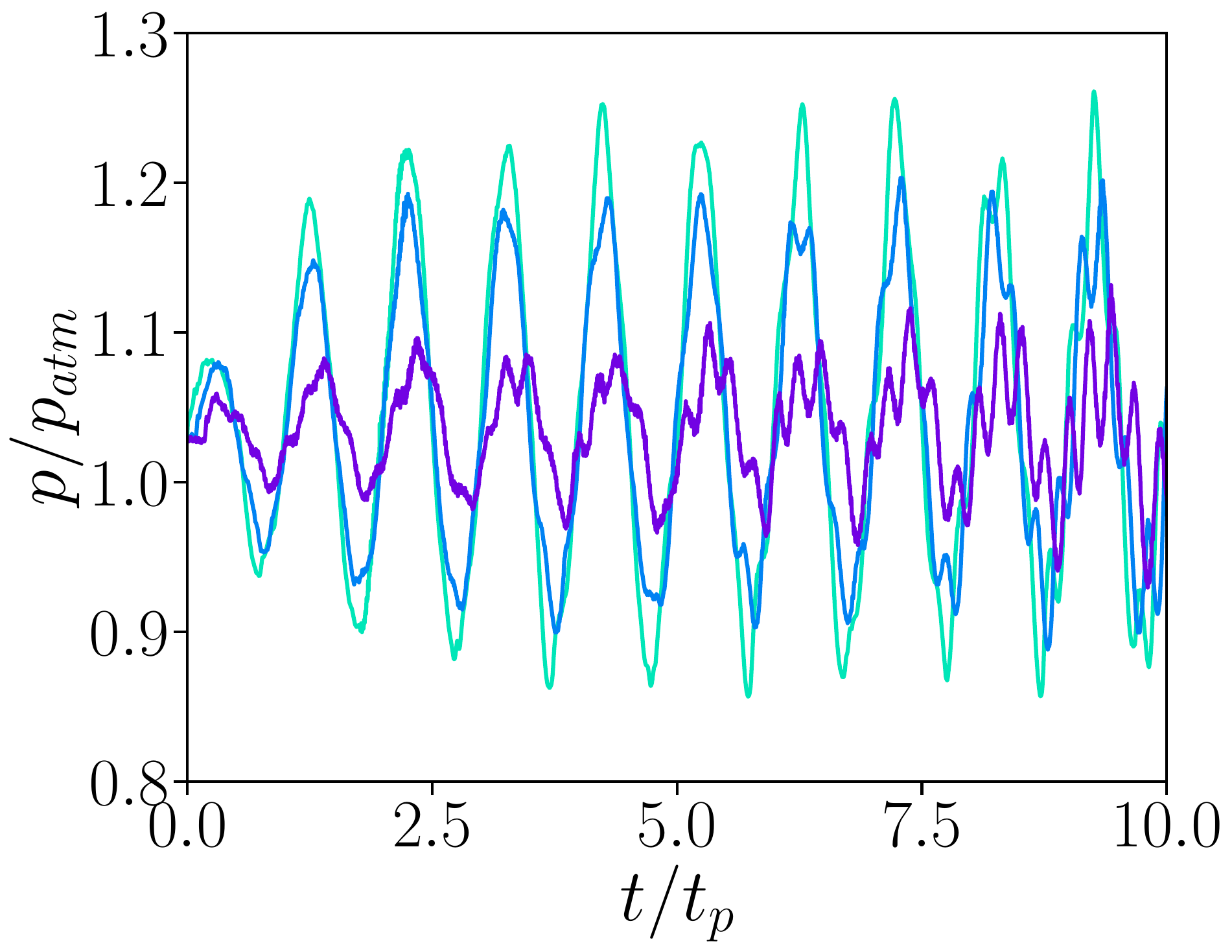}
        \subcaption{}
        \label{fig:portEntry_p_131M}
    \end{subfigure}
    \caption{Evolution of (a) the normalized axial velocity $u_x / u_b$ and (b) the normalized pressure $p / p_{atm}$ for the run LES-HF. The values are given for $3$ sensors located along the inlet duct for $x = 0$ (light blue), $x=\SI{0.046}{\m}$ (dark blue) and  $x=\SI{0.092}{\m}$ (purple).}
    \label{fig:evolution_Inlet_Ref}
\end{figure}

The evolution of the velocity and the pressure fields in the intake pipe are investigated using probes positioned at the centreline (see Fig. \ref{fig:SensorsInlet}). The three sensors are positioned at the inlet, half-way through the intake pipe and in proximity of the valve. Fig. \ref{fig:evolution_Inlet_Ref} shows the evolution of velocity and pressure on the three sensors investigated. Light blue data, which is sampled on the inlet boundary condition, clearly shows the periodic behaviour at the centreline. Moving downstream, one can also see that the velocity fluctuations increase in magnitude (up to $60-70\%$ of the bulk velocity) and the pressure exhibits higher frequency fluctuations. The pressure fluctuations sampled at the end of the intake pipe also seem to indicate that the signature of the initial conditions seem to last for the first six cycles. Therefore, DA analyses in Sec. \ref{sec:inletCalibration} and \ref{sec:localSynchronization} are performed using observation sampled from cycles $7$ to $10$, in order to avoid transient effects which could have an impact on the conclusions drawn.

\begin{figure}
    \centering
    \includegraphics[width=0.8\textwidth]{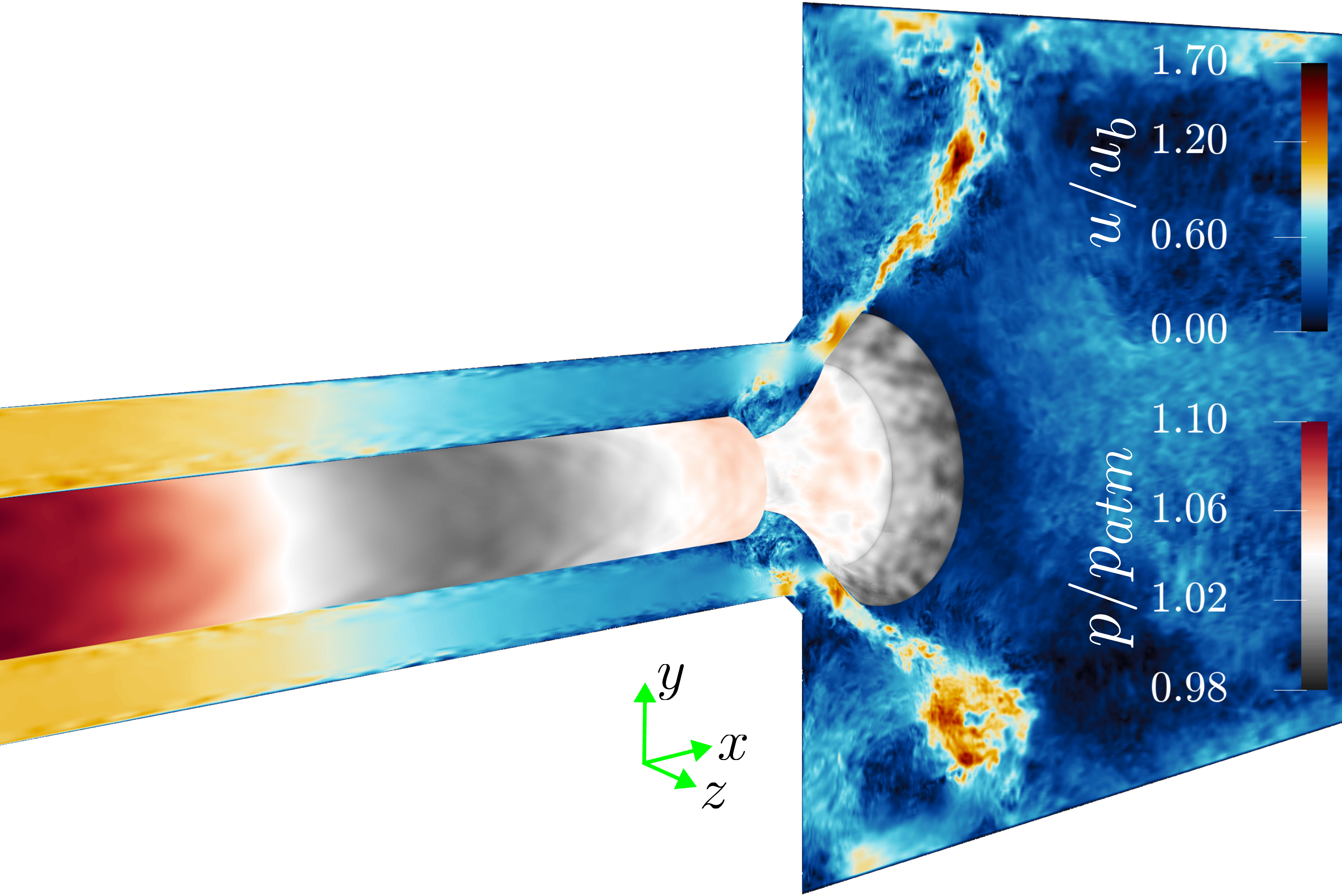}
    \caption{Instantaneous velocity field for the run LES-HF for $t=9 t_p$ on the section $z = 0$. The pressure field is shown on the surface of the valve.}
    \label{fig:velocityField_ref}
\end{figure}

Fig. \ref{fig:velocityField_ref} shows the instantaneous velocity field in the domain for a plane sampled for $z = 0$. In addition, the normalized pressure is shown on the surface of the guide and the valve. Values have been selected from a flow snapshot selected during the 9th cycle. An annular recirculation region is visible in the proximity of the valve at the end of the intake pipe. 
The flow moving through the valve curtain behaves similarly to a flow through a contraction, leading to acceleration. Flow separation at the valve head alters the flow structure~\citep{annand1974gas}. Downstream of the valve, the axi-symmetric flow structure is similar to the one of a confined bluff-body flow evolving through a sudden expansion~\citep{massey2019scaling,ayache2010experiments}. It features a large inner recirculation zone formed by two counter-rotating vortices. Additionally, upon impacting the wall, the jet generates a wall-bounded recirculation zone, also known as an outer recirculation zone. Sharp axial and tangential velocity gradients exist between the recirculation zones and the jet, creating regions of high shear that reattach to the wall. A detailed description of the flow topology for this specific configuration is provided in~\citep{hassan2019development,thobois2005large,Thobois2004_saefl}. The effects of the periodic inlet boundary conditions are also visible in the intake pipe, with important gradients of the velocity field in the $x$ direction.

\begin{figure}
    \centering
    \includegraphics[width=0.6\linewidth]{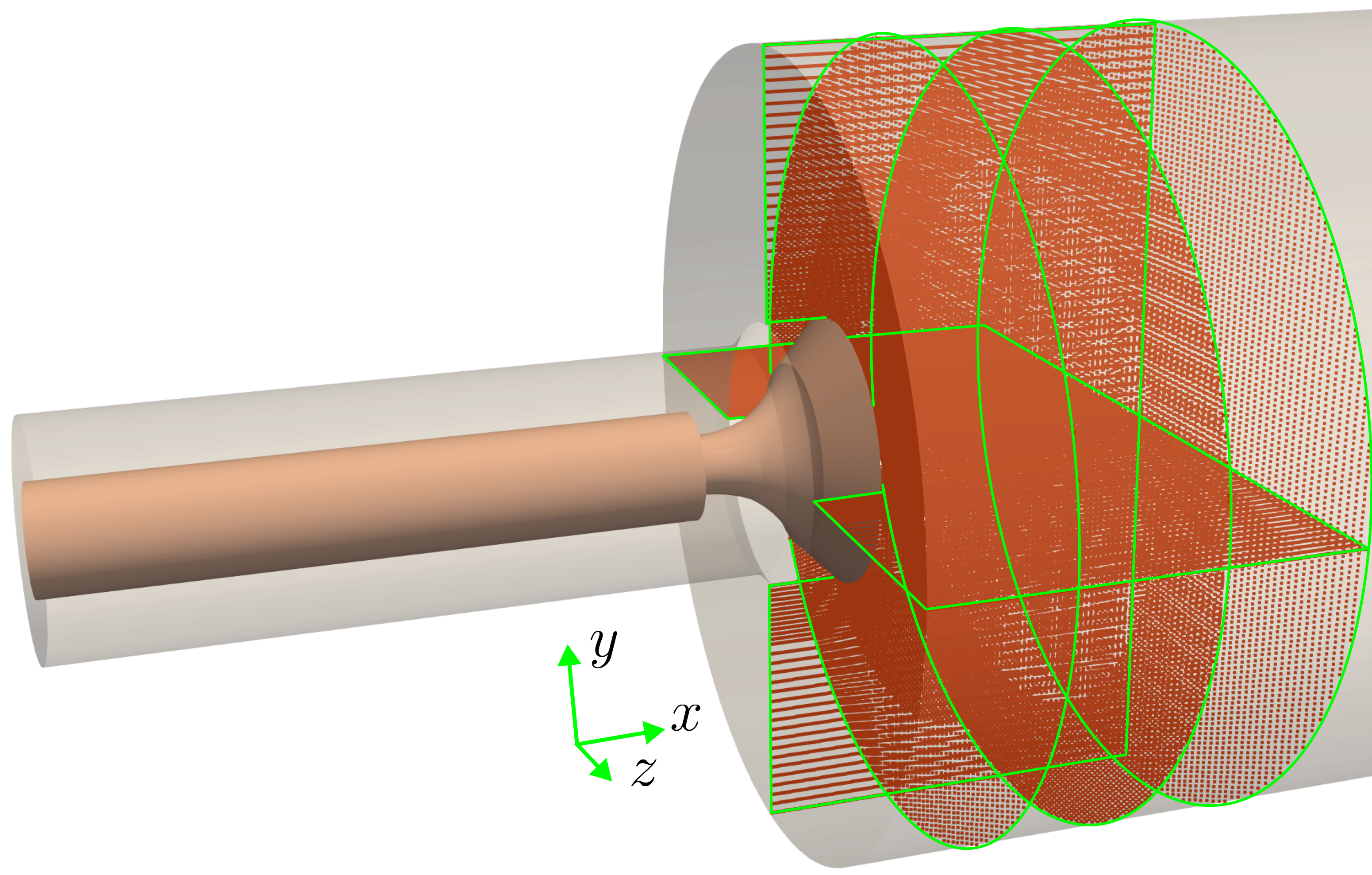}
    \caption{Sensors used for the sampling of observation. They are positioned over $5$ planes: $x = \SI{0.124}{\m}$, $x=\SI{0.144}{\m}$, $x=\SI{0.164}{\m}$, $y=0$ and $z=0$.}
    \label{fig:positionSensorsAD}
\end{figure}

As indicated at the beginning of this section, the simulation LES-HF has two main purposes. First, instantaneous data is sampled during the cycles seven to ten to create observation. Most of the measurements are performed in the cylinder region and in particular sampling is performed using the two-dimensional planes for $x = \SI{0.124}{\m}$, $x=\SI{0.144}{\m}$, $x=\SI{0.164}{\m}$, $y=0$ and $z=0$. In order to have the possibility to test several configurations, a total of $56 \, 226$ sensors are used for the measurements, whose distribution is shown in Fig. \ref{fig:positionSensorsAD}. While not all the data stored during the realization of this simulation has been used for the present DA runs, the database will be used in future analysis and it will be made available using dedicated platforms. Details about the sampling strategy are now provided. In terms of spatial distribution, the sensors' density is higher in the region of the jet exiting the intake duct. The velocity is sampled at the sensor positions every $\Delta_{tm} = 50\Delta_t = \SI{1e-6}{\second}$. This corresponds to $1 \, 250$ samples per cycle, therefore  $t_p = 1 \, 250 \Delta_{tm} = 62 \, 500 \Delta_t$. 
A qualitative study of the flow fields seems to indicate that the size of the resolved eddies in the valve region are of the order of $\SI{0.0005}{\metre}$ up to $\SI{0.005}{\metre}$ for the largest coherent structures. Approximated values for the turnover time obtained via classical theoretical and dimensional arguments seem to indicate that the chosen sampling frequency correspond to $8-80$ measurements for the resolved turbulent structures. The sampling frequency is therefore sufficient to capture the dynamic structures observed. Therefore, problems in respecting the Nyquist-Shannon sampling theorem which could have an impact over the accuracy of the DA procedure \citep{Meldi2018_ftc} should be here excluded.

\subsection{Numerical model for ensemblistic DA: low-fidelity simulation (LES-LF)}
\label{sec:LES-LF}
\begin{figure}
    \centering
    \begin{subfigure}{.45\textwidth}
        \centering
        \includegraphics[width=1\textwidth]{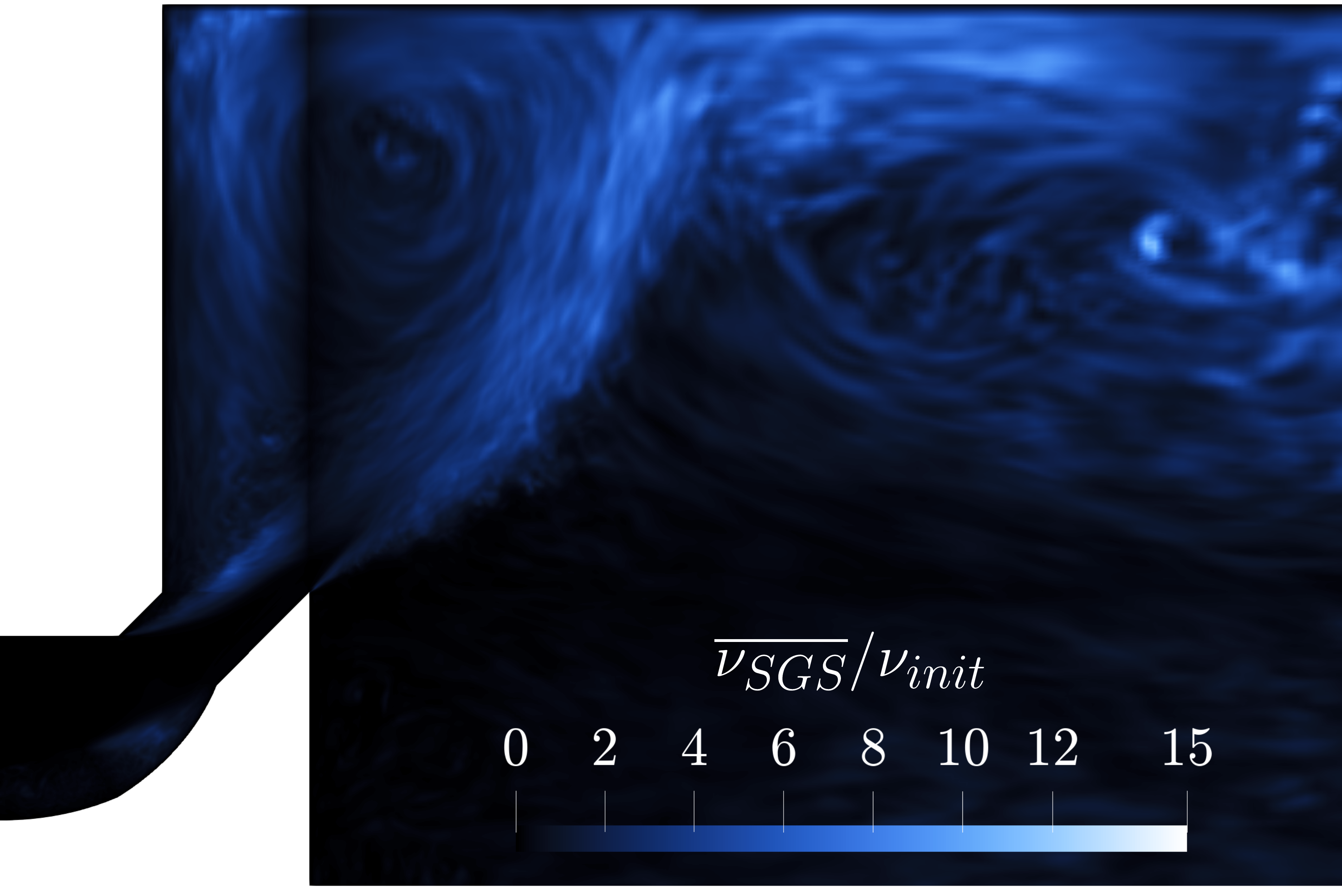}
        \label{fig:nut_field_LES-HF}
        \caption{LES-HF}
    \end{subfigure}
    \hspace{2em}
    \begin{subfigure}{.45\textwidth}
        \centering
        \includegraphics[width=1\textwidth]{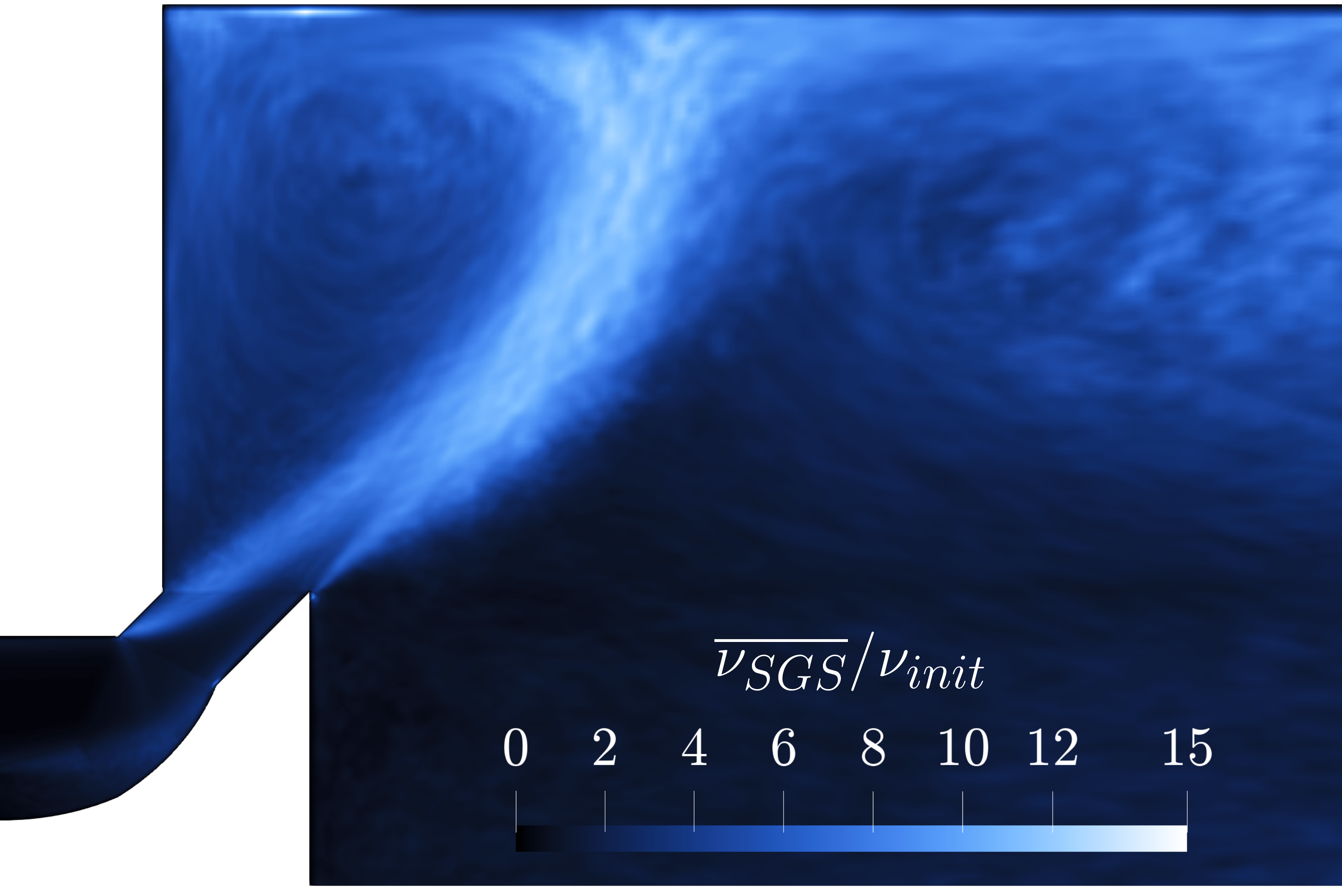}
        \label{fig:nut_field_LES-LF}
        \caption{LES-LF}
    \end{subfigure}
    \caption{Mean subgrid-scale viscosity fields normalized by the initial kinematic viscosity $\nu_{SGS}/\nu_{init}$ on a 2D plane at $z=0$ for (a) the LES-HF run and (b) the LES-LF run.}
    \label{fig:nut_fields}
\end{figure}

The numerical solver used as a model for DA purposes is now presented. This model is going to use the same flow solver with the same specifications of LES-HF in terms of numerical schemes, subgrid scale closure and boundary conditions, but it will be run on a significantly coarser grid. The reason why are two. First, an ensemble of realizations as required by the EnKF would require prohibitive computational costs when using the grid prepared for the run LES-HF. Second, one of the points of interest of this work is to investigate how DA's parametric investigation and state estimation are affected by different levels of bias in the model and in the observation. This low-fidelity model will be referred to as LES-LF (\textit{low fidelity}).  The grid used for calculation consists of $8 \, 764 \, 400$ elements. The structure of the grid is very similar to the one used for the run LES-HF, but the number of elements is reduced by approximately a factor of $15$. This time, the size of the near-wall grid elements is adjusted so that $\Delta r_1^+ = r/\delta_{\nu} = r u_{\tau}/\nu \approx 4-5$, meaning the simulation in the intake pipe is not any more wall resolved. Grid elements are also significantly less resolved in the valve and cylinder area, precluding therefore the direct resolution of small eddies and relying on a stronger contribution of the WALE closure. This can be clearly observed comparing the mean subgrid scale viscosity between the runs LES-HF and LES-LF, which is shown in Fig. \ref{fig:nut_fields}. The mean subgrid-scale viscosity field $\overline{\nu_{SGS}}$ normalized by the initial kinematic viscosity $\nu_{init}$ is approximately twice as high downstream of the valve for LES-LF compared to LES-HF. Consequently, the flow dynamics differ due to the stronger diffusive effect of the subgrid scale closure.

\begin{figure}
    \centering
    \begin{subfigure}{.425\textwidth}
        \centering
        \includegraphics[width=1\textwidth]{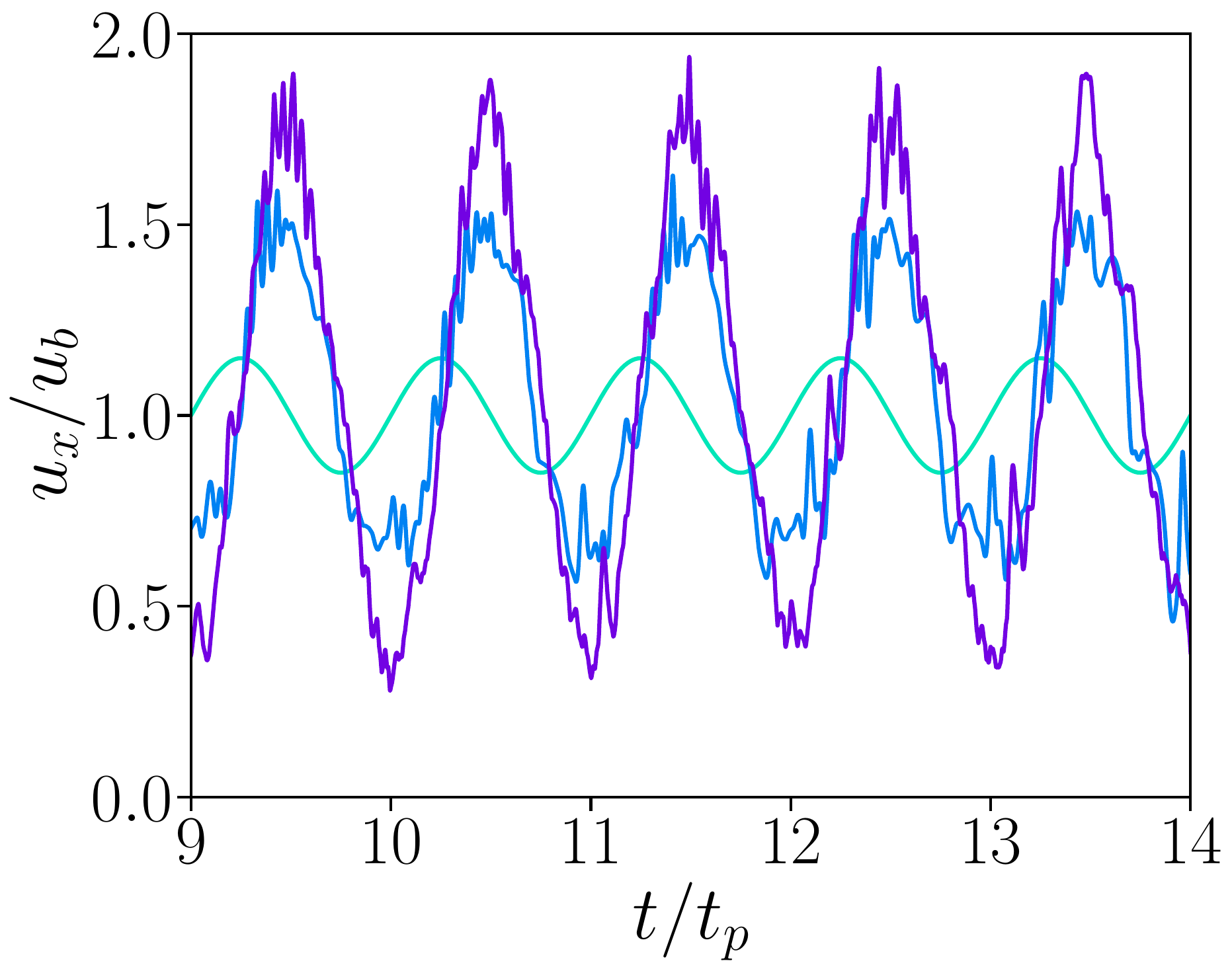}
        \subcaption{}
        \label{fig:portEntry_U_8M}
    \end{subfigure}
    \hspace{2em}
    \begin{subfigure}{.425\textwidth}
        \centering
        \includegraphics[width=1\textwidth]{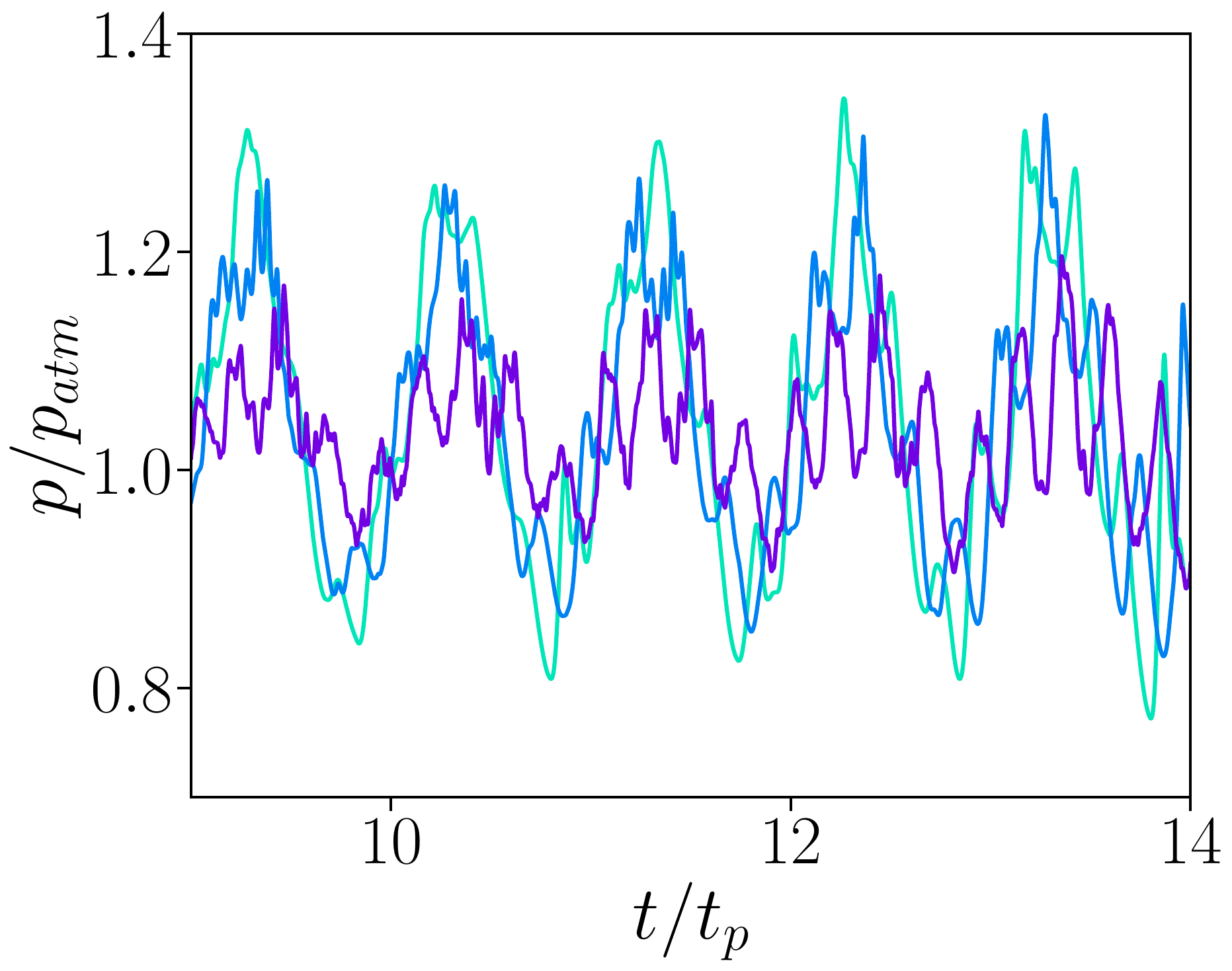}
        \subcaption{}
        \label{fig:portEntry_p_8M}
    \end{subfigure}
    \caption{Evolution of (a) the normalized axial velocity $u_x / u_b$ and (b) the normalized pressure $p / p_{atm}$ for the run LES-LF. The values are given for $3$ sensors located along inlet duct for $x = 0$ (light blue), $x=\SI{0.046}{\m}$ (dark blue) and  $x=\SI{0.092}{\m}$ (purple).}
    \label{fig:evolution_Inlet_baseline}
\end{figure}

Differences between LES-HF and LES-LF are also observed comparing data sampled from the three sensors positioned in the intake pipe. Once the effects of the initial conditions are dissipated, the velocity oscillations in the intake duct exhibit higher variations for the case LES-LF, as shown in Fig. \ref{fig:evolution_Inlet_baseline}. This time, the maximum velocity reached at the end of the intake duct exhibits oscillations up to $+ 100\%$ relative to the bulk velocity. On the other hand, the minimum of the oscillation goes down to $50-60\%$ of $u_b$. The maximum and minimum values for the velocity field are observed at the sensor located at the end of the intake pipe, just before the valve region. It also seems that, on average, the flow at the centreline exhibits a non-negligible acceleration moving downstream from the inlet, which was not observed for the run LES-HF.  This important dynamic difference affects the development of the jet moving from the valve to the cylinder, and it confirms the strong sensitivity of LES to changes in the grid used \citep{Sagaut2006_springer}. This point will also be extensively investigated via the analysis of the DA results.

\section{Calibration of inlet boundary conditions via DA}
\label{sec:inletCalibration}

The EnKF algorithm is here used to calibrate the features of the inlet boundary conditions for an ensemble of numerical simulations based on the model LES-LF. More precisely, the free coefficients governing the parametric description of the inlet velocity field will be optimized in an online procedure to reduce the discrepancy of the prediction of the ensemble run with local data sampled from simulation LES-HF. The inlet boundary condition of the latter is supposed to be unknown and its effects are only measured by the measurements that are obtained downstream at the sensors and used as observation. It is important to stress here that multiple analysis phases are performed comparing instantaneous flow snapshots, permitting for an optimization process which can naturally evolve in time with the sequential analysis phases. 

\subsection{Data assimilation experiment}
The first assumption here performed is that the inlet velocity profile of Eq. \ref{eq:InletBF-HF} is unknown and only limited measurements from the physical representation of LES-HF are available. Therefore, the calibration performed via the DA procedures aims to reconstruct global information from an ensemble of numerical realizations and limited observed measurements. The optimization of the features of the inlet is performed using the hyper-localized EnKF algorithm presented in Sec. \ref{sec:HLEnKF}. The DA procedure performed here aims for a parametric optimization of the inlet boundary condition without including a state update. This approach is intended to provide insights into the performance of DA optimization when using instantaneous data at high Reynolds numbers, where strong fluctuations are present. The inclusion of state estimation would have significantly increased the complexity of the procedure, not only in terms of the features to be considered but also in ensuring the stability of the numerical system.. The ingredients used in the present DA procedure are now introduced.
\begin{itemize}
    \item \textit{A model} is used in the forecast steps to propagate the solutions of the ensemble members in time. The numerical model employed for this purpose is the same as that used for the LES-LF simulation. In particular, a coarse grid of about $8 \cdot 10^{6}$ elements is used for the calculations. Each of the ensemble realizations is run with a different parametric description of the inlet, which is updated at each DA analysis phase.
    \item \textit{Observation}, in the form of instantaneous values of the velocity field sampled from the run LES-HF, is coupled with the results from the ensemble runs in the EnKF formalism. For this investigation 400 sensors are selected among the $56 \, 226$ available and therefore 1200 local velocity values ($3$ components) are provided to the EnKF at each analysis phase.  
    \item The open-source library \textit{CONES} is used to perform an online connection between the $N_e$ CFD realizations using the OpenFOAM solver rhoPimpleFoam 
 and an in-house code performing the tasks required for the HLEnKF.
\end{itemize}

Additional details are now provided for each of the key elements used. The size of the ensemble selected for this analysis is $N_e=35$, which is consistent with similar analyses in the literature for application with LES \citep{Moldovan2022_cf,Villanueva2024_ijhff}. Each member of the ensemble is initially run with a prescribed inlet condition for the velocity, which is described by the following equation: 
\begin{equation}
    U_{inlet} = \sum_{i=1}^4 a_i sin(2\pi f_i t + \phi_i) + U_o.
    \label{eq:InletBC-LF}
\end{equation}
One can see that the velocity imposed at the inlet is given by a constant value $U_o$, which will be referred to as base velocity, plus four sinus functions. These functions govern the time evolution of the velocity at the inlet and they are characterized by an amplitude $a$, a frequency $f$ and a phase $\phi$. One can see that, while Eq. \ref{eq:InletBC-LF} can span a significantly large space in terms of possible velocity conditions at the inlet, the velocity behaviour of Eq. \ref{eq:InletBF-HF} can be exactly represented by such a system. It is therefore interesting to investigate whether the optimization of the inlet parameters when using a coarse grid will converge to the same behaviour of Eq. \ref{eq:InletBF-HF} or if numerical and modelling errors due to the different grid used will affect the final result. Such DA optimization will be then performed tuning the $N_\gamma=13$ free parameters available in Eq. \ref{eq:InletBC-LF}. The initial values imposed for the parameters at the very beginning of the DA run are important, because they can lead the convergence of the algorithm towards local optima, therefore reducing the potential accuracy of the DA tool \citep{Asch2016_SIAM}. Considering the importance of this element for the analysis of turbulent flows, two prior states are investigated, whose features are reported in Tab. \ref{tab:priorsTable}.  

\begin{figure}
    \centering
    \begin{subfigure}{.425\textwidth}
        \centering
        \includegraphics[width=1\textwidth]{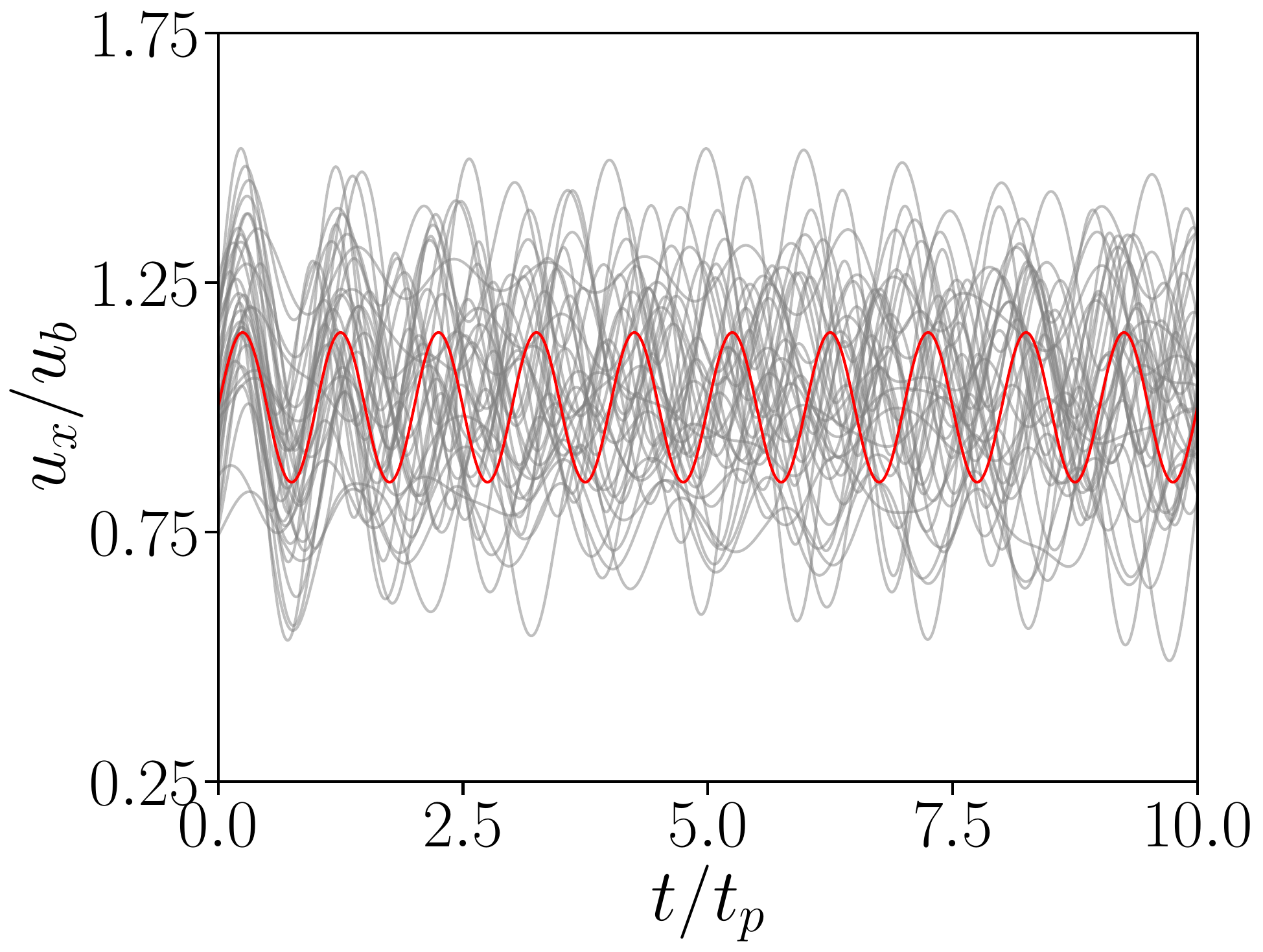}
        \subcaption{}
        \label{fig:initConditions_OFR_DA_v2}
    \end{subfigure}
    \hspace{2em}
    \begin{subfigure}{.425\textwidth}
        \centering
        \includegraphics[width=1\textwidth]{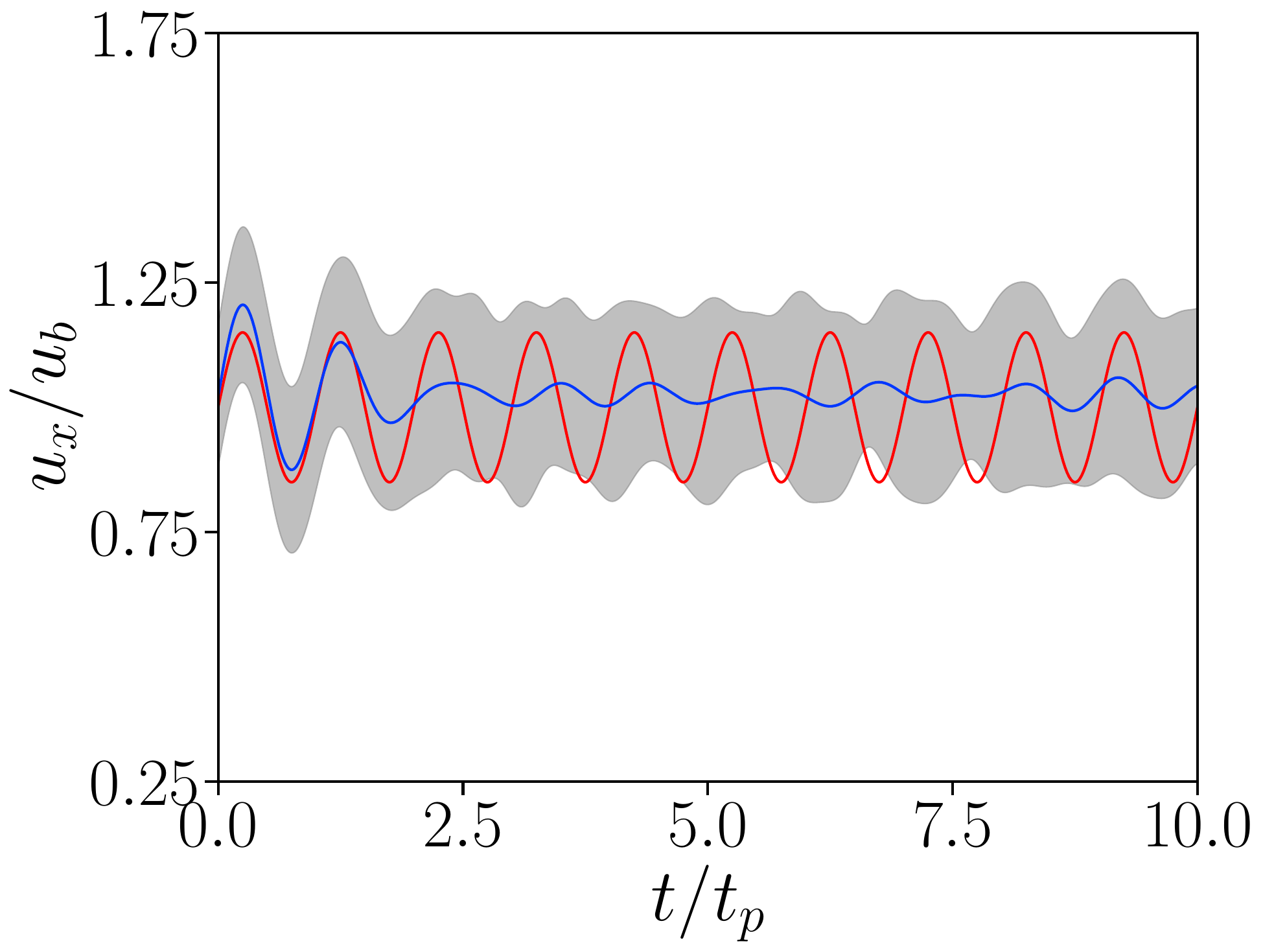}
        \subcaption{}
        \label{fig:initConditions_mean_OFR_DA_v2}
    \end{subfigure}
    \begin{subfigure}{.425\textwidth}
        \centering
        \includegraphics[width=1\textwidth]{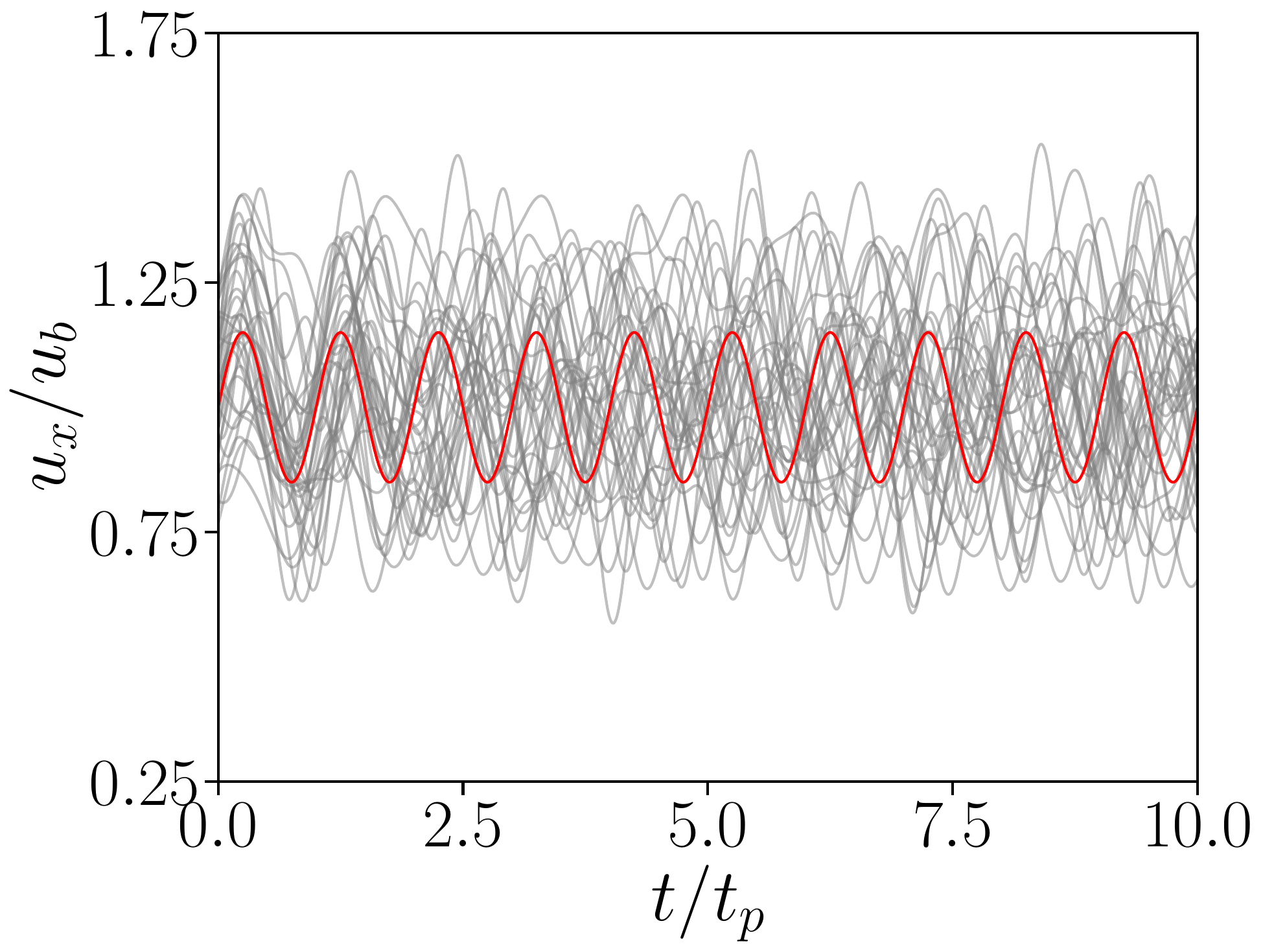}
        \subcaption{}
        \label{fig:initConditions_OFR_DA_v3}
    \end{subfigure}
    \hspace{2em}
    \begin{subfigure}{.425\textwidth}
        \centering
        \includegraphics[width=1\textwidth]{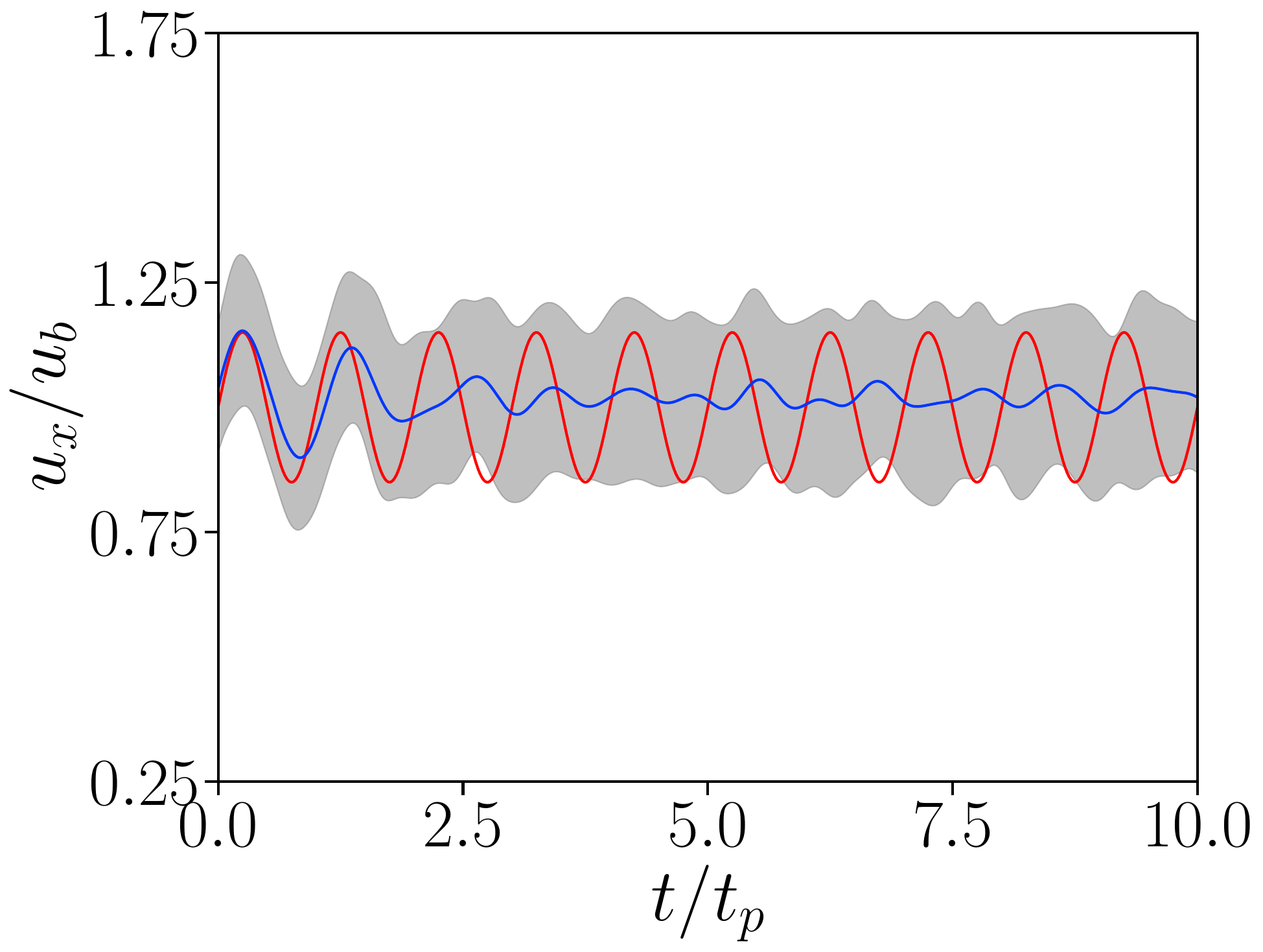}
        \subcaption{}
        \label{fig:initConditions_mean_OFR_DA_v3}
    \end{subfigure}
    \caption{Evolution of the axial velocity inlet condition for (top line) the distribution \textbf{\textit{prior state} 1} and (bottom line) the distribution \textbf{\textit{prior state} 2}. The red curve corresponds to the reference inlet condition used for LES-HF. The gray curves correspond to realizations for the ensemble members. The blue curve corresponds to the mean of the ensemble members. The gray area indicates the standard deviation.}
    \label{fig:prior}
\end{figure}

\begin{table}
    \centering
    \begin{tabular}{ccc}
        \hline
         \textbf{Distributions} & \textit{prior state} 1 & \textit{prior state} 2 \\
         \hline
         \textbf{Amplitudes} & 
                      \makecell{$a_1 \thicksim \mathcal{N}(8,1)$, $a_2 \thicksim \mathcal{N}(5,2)$ \\
                      $a_2 \thicksim \mathcal{N}(2,4)$, $a_4 \thicksim \mathcal{N}(0,5)$} & 
                      \makecell{$a_1 \thicksim \mathcal{N}(9.75,1)$, $a_2 \thicksim \mathcal{N}(0,2)$ \\
                      $a_2 \thicksim \mathcal{N}(0,3)$, $a_4 \thicksim \mathcal{N}(0,4)$} \\
          \hline
         \textbf{Frequencies} & $f_i \thicksim \mathcal{N}(800,100)$ & 
                      \makecell{$f_1 \thicksim \mathcal{N}(700,100)$, $f_2 \thicksim \mathcal{N}(900,100)$ \\
                      $f_3 \thicksim \mathcal{N}(1100,100)$, $f_4 \thicksim \mathcal{N}(1300,100)$}\\
         \hline
         \textbf{Phases} & $\phi_i \thicksim \mathcal{N}(0,0.5)$ & $\phi_i \thicksim \mathcal{N}(0,0.5)$ \\
         \hline
         \textbf{Base velocity} & $U_o \thicksim \mathcal{N}(65,10)$ & $U_o \thicksim \mathcal{N}(65,10)$ \\
         \hline
    \end{tabular}
    \caption{Distributions of the parameters describing the inlet boundary condition used  as prior states for the DA parametric optimization.}
    \label{tab:priorsTable}
\end{table}
First of all, the prior state for each free coefficient is described using random Gaussian distributions, but the values are truncated to $\pm 2\sigma$ in order to avoid ill-conditioned problems.  The distribution referred to as \textit{prior state} 1 is selected to examine in detail the variability of the amplitude parameters. The latter directly controls the velocity levels of the inlet and therefore a high sensitivity is observed to its variations. For this reason, the distributions of the four parameters $a_i$ are deliberately chosen to be significantly different from one another. Similarly, the investigation using the \textit{prior state} 2 focuses on assessing the sensitivity of the optimization to large variations in the examined frequencies, the distribution of which is given in Tab.~\ref{tab:priorsTable}. In both cases, the parameters describing the phase and the base velocity are initially obtained using a Gaussian distribution centred around the true value imposed in the run LES-HF. Although the parameters are still optimized, this simplification has been intentionally chosen to focus on the investigation of the amplitude and frequency of the velocity fluctuations. Once the initial coefficients are determined for each ensemble member and for the two configurations, one can observe the behaviour of the inlet conditions during the forecast step simply tracing the curves obtained by Eq. \ref{eq:InletBC-LF} and without any need to perform simulations. These diagrams are shown in Fig. \ref{fig:prior}. The first row provides information about the \textit{prior state} 1, while the second row includes results from the \textit{prior state} 2. On the left column, one can see the inlet behaviour of the simulation LES-HF (in red) with the same condition initially prescribed for the ensemble members. The right column does not include information of each simulation of the ensemble, but it provides its mean (blue line) and its standard deviation (grey region). As expected, the \textit{prior state} 1 exhibits higher amplitude variations when compared with the \textit{prior state} 2. The latter, however, shows higher frequency dynamics. At last, for every simulation of the ensemble, the initial velocity and pressure fields are taken from the simulation LES-LF for $t = 10 t_p$. 

\begin{figure}
    \centering
    \includegraphics[width=0.6\linewidth]{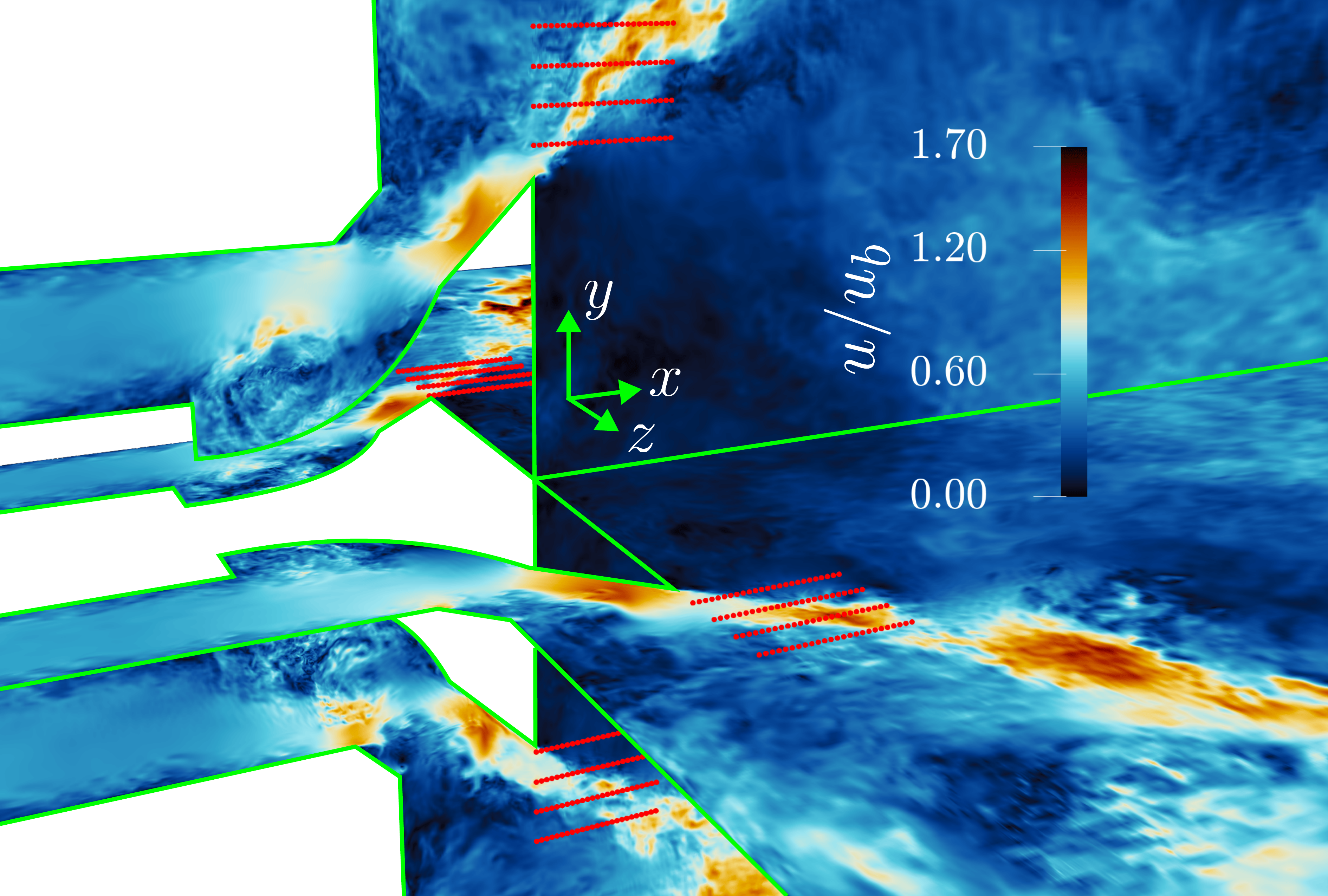}
    \caption{Velocity field of the run LES-HF for $t=9 t_p$ shown on two sections for $y=0$ and $z = 0$. The red dots correspond to the observation sensors selected to calibrate the inlet condition. }
    \label{fig:observationLocation}
\end{figure}

As previously mentioned, data is sampled at $400$ sensors positioned in the physical domain for the simulation LES-HF. These sensors measure the instantaneous values for the three velocity components for a total of $1200$ observations for each analysis phase. The sensors are positioned in the jet area at the outlet of the intake duct, on the planes $y = 0$ and $z = 0$ presented in Sec. \ref{sec:LES-HF}. The selected positions are shown in Fig. \ref{fig:observationLocation} for reference. Data is observed for the cycles $7$ to $10$ of the simulation LES-HF to eliminate effects associated with the initial conditions, providing a total of four cycles for analysis. To extend the time window for data assimilation, the observation is restarted from the beginning of cycle $7$ after the last snapshot from cycle $10$ is used. It has been verified that these two instantaneous solutions remain similar, and, in particular, that sensor data do not exhibit significant differences. This verification ensures that resampling the observation from cycle 7 does not introduce strong discontinuities in the flow field, preventing convergence issues in the algorithm.

At last, the DA strategy and the selected hyperparameters are discussed. The main feature of the HLEnKF is that multiple analyses are performed on smaller domains corresponding to a sufficiently large volume around each sensor. The size of each domain, further explained in Sec. \ref{sec:localSynchronization}, is approximately $\SI{3.4d-3}{\m}$. Considering that the latter are clustered in four main regions and they are close together, some regions are overlapping. This overlap is not important for the present DA analysis because no state update is performed and the only target is the optimization of the inlet boundary conditions. To this purpose, the optimized parameters obtained by each local EnKF are averaged before the updated values are sent back to the ensemble members, updating the inlet.

The size of the state matrix $\boldsymbol{\mathcal{U}}$ is $[N_{ext}\, , \, N_e]$  where $N_e=35$ is the number of ensemble members and $N_{ext} = N + N_{\gamma}$ represents the total number of degrees of freedom considered for each analysis phase. $N$ characterizes the features of the flow field in the local region of the analysis and it is usually equal to $N = 3 \, n_{cells}$. Indeed, the local HLEnKF procedure considers all three components of the velocity field measured at each grid element. However, since no state estimation is  performed in this case and the flow field information  is already incorporated in the terms $s_i$, the value of N is set to zero. Consequently, $N_{ext}$ and $N_\gamma$ are both equal to $13$, which is the number of free parameters optimized by the DA procedure.

\begin{figure}
    \centering
    \includegraphics[width=0.8\linewidth]{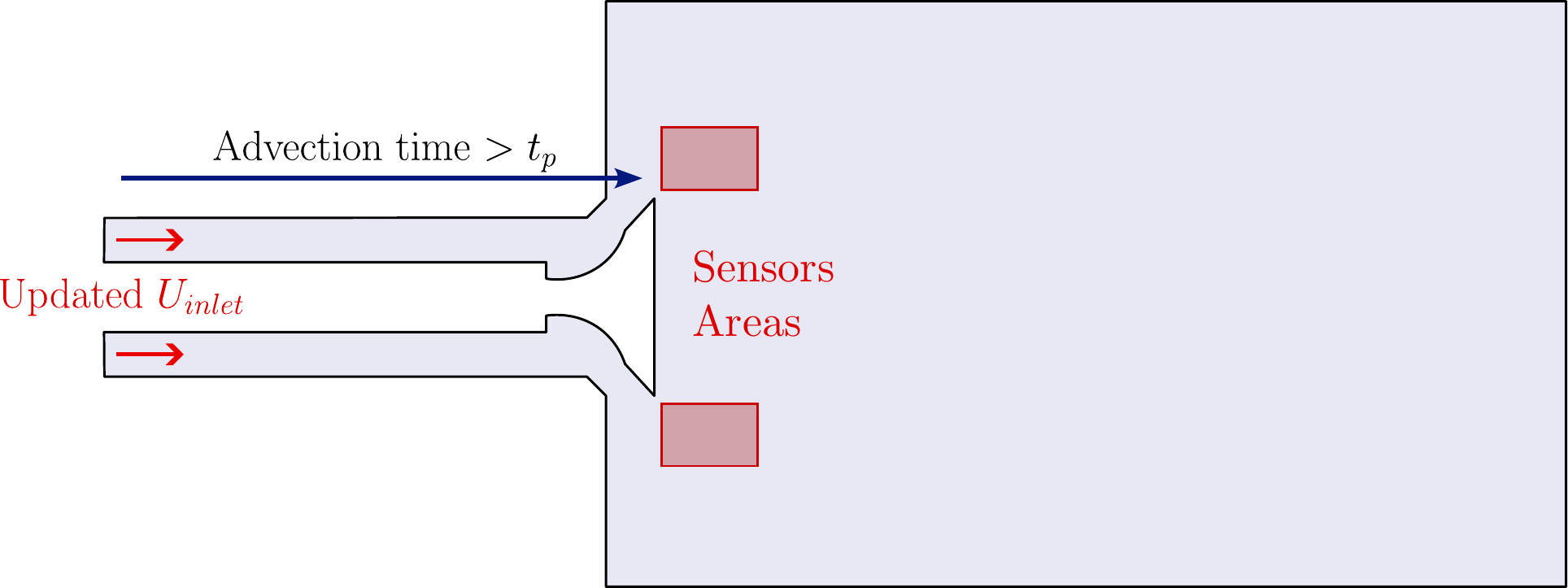}
    \caption{Scheme illustrating the time needed to advect downstream the updated inlet information to the sensors.}
    \label{fig:OFR_advectionScheme}
\end{figure}

The time frequency of the analysis phases is particularly important because, once the parametric description of the inlet is updated via the DA procedure, sufficient time must be allowed for the signal to propagate and influence the observation region before conducting a new analysis. This time can be estimated considering the advection by $u_b$, which provides the formula $t_A^{inletDA} = (L_e + L_{vh})/u_b \approx \SI{1.75d-3}{\s} > t_p = \SI{1.25d-3}{\s}$, see Fig. \ref{fig:OFR_advectionScheme} for a qualitative description. Clearly, analysis over a short time span would operate using a wrong correlation between the observed results at sensors and the parametric description of the inlet, potentially leading to the divergence of the filter. One possible solution to avoid this constraint is to rely on an \textit{ensemble Kalman smoother} \citep{Evensen2000_mwr} algorithm, which enables  optimization over larger time windows that include multiple analyses. However, computational and storage costs for this case would be prohibitive. The preliminary tests performed show that in this case, the very first analysis phase is critical. The following DA optimizations are progressively less critical as the offset between the data observed at the sensors and the parametric description of the inlet reduces as the free parameters reach convergence. Therefore, the following DA has been performed. Once the procedure starts, the first forecast phase runs for a duration of $t_{f,1}=2.2t_p$, allowing  the flow field obtained with the initial inlet conditions to propagate downstream for all 35 ensemble members. After this first analysis phase, the length of the forecast phase is reduced to $t_{f}=0.2 t_p$ i.e. $5$ analysis steps are performed for each cycle. The DA runs are performed for a total time of $12t_p$, for a total of $50$ analysis steps.

The last set of hyperparameters to be discussed is the confidence level imposed for the observations and for the model. The measurement error covariance matrix $\mathbf{R}$ in Eq. \ref{eqn:EnKF_gain_R} is  diagonal and expressed as $\mathbf{R} = \sigma_m^2\mathbf{I}$, where $\sigma_m$ quantifies the uncertainty of the measurements. A multiplicative uncertainty of $17.5\%$ is applied to the values of each observation during the first three inferred cycles. Thereafter, $\sigma_m = 5\%$. This choice has been performed to obtain a robust convergence of the parametric optimization. Stochastic inflation (see Sec. \ref{sec:inflation}) is used to increase the variance of the parametric description during the calculation. 
It is reminded that inflation is an excellent tool to avoid the collapse of the variance of optimized parameters. This phenomenon, which is common using the classical formulation of the EnKF, can preclude the identification of efficient regions in the parameter's space once a local optimum is identified. Low values for inflation value is hence used after the first inferred cycle, i.e. $\lambda = 0.5\%$. After the 9th cycle, the coefficient controlling the multiplicative inflation is again lowered to $\lambda = 0.375\%$.

\subsection{Results obtained for the DA parametric optimization}
\begin{figure}[h]
    \centering
    \begin{subfigure}{.33\textwidth}
        \centering
        \includegraphics[width=1\textwidth]{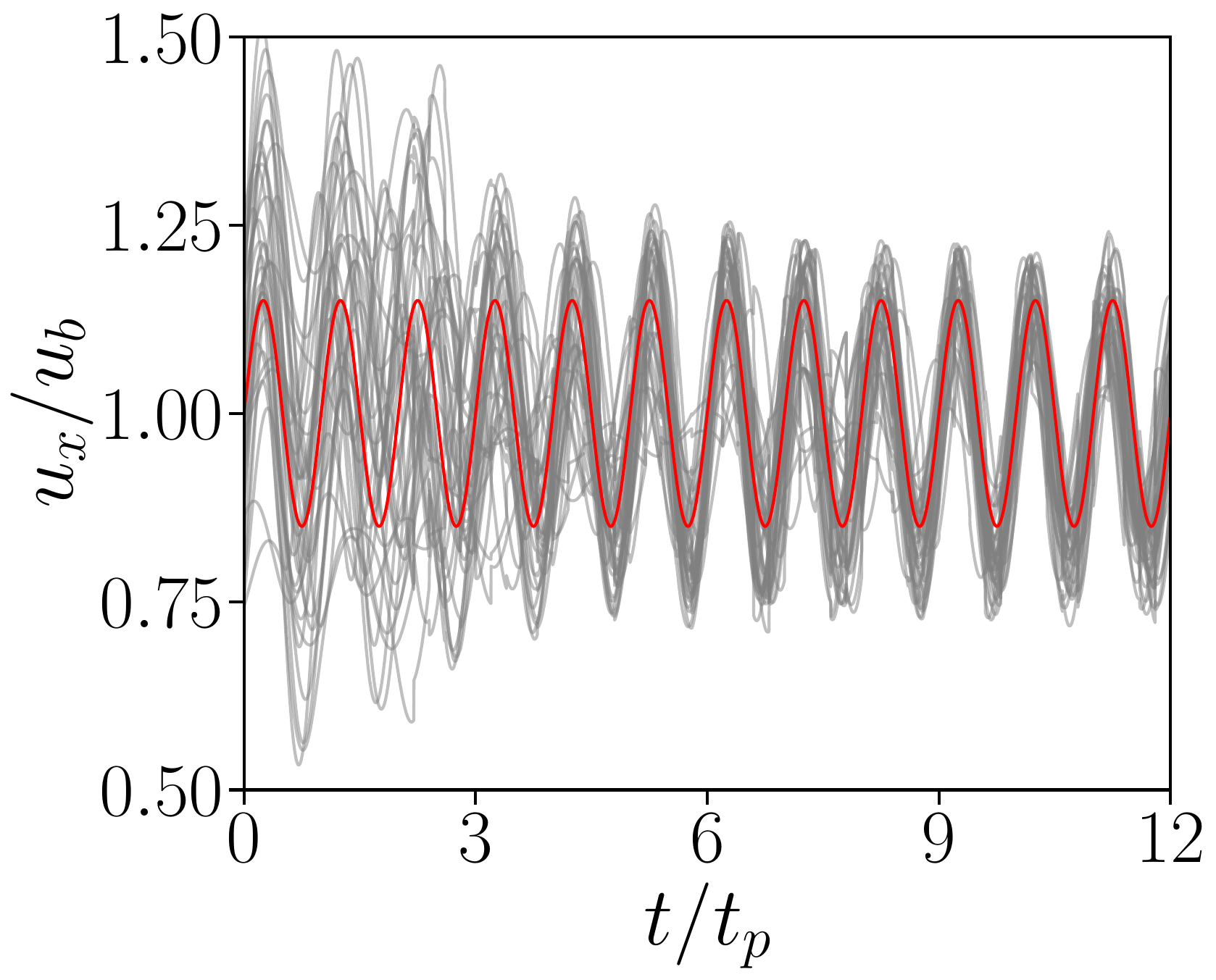}
        \caption{}
        \label{fig:inletVel_DA_OFR2_IR_conf05}
    \end{subfigure}
    \hfill
    \begin{subfigure}{.32\textwidth}
        \centering
        \includegraphics[width=1\textwidth]{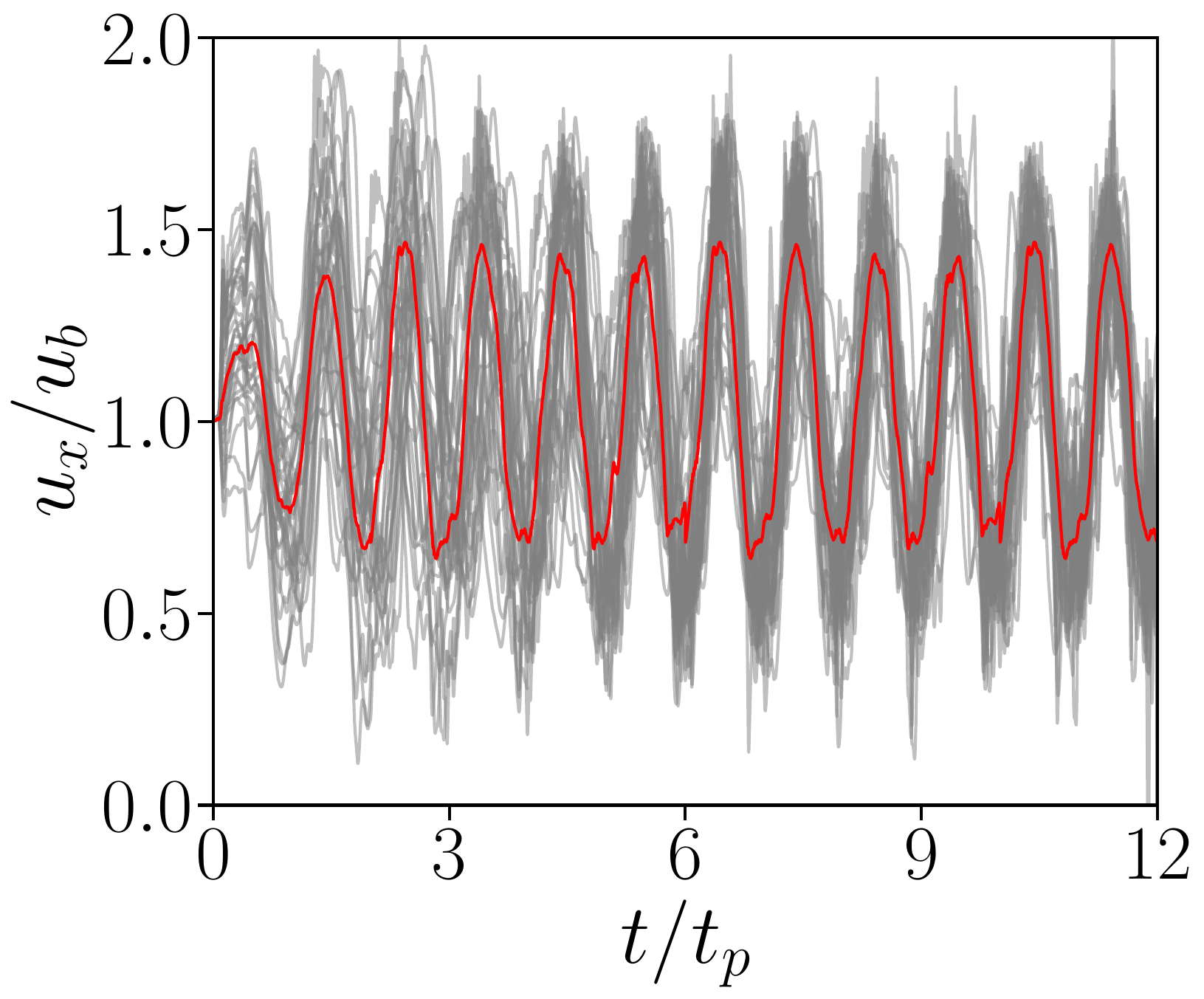}
        \caption{}
        \label{fig:middleVel_DA_OFR2_IR_conf05}
    \end{subfigure}
    \hfill
    \begin{subfigure}{.32\textwidth}
        \centering
        \includegraphics[width=1\textwidth]{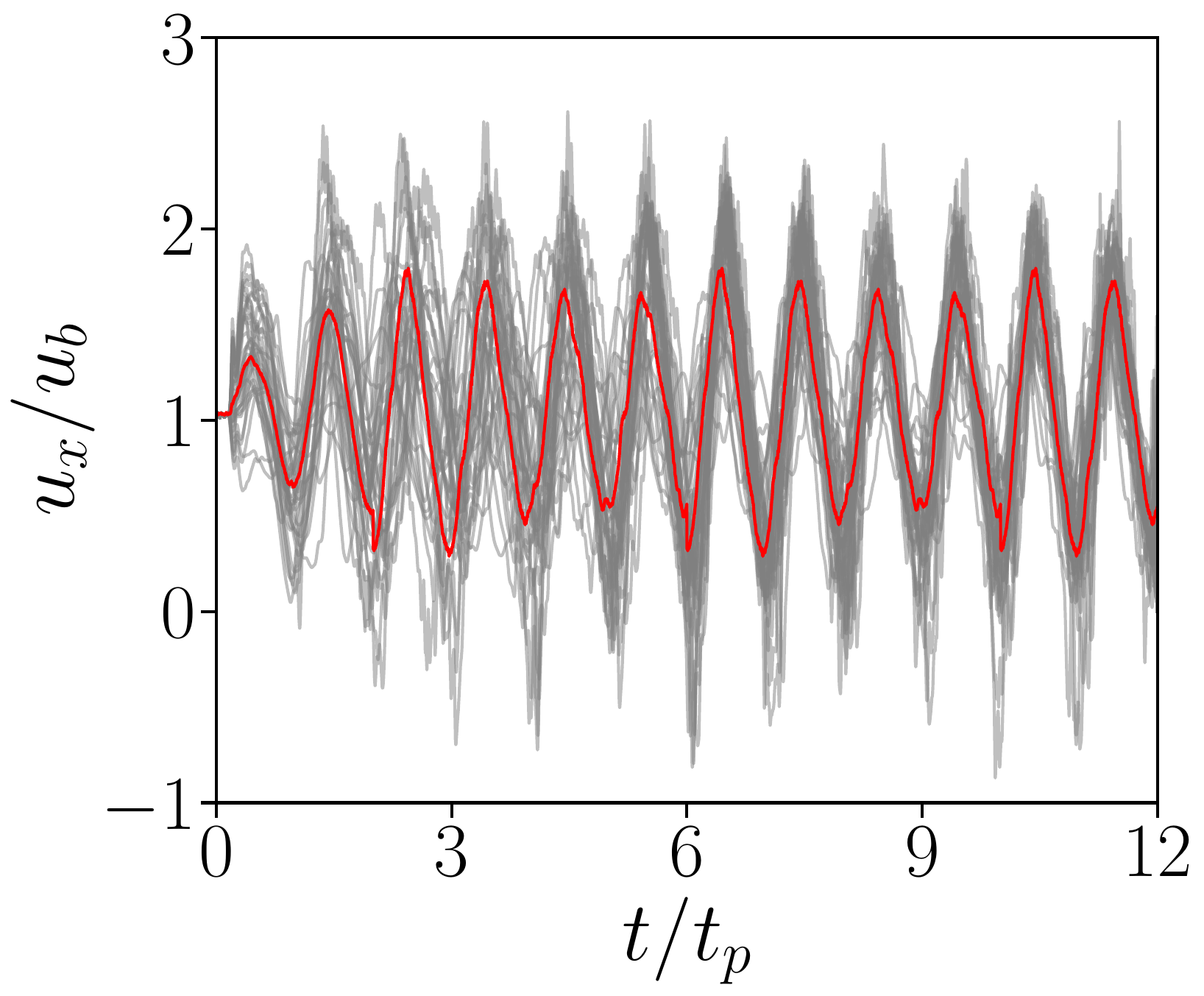}
        \caption{}
        \label{fig:endVel_DA_OFR2_IR_conf05}
    \end{subfigure}
    \hfill
    \begin{subfigure}{.33\textwidth}
        \centering
        \includegraphics[width=1\textwidth]{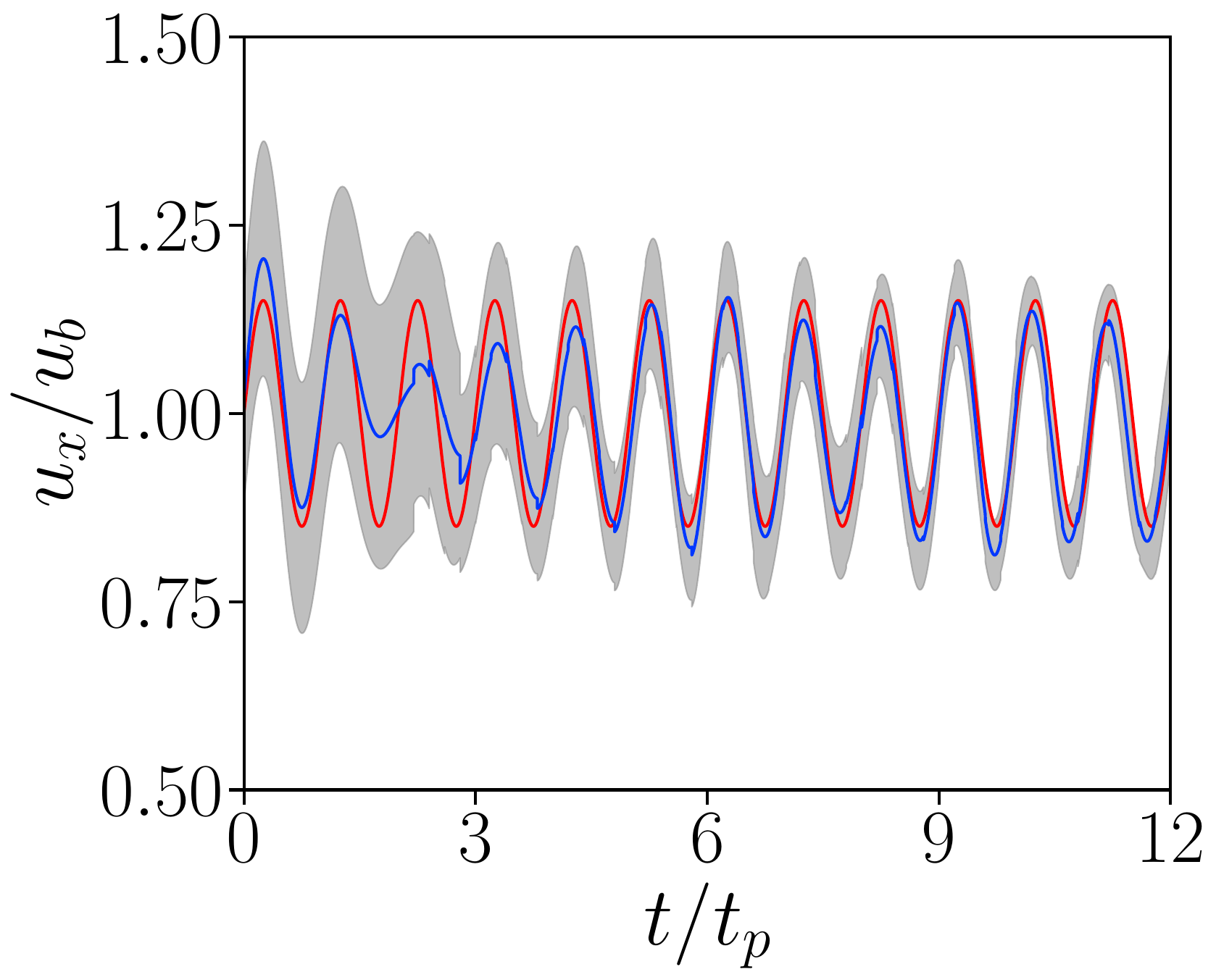}
        \caption{}
        \label{fig:inletVel_DA_OFR2_IR_MeanStd_conf05}
    \end{subfigure}
    \hfill
    \begin{subfigure}{.32\textwidth}
        \centering
        \includegraphics[width=1\textwidth]{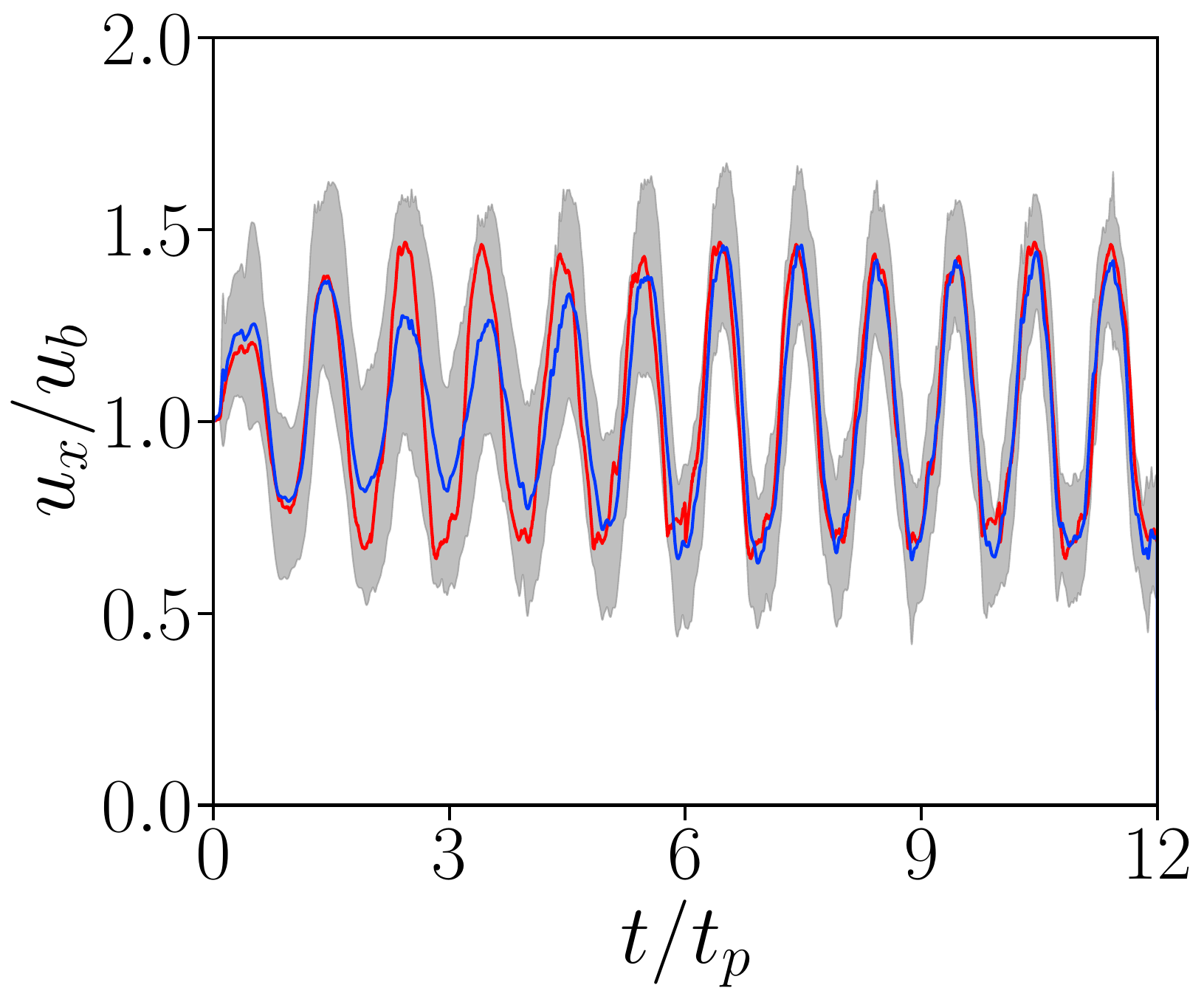}
        \caption{}
        \label{fig:middleVel_DA_OFR2_IR_MeanStd_conf05}
    \end{subfigure}
    \hfill
    \begin{subfigure}{.32\textwidth}
        \centering
        \includegraphics[width=1\textwidth]{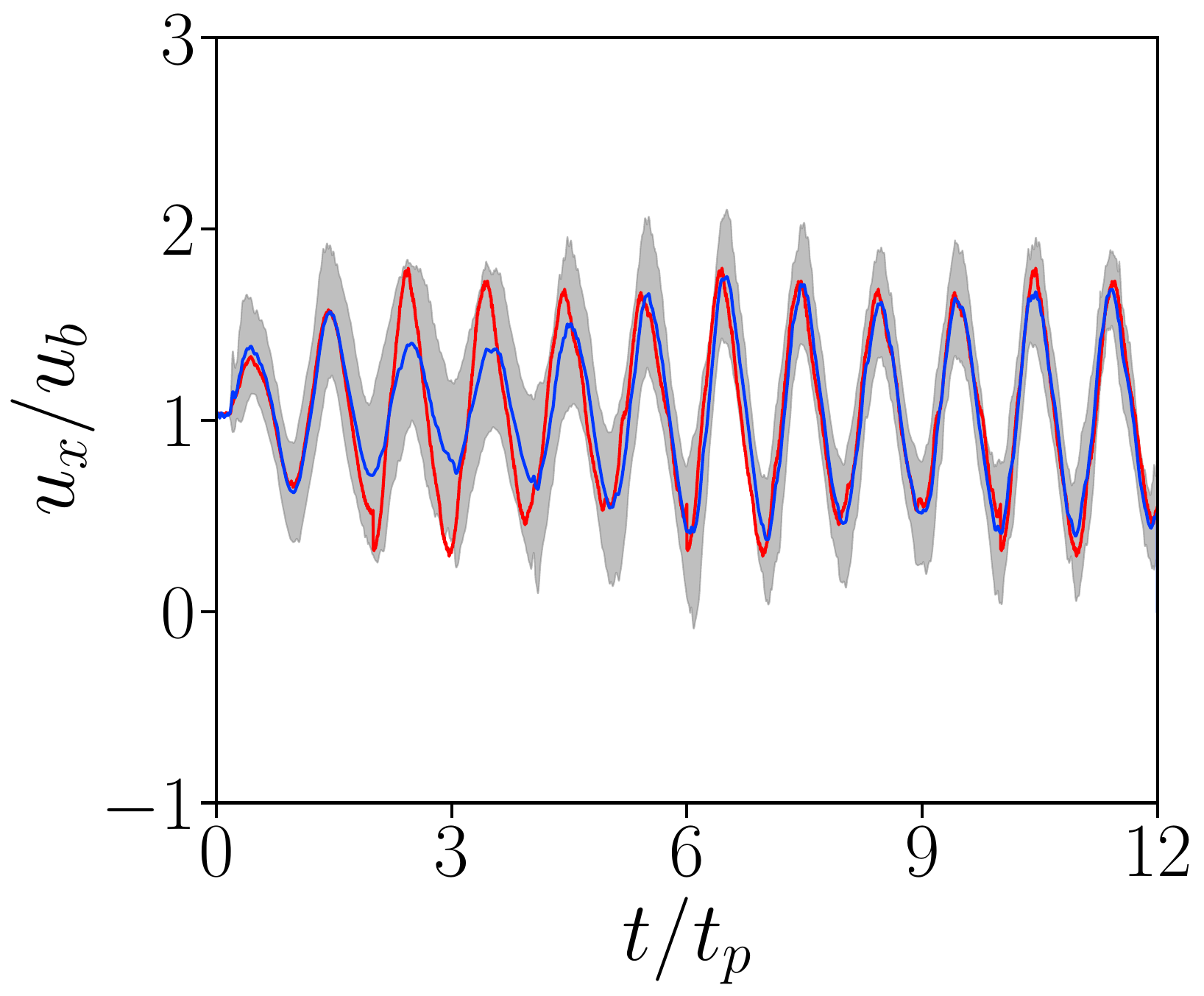}
        \caption{}
        \label{fig:endVel_DA_OFR2_IR_MeanStd_conf05}
    \end{subfigure}
    \caption{Evolution of the axial velocity for the distribution \textbf{\textit{prior state} 1}. Red curve: LES-HF. Grey curves: DA ensemble members. Blue curve: average of ensemble members. Shaded area: standard deviation from the mean. Data is sampled from sensors at (left column) $x=0$, (central column) $x=0.046$ and (right column) $x=0.092$ shown in Fig.~\ref{fig:SensorsInlet}.}
    \label{fig:prior1_optim}
\end{figure}

\begin{figure}
    \centering
    \begin{subfigure}{.33\textwidth}
        \centering
        \includegraphics[width=1\textwidth]{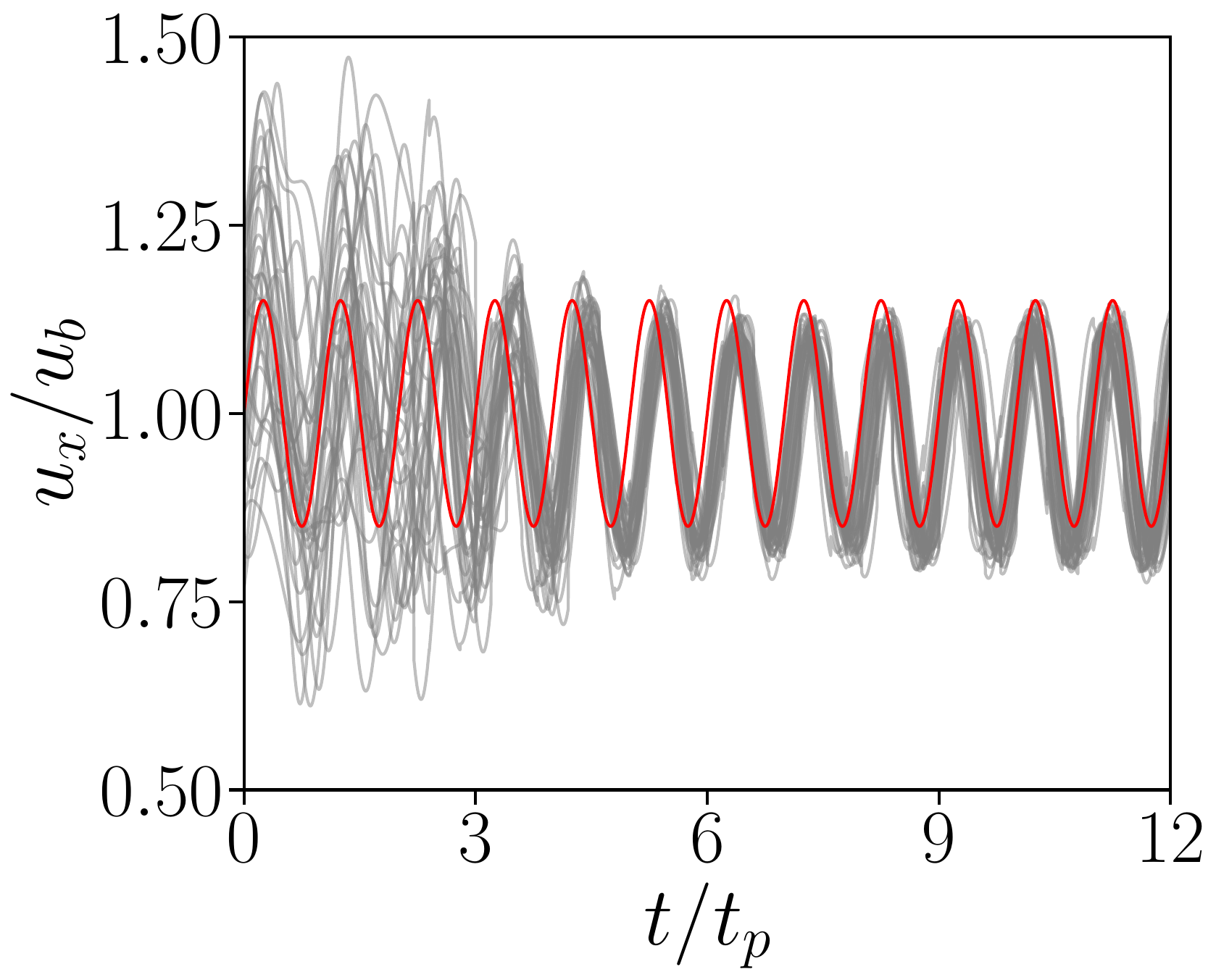}
        \caption{}
        \label{fig:inletVel_DA_OFR3_IR_conf05_Pi00375}
    \end{subfigure}
    \hfill
    \begin{subfigure}{.32\textwidth}
        \centering
        \includegraphics[width=1\textwidth]{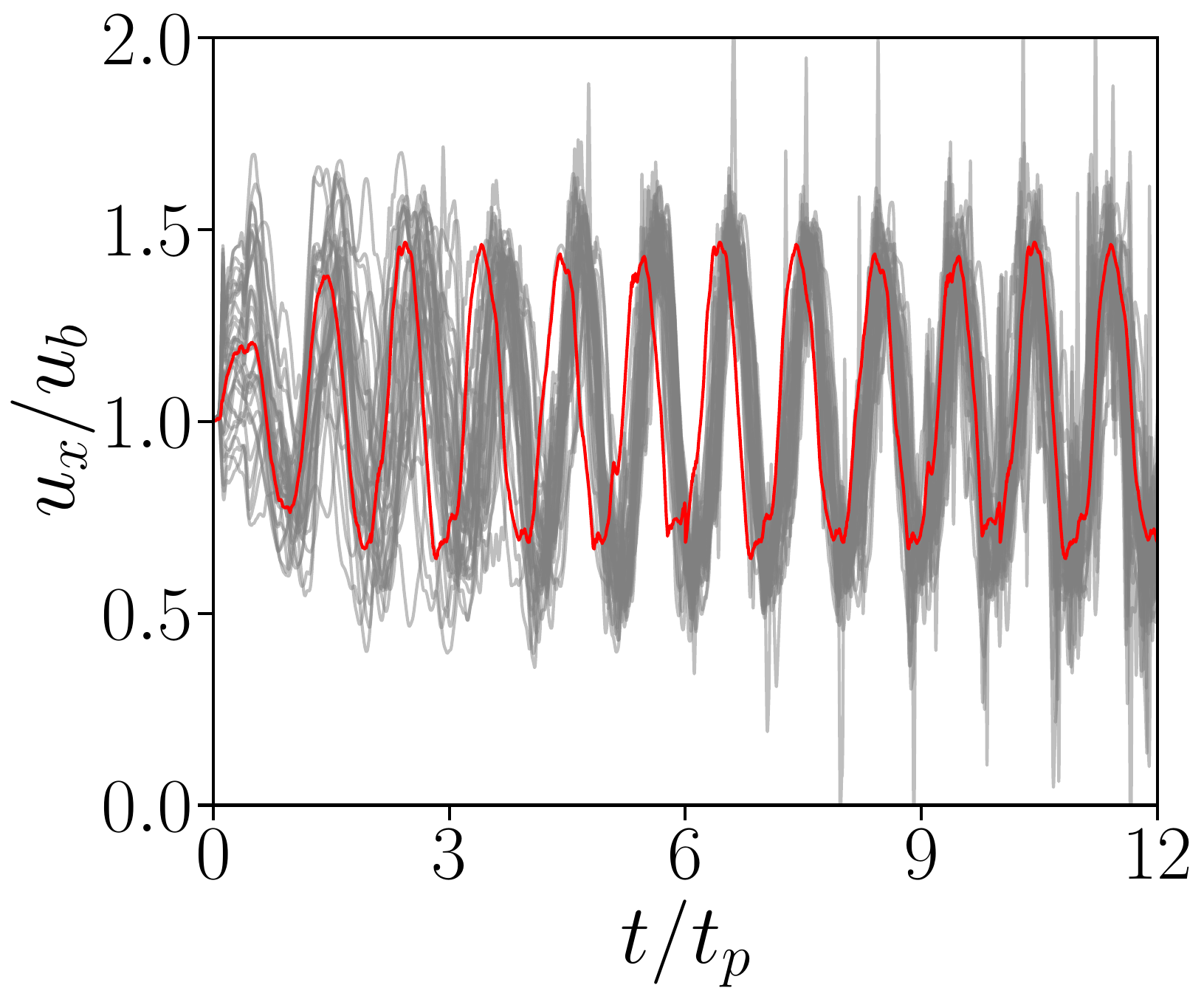}
        \caption{}
        \label{fig:middleVel_DA_OFR3_IR_conf05_Pi00375}
    \end{subfigure}
    \hfill
    \begin{subfigure}{.32\textwidth}
        \centering
        \includegraphics[width=1\textwidth]{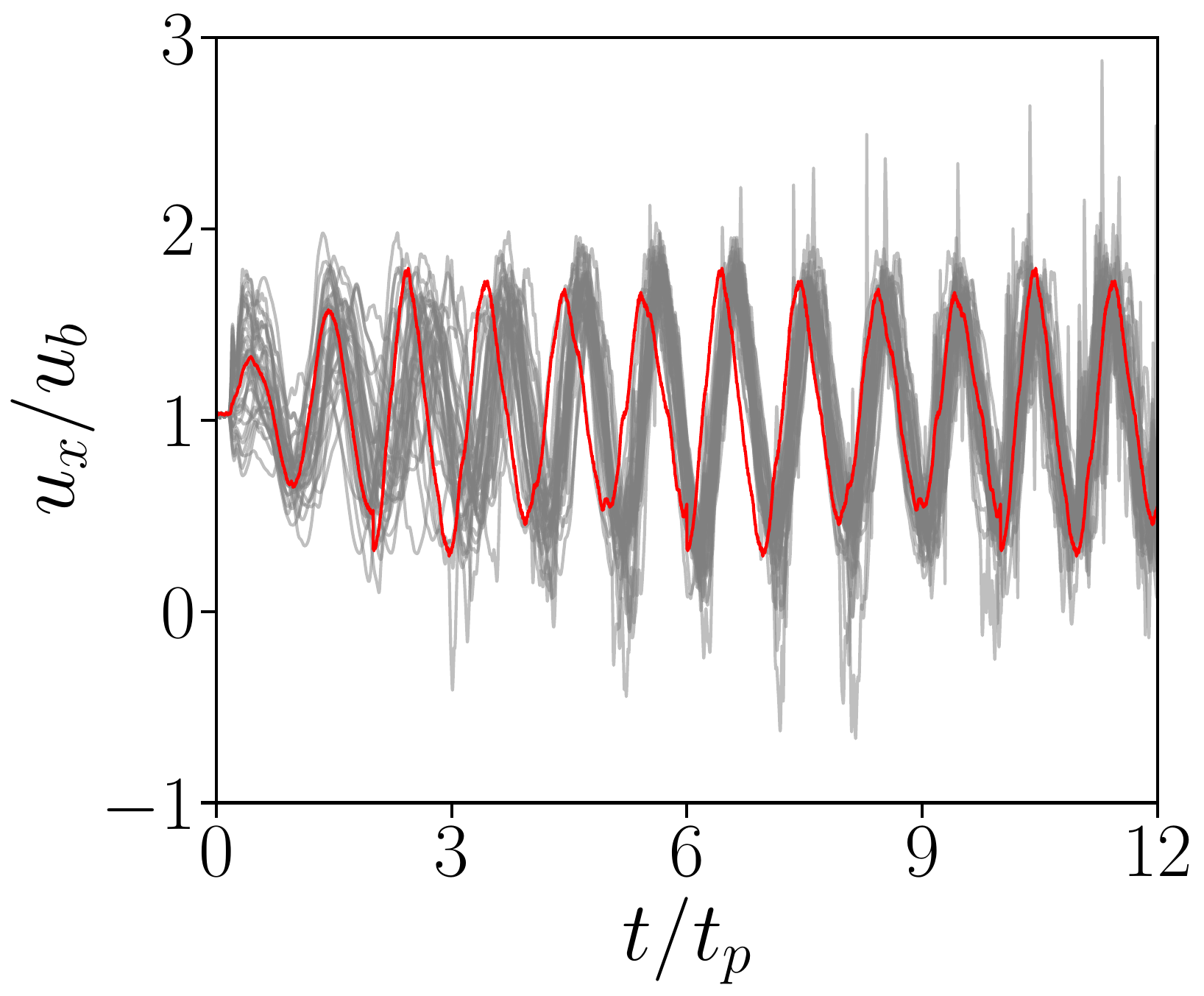}
        \caption{}
        \label{fig:endVel_DA_OFR3_IR_conf05_Pi00375}
    \end{subfigure}
    \hfill
    \begin{subfigure}{.33\textwidth}
        \centering
        \includegraphics[width=1\textwidth]{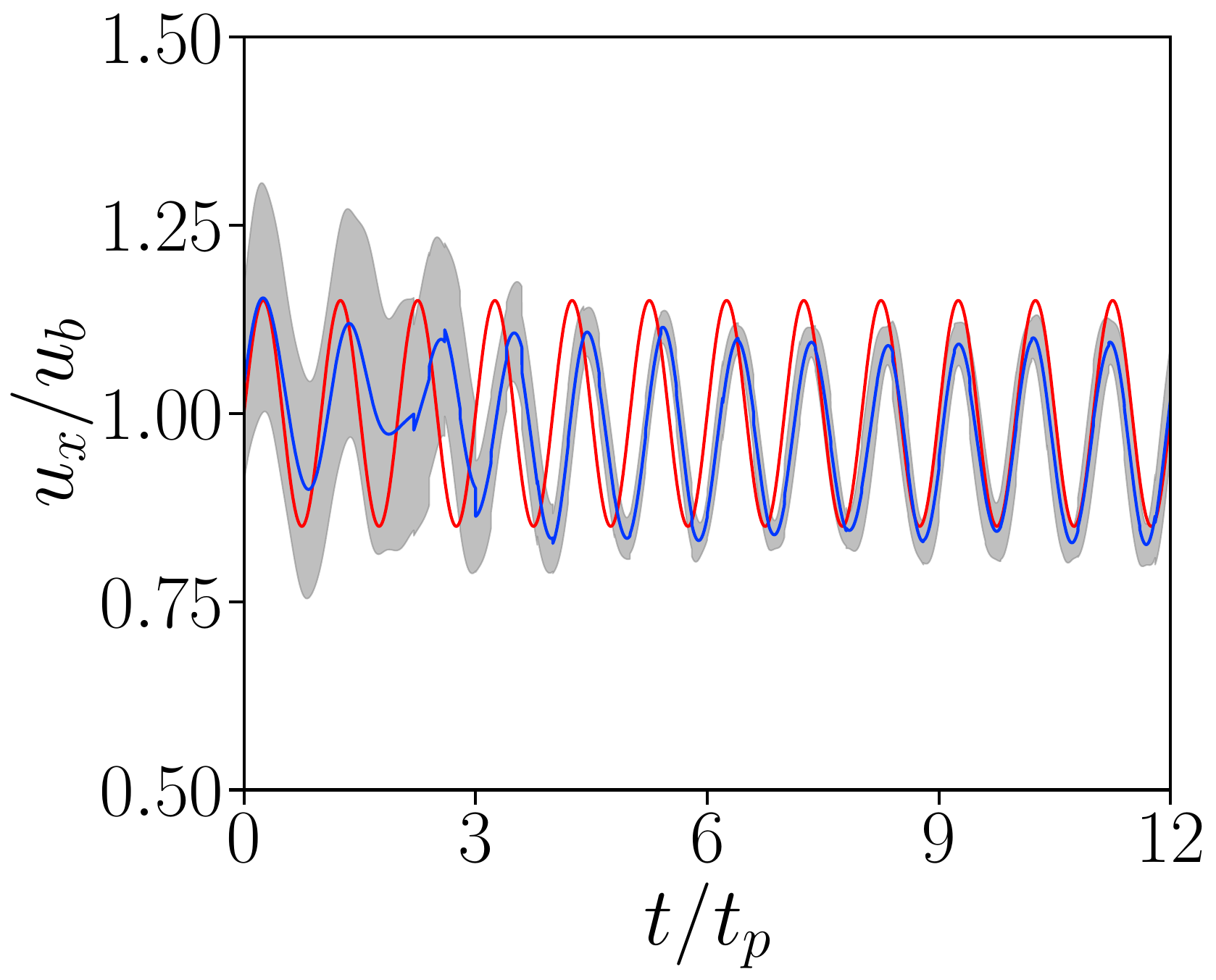}
        \caption{}
        \label{fig:inletVel_DA_OFR3_IR_MeanStd_conf05_Pi00375}
    \end{subfigure}
    \hfill
    \begin{subfigure}{.32\textwidth}
        \centering
        \includegraphics[width=1\textwidth]{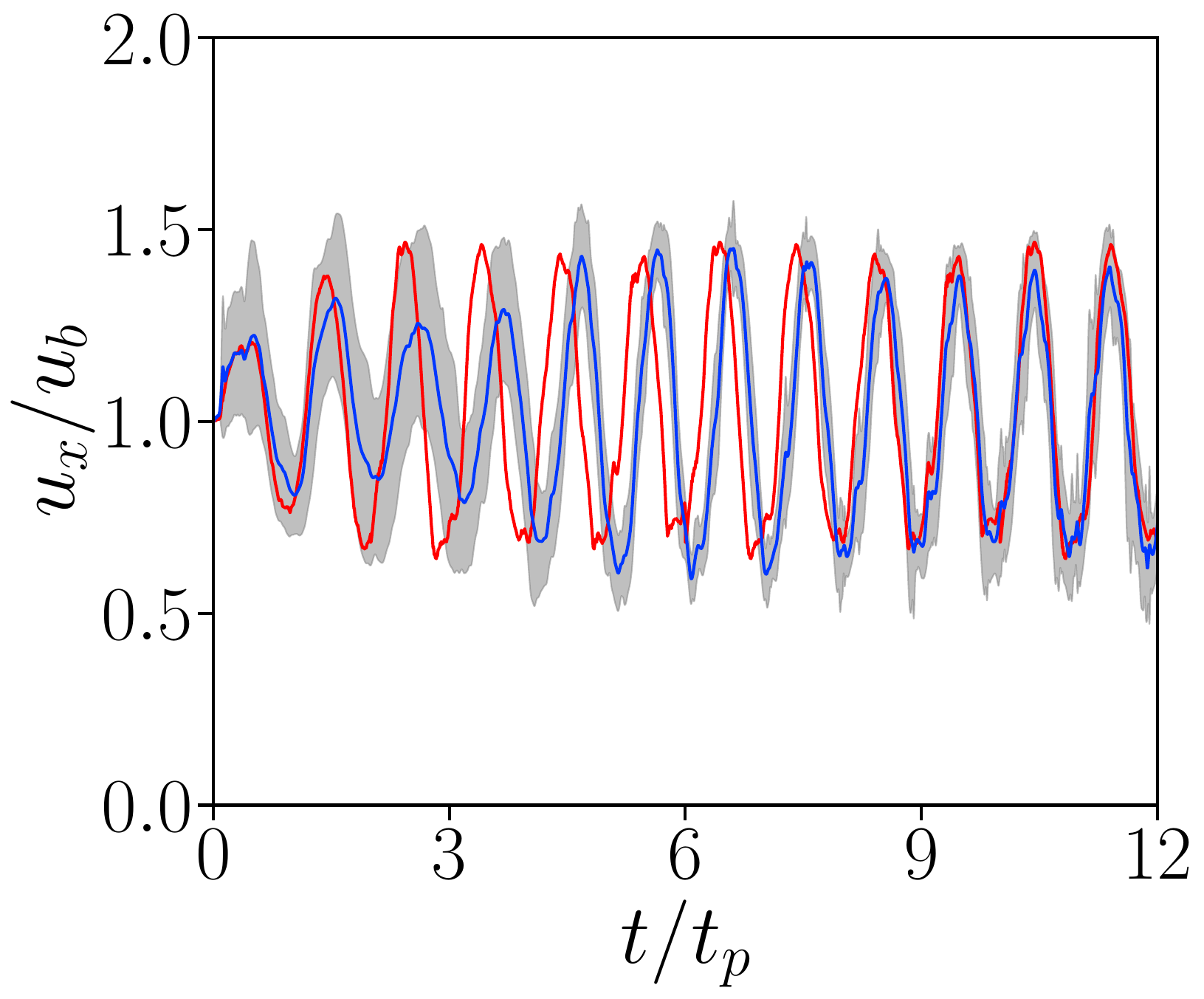}
        \caption{}
        \label{fig:middleVel_DA_OFR3_IR_MeanStd_conf05_Pi00375}
    \end{subfigure}
    \hfill
    \begin{subfigure}{.32\textwidth}
        \centering
        \includegraphics[width=1\textwidth]{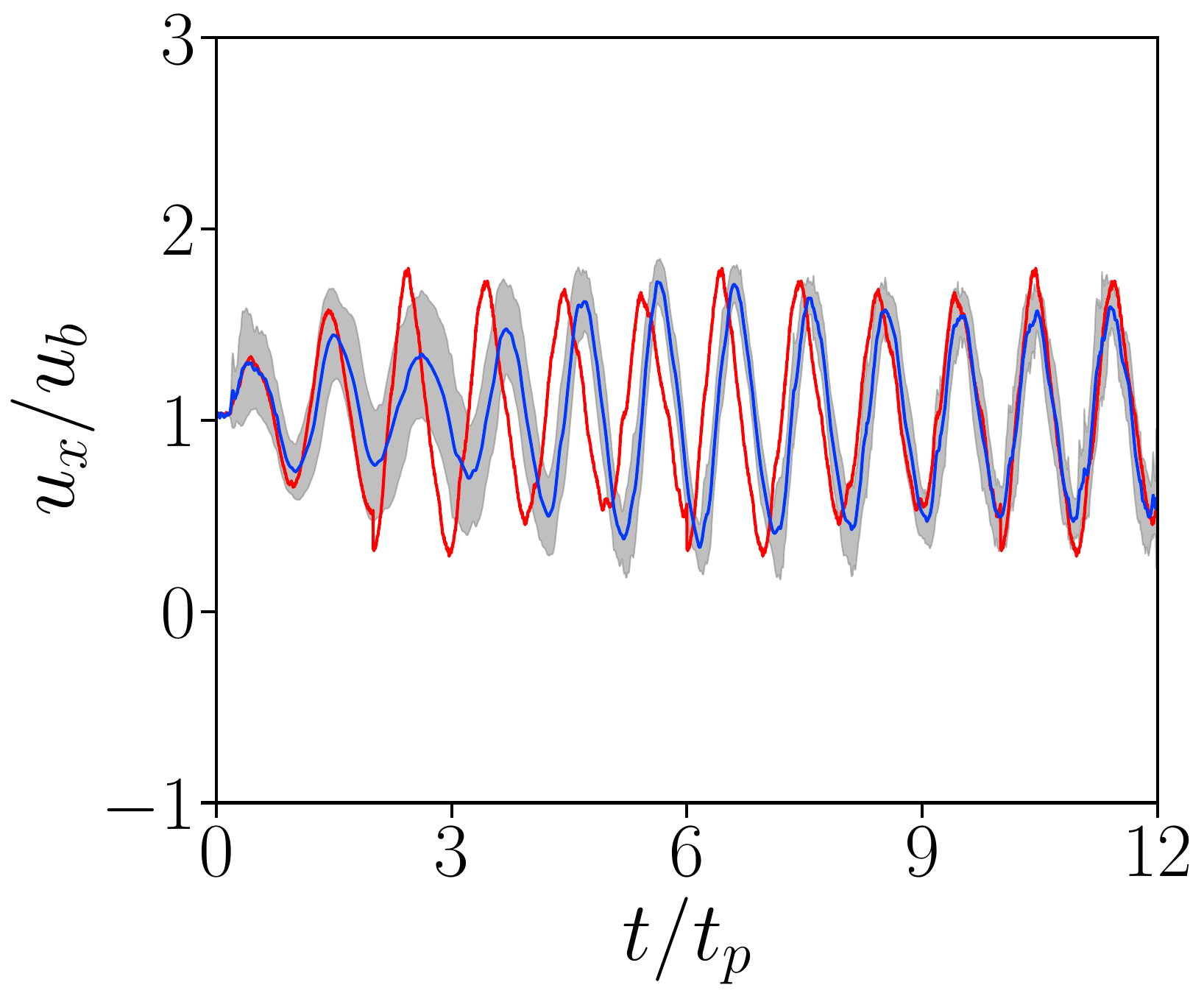}
        \caption{}
        \label{fig:endVel_DA_OFR3_IR_MeanStd_conf05_Pi00375}
    \end{subfigure}
    \caption{Evolution of the axial velocity for the distribution \textbf{\textit{prior state} 2}. Red curve: LES-HF. Grey curves: DA ensemble members. Blue curve: average of ensemble members. Shaded area: standard deviation from the mean. Data is sampled from sensors at (left column) $x=0$, (central column) $x=0.046$ and (right column) $x=0.092$ shown in Fig.~\ref{fig:SensorsInlet}.}
    \label{fig:prior2_optim}
\end{figure}

The results of the calibration of the free parameters of the inlet boundary condition for the \textit{prior state} 1 and the \textit{prior state} 2 are discussed in this section. Figures \ref{fig:prior1_optim} and \ref{fig:prior2_optim} show the velocity field for the ensemble realizations measured at the sensors positioned in the inlet duct (shown in Fig. \ref{fig:SensorsInlet}) which were used to study the features of the LES-HF run. The left column in the figures corresponds to the sensor located at the inlet boundary condition. The middle and right columns correspond to the sensors located in the middle of the intake pipe and in the proximity of the valve region, respectively. Comparing the present results with those reported in Fig. \ref{fig:prior} for the non-optimized inlet conditions, one can see that the modifications performed by the HLEnKF are visible from the second-third cycle, as expected. A slight delay can be observed for the two sensors in the middle and at the end of the intake pipe, caused by the time required to transport the updated solution downstream.. For both DA runs, the optimization procedure provides an adequate reconstruction of the flow features. In particular, the variance of the ensemble reaches a converged behaviour around cycle 5-6. One interesting point here is that, despite significant differences in the initial conditions for the two \textit{prior states} investigated, the DA algorithm successfully reconstructs the global dynamics, even when relying on localized instantaneous flow observations. This result highlights the robustness of the algorithm to variations of the prior conditions.

One interesting feature that is observed in both runs is that the amplitude of the velocity signal obtained via DA (blue line) is very similar to the one sampled for the simulation LES-HF for the sensor downstream (right column). Moving upstream, one can see that the signal follows the same dynamics, but it is, on average, lower. This is particularly visible at the inlet (left column) for the \textit{prior state} 2. This result is justified by the numerical and modelling errors observed in the simulation LES-LF using a coarse grid. One consequence of these errors is that the velocity at the end of the intake pipe is higher than for the case LES-HF. Therefore, one can see that the DA optimization here decreases the velocity at the inlet, in order to match the observed velocity in the valve region. The DA calibration of the inlet compensates the effects of the numerical errors over the flow prediction in the intake pipe, so that the prediction in the valve region between model and observation is the same.

\begin{figure}
    \centering
    \begin{subfigure}{.24\textwidth}
        \centering
        \includegraphics[width=1\textwidth]{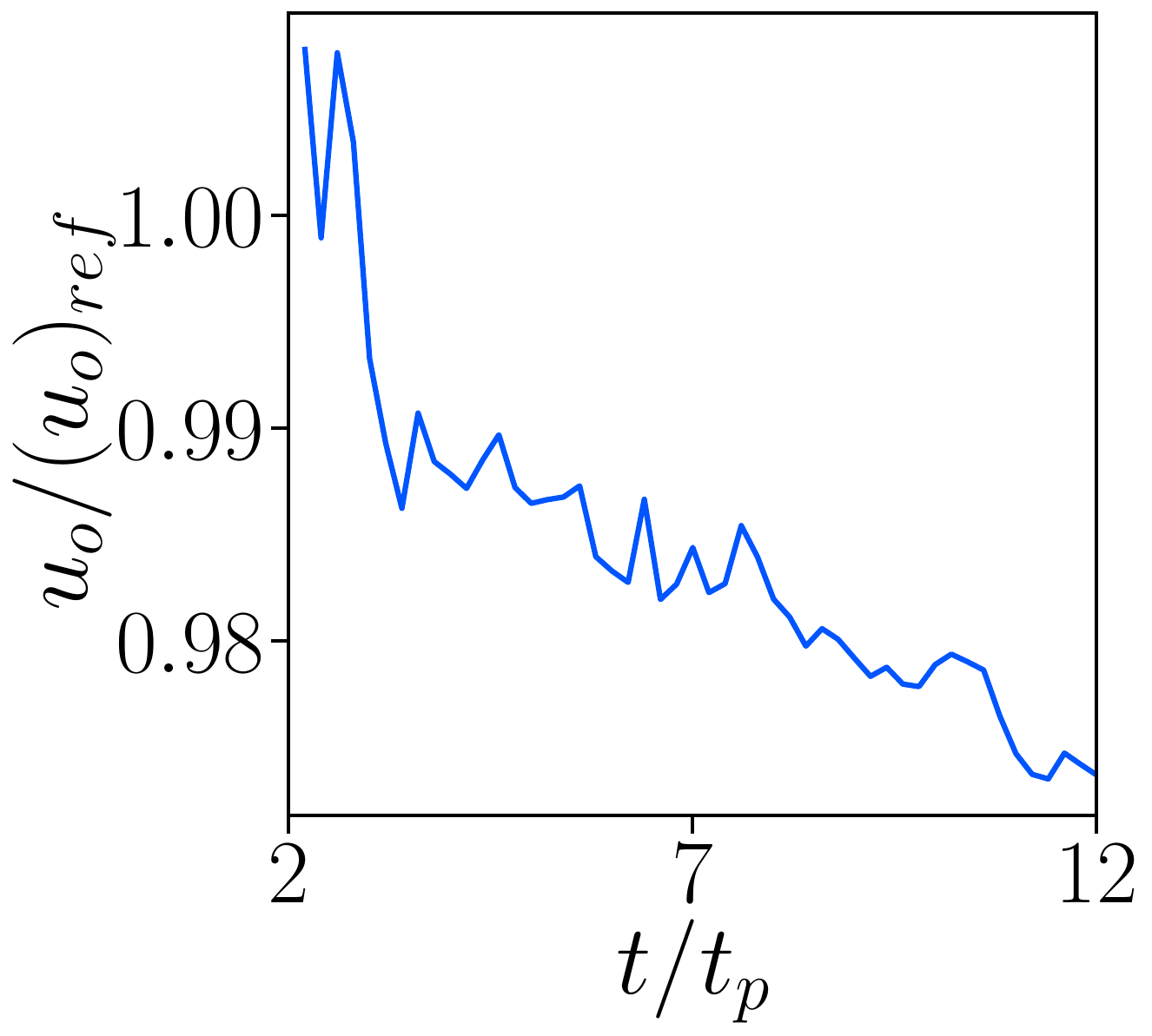}
        \caption{}
        \label{fig:prior1_offset}
    \end{subfigure}
    \hfill
    \begin{subfigure}{.24\textwidth}
    \centering
        \includegraphics[width=1\textwidth]{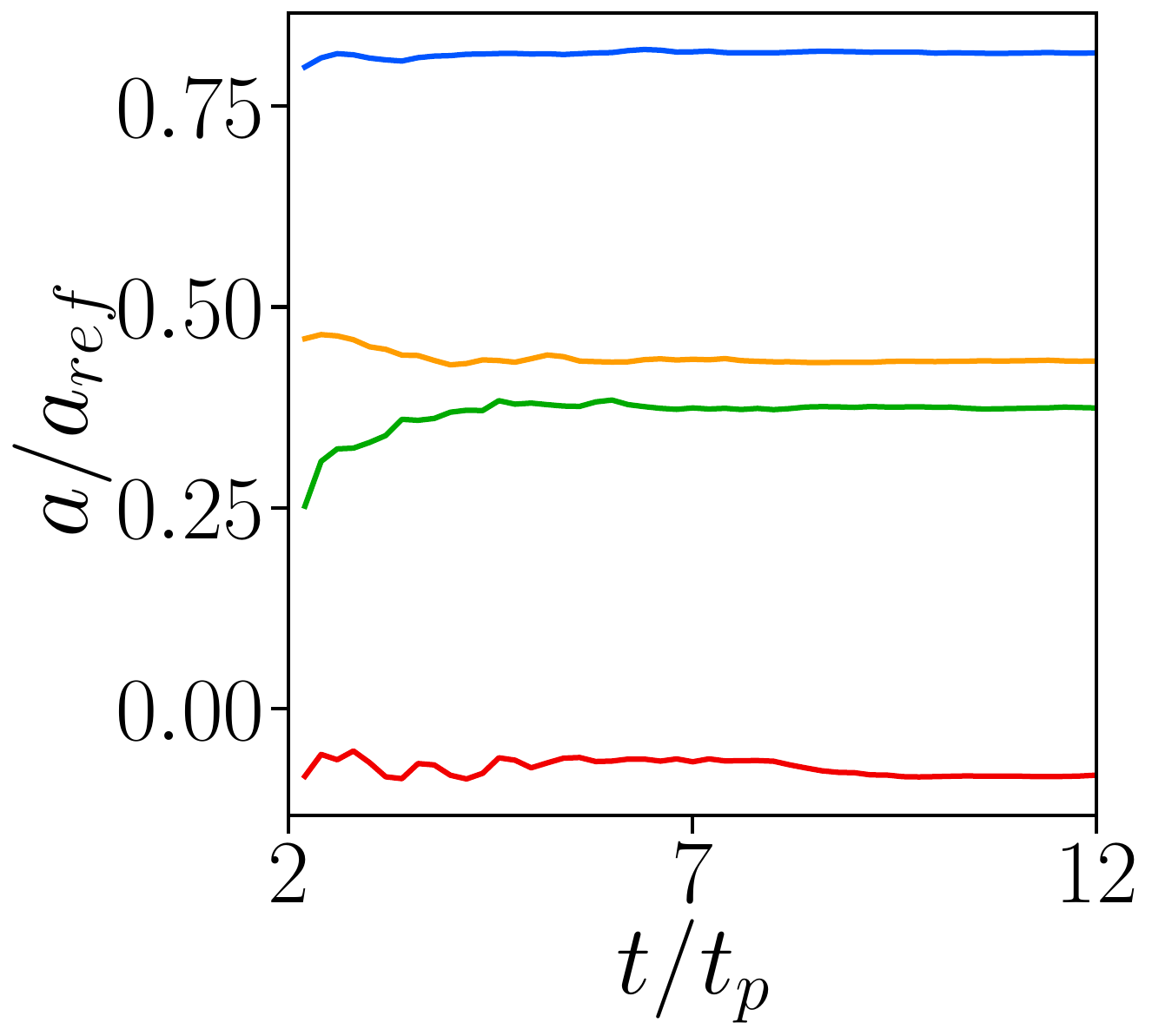}
        \caption{}
        \label{fig:prior1_amplitudes}
    \end{subfigure}
    \hfill
    \begin{subfigure}{.24\textwidth}
    \centering
        \includegraphics[width=1\textwidth]{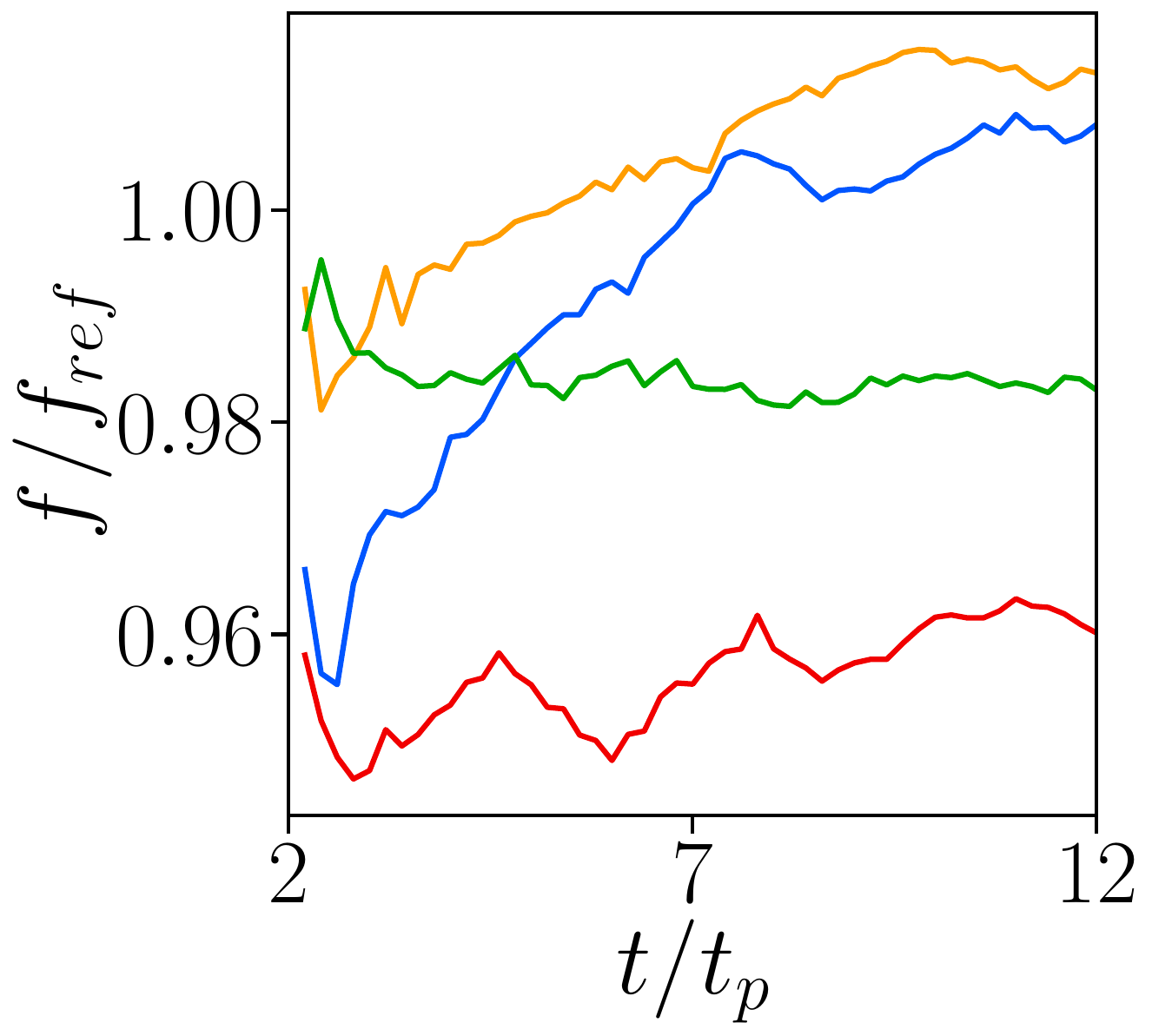}
        \caption{}
        \label{fig:prior1_frequencies}
    \end{subfigure}
    \hfill
    \begin{subfigure}{.251\textwidth}
    \centering
        \includegraphics[width=1\textwidth]{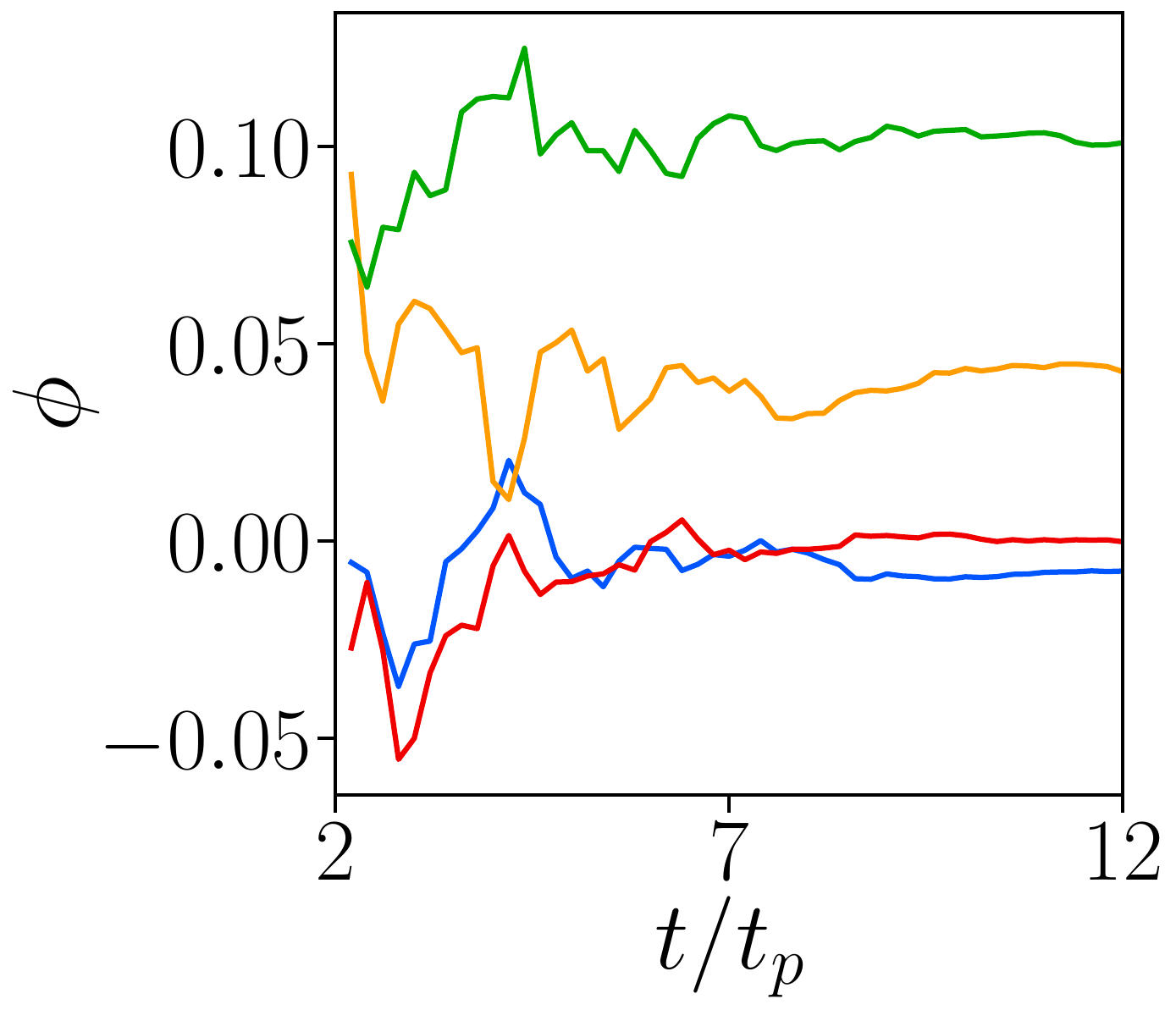}
        \caption{}
        \label{fig:prior1_phases}
    \end{subfigure}
    \caption{Evolution of the parameters describing the inlet condition during the assimilation procedure for the distribution \textbf{\textit{prior state } 1}. Blue: sine function 1. Yellow: sine function 2. Green : sine function 3. Red: sine function 4.}
    \label{fig:paramEvolution_prior1}
\end{figure}

\begin{figure}
    \centering
    \begin{subfigure}{.248\textwidth}
        \centering
        \includegraphics[width=1\textwidth]{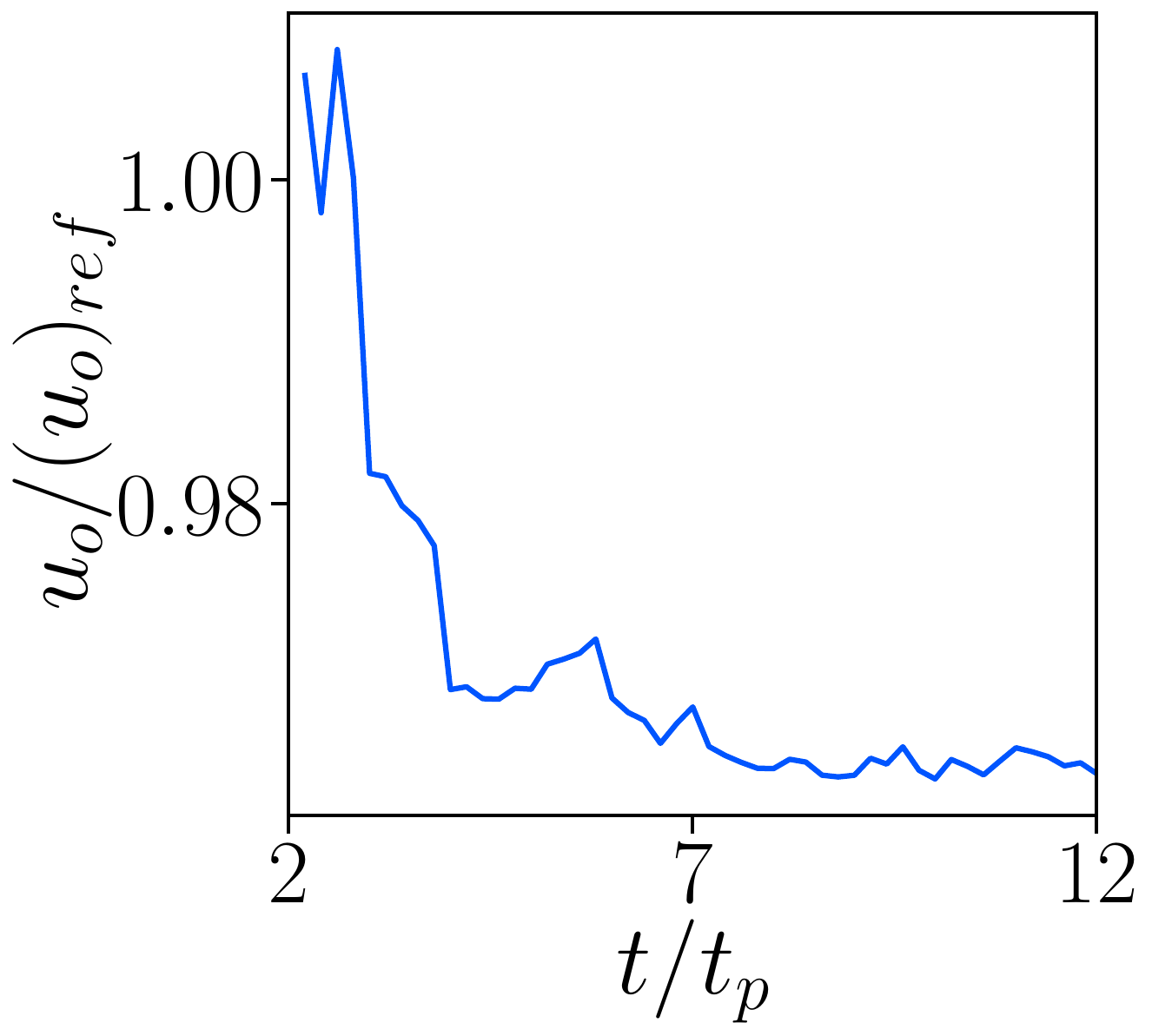}
        \caption{}
        \label{fig:prior2_offset}
    \end{subfigure}
    \hfill
    \begin{subfigure}{.236\textwidth}
    \centering
        \includegraphics[width=1\textwidth]{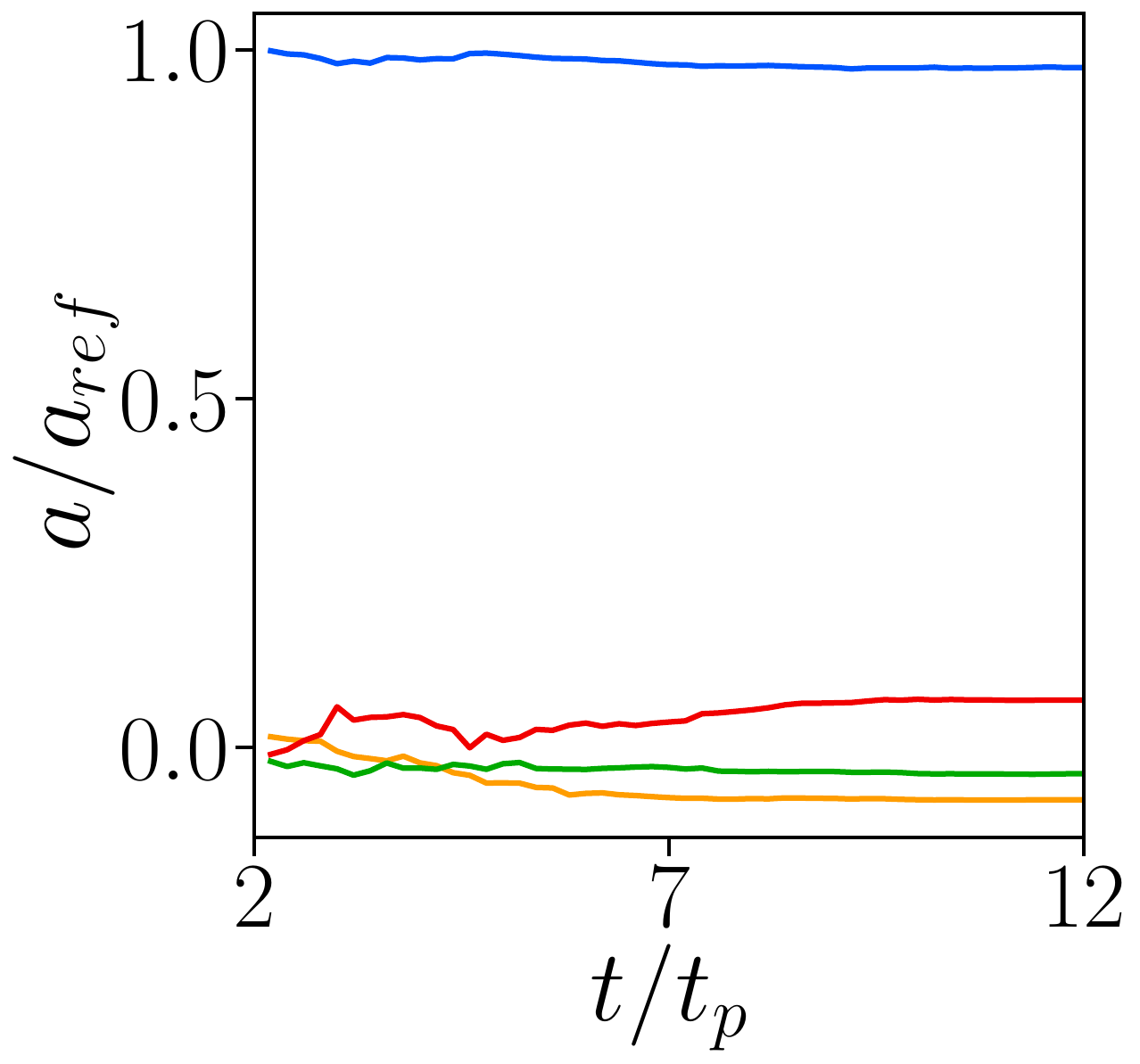}
        \caption{}
        \label{fig:prior2_amplitudes}
    \end{subfigure}
    \hfill
    \begin{subfigure}{.245\textwidth}
    \centering
        \includegraphics[width=1\textwidth]{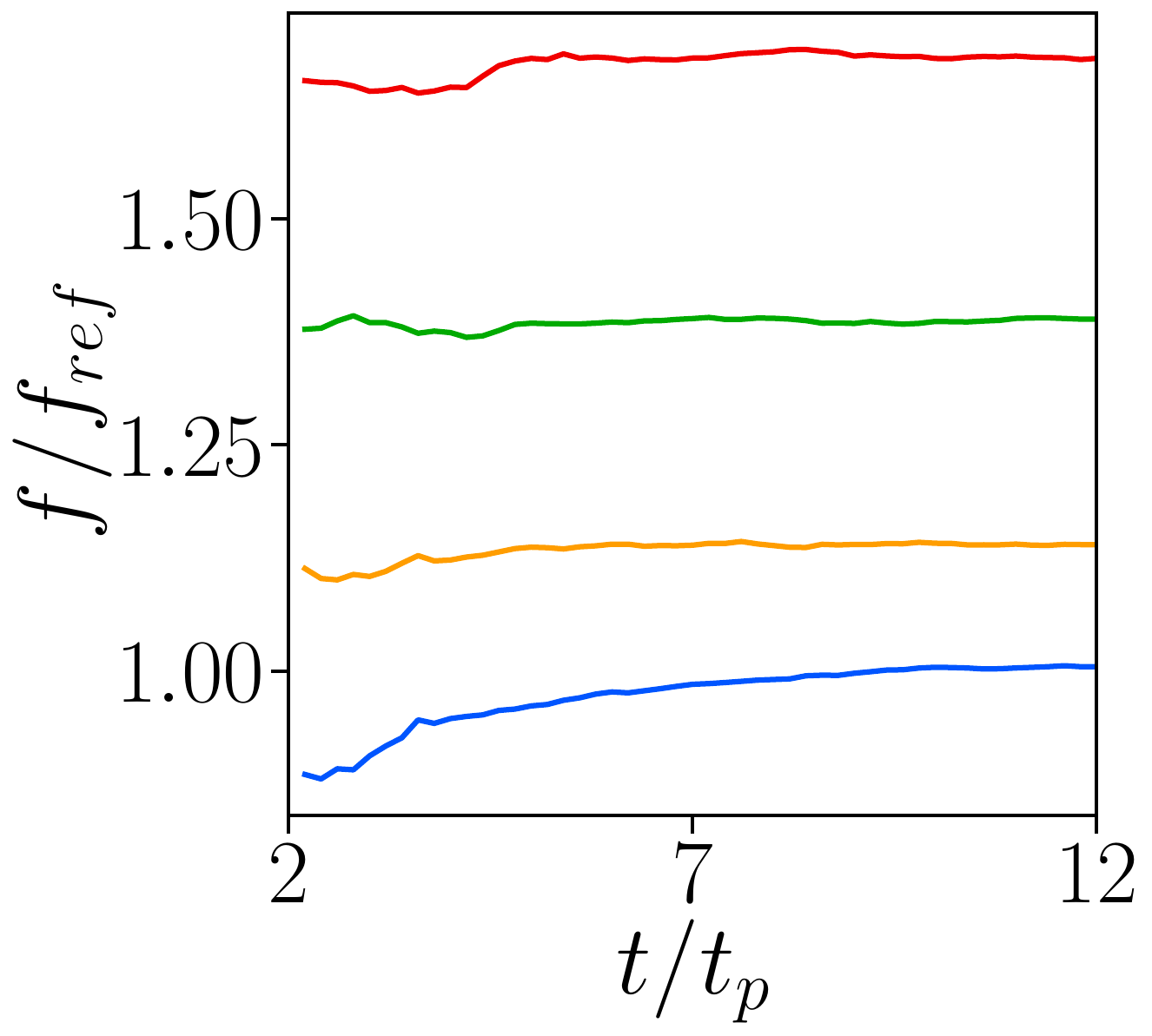}
        \caption{}
        \label{fig:prior2_frequencies}
    \end{subfigure}
    \hfill
    \begin{subfigure}{.25\textwidth}
    \centering
        \includegraphics[width=1\textwidth]{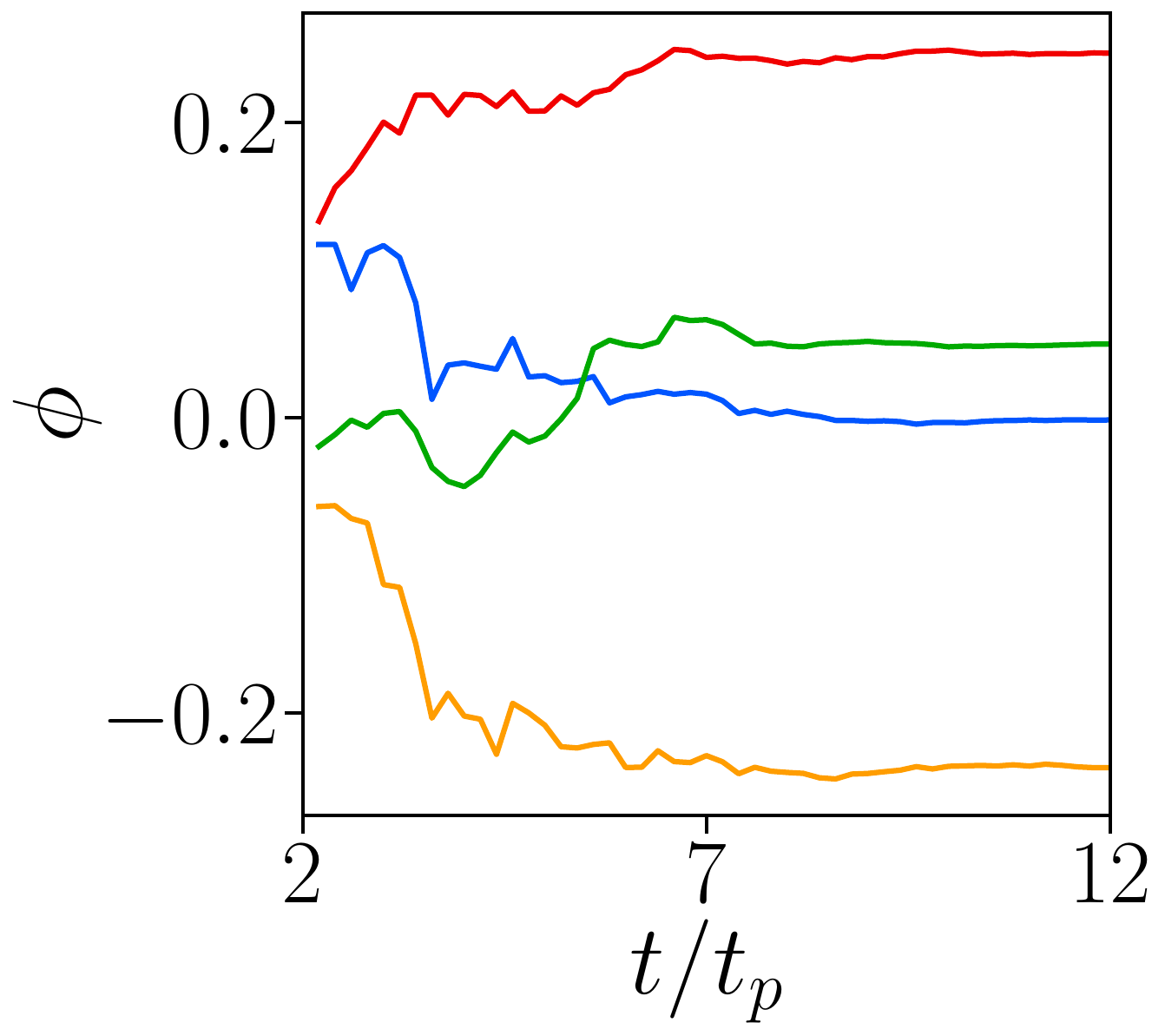}
        \caption{}
        \label{fig:prior2_phases}
    \end{subfigure}
    \caption{Evolution of the parameters describing the inlet condition during the assimilation procedure for the distribution \textbf{\textit{prior state } 2}. Blue: sine function 1. Yellow: sine function 2. Green : sine function 3. Red: sine function 4.}
    \label{fig:paramEvolution_prior2}
\end{figure}

The same conclusions can be drawn via the investigation of the time evolution of the ensemble average of the free parameters describing the inlet velocity, which are shown in Fig. \ref{fig:paramEvolution_prior1} and \ref{fig:paramEvolution_prior2} for both DA runs. The base velocity (a), the amplitudes (b) and the frequencies (c) are adimensionalized by the reference values used for the simulation LES-HF. The base velocity seems to converge to a value $2$ to $4\%$ lower than the reference value for both DA runs. This result supports previous discussion about the compensation of numerical and modelling errors due to the usage of the coarse grid. The amplitude parameters are now investigated. A perfect DA calibration would have here provided the convergence of $3$ amplitude parameters to $0$ while the last one would have converged to the value imposed for the LES-HF run. Differences are here observed for the cases \textit{prior state} 1 and 2. For the latter, the amplitude coefficients actually converge to the desired values pretty quickly with a tolerance of $\approx5\%$, and they exhibit very mild variations in time. It is however important to stress that the amplitude parameters here converge to the mean value of the Gaussian distribution imposed as initial condition. For the case \textit{prior state} 1, a mild evolution is observed in particular for the parameters $a_2$ and $a_3$, which appear to compensate for each other (see Fig. \ref{fig:prior1_amplitudes}). Therefore, the DA algorithm yields two distinct optimization strategies. For \textit{prior state 1}, mild variations in amplitudes are observed, which are compensated by occasionally significant changes in phase, frequencies and base velocity. For the latter two, one could even argue that no convergence is reached, as the algorithm continues to adjust these quantities over time to compensate for the inaccurate convergence observed in the amplitude parameters. On the other hand, the better guess in terms of initial estimation of the amplitude coefficients for \textit{prior state} 2 provides a rapid convergence towards expected values for all the parameters at play, see in particular the blue lines for the frequency and the phase in Fig. \ref{fig:paramEvolution_prior2}.

\begin{figure}[h]
    \centering
    \begin{subfigure}{.32\textwidth}
        \centering
        \includegraphics[width=1\textwidth]{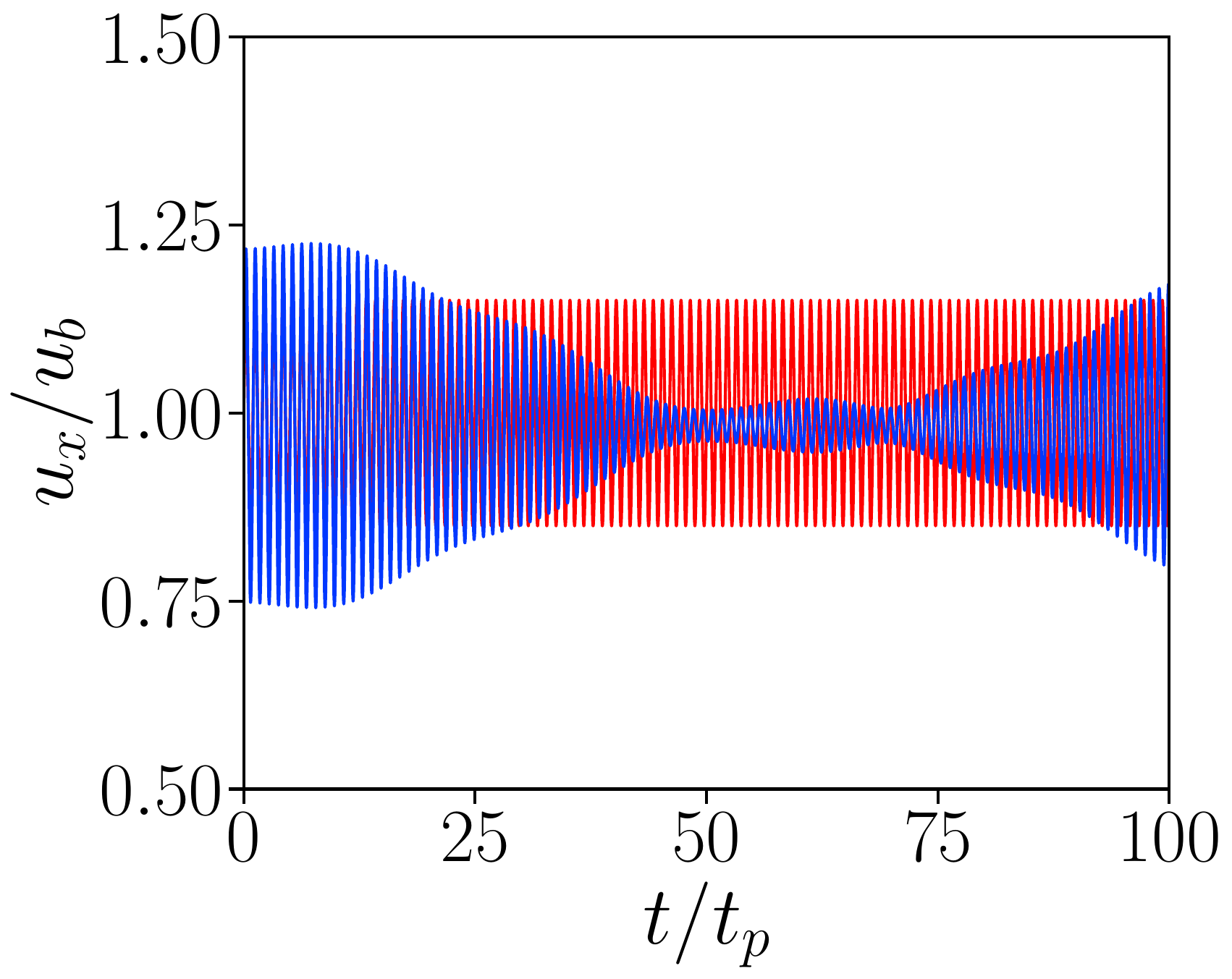}
        \caption{}
        \label{fig:prior2_0075}
    \end{subfigure}
    \hfill
    \begin{subfigure}{.32\textwidth}
        \centering
        \includegraphics[width=1\textwidth]{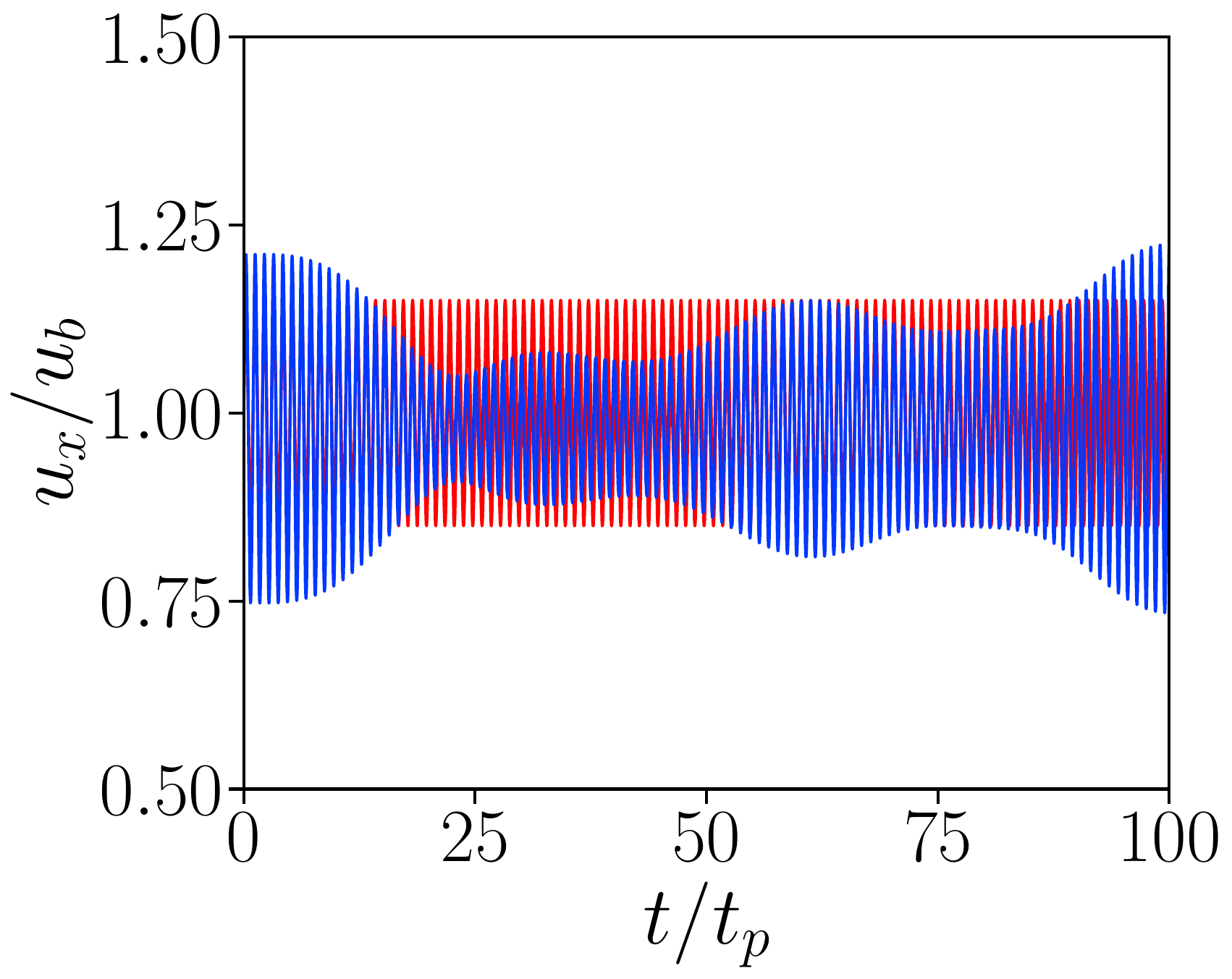}
        \caption{}
        \label{fig:prior2_01125}
    \end{subfigure}
    \hfill
    \begin{subfigure}{.32\textwidth}
        \centering
        \includegraphics[width=1\textwidth]{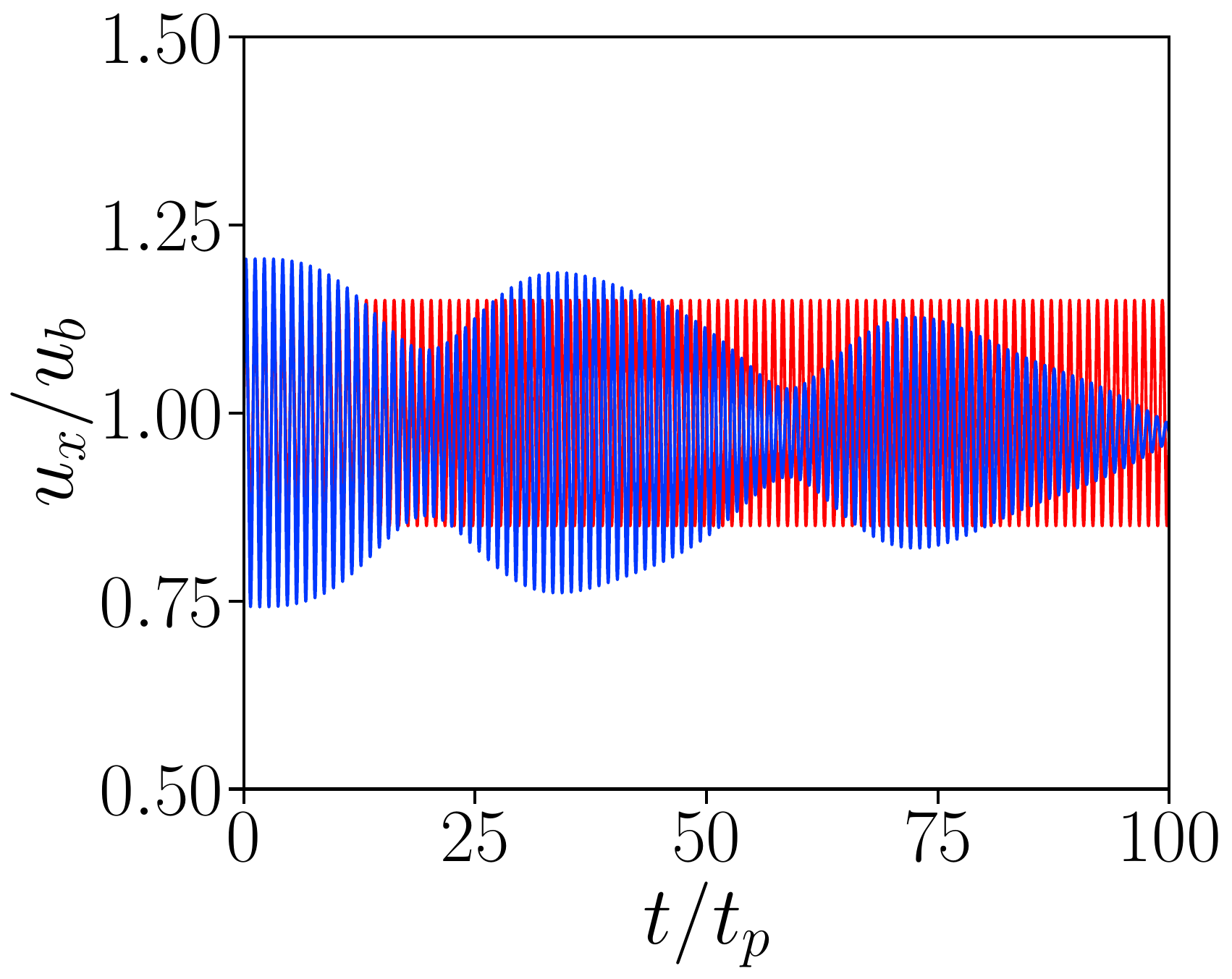}
        \caption{}
        \label{fig:prior2_015}
    \end{subfigure}
    \hfill
    \begin{subfigure}{.32\textwidth}
        \centering
        \includegraphics[width=1\textwidth]{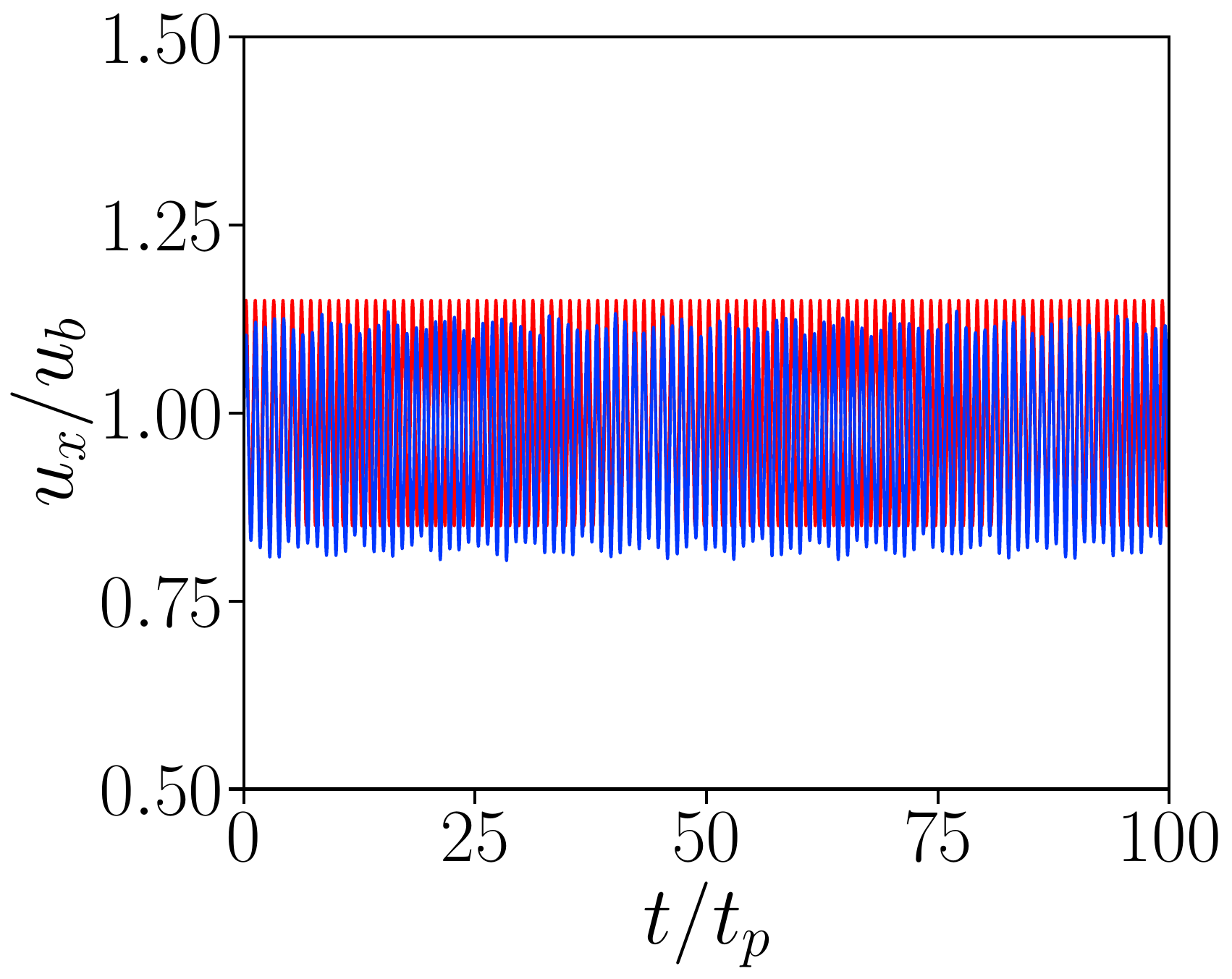}
        \caption{}
        \label{fig:prior3_0075}
    \end{subfigure}
    \hfill
    \begin{subfigure}{.32\textwidth}
        \centering
        \includegraphics[width=1\textwidth]{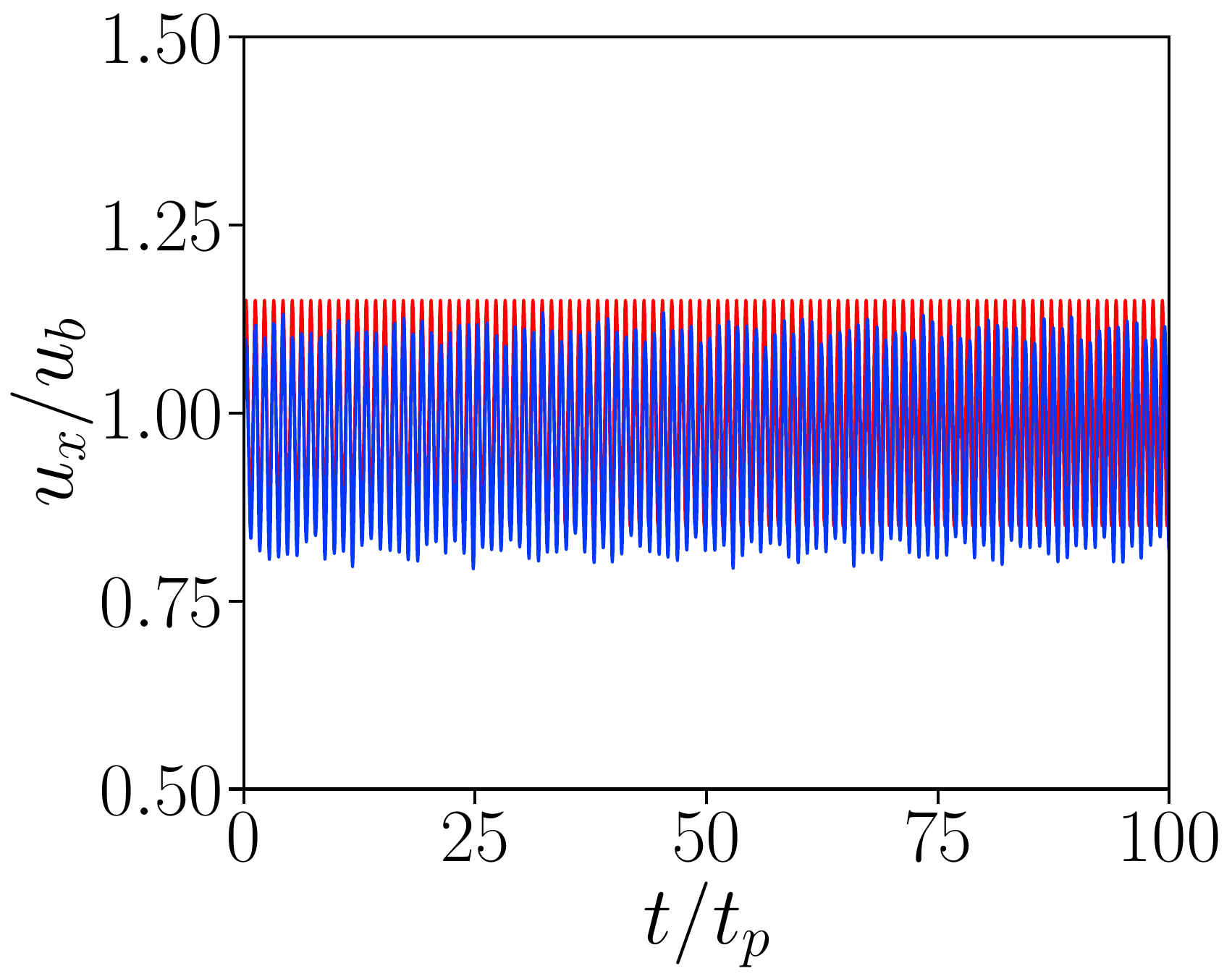}
        \caption{}
        \label{fig:prior3_01125}
    \end{subfigure}
    \hfill
    \begin{subfigure}{.32\textwidth}
        \centering
        \includegraphics[width=1\textwidth]{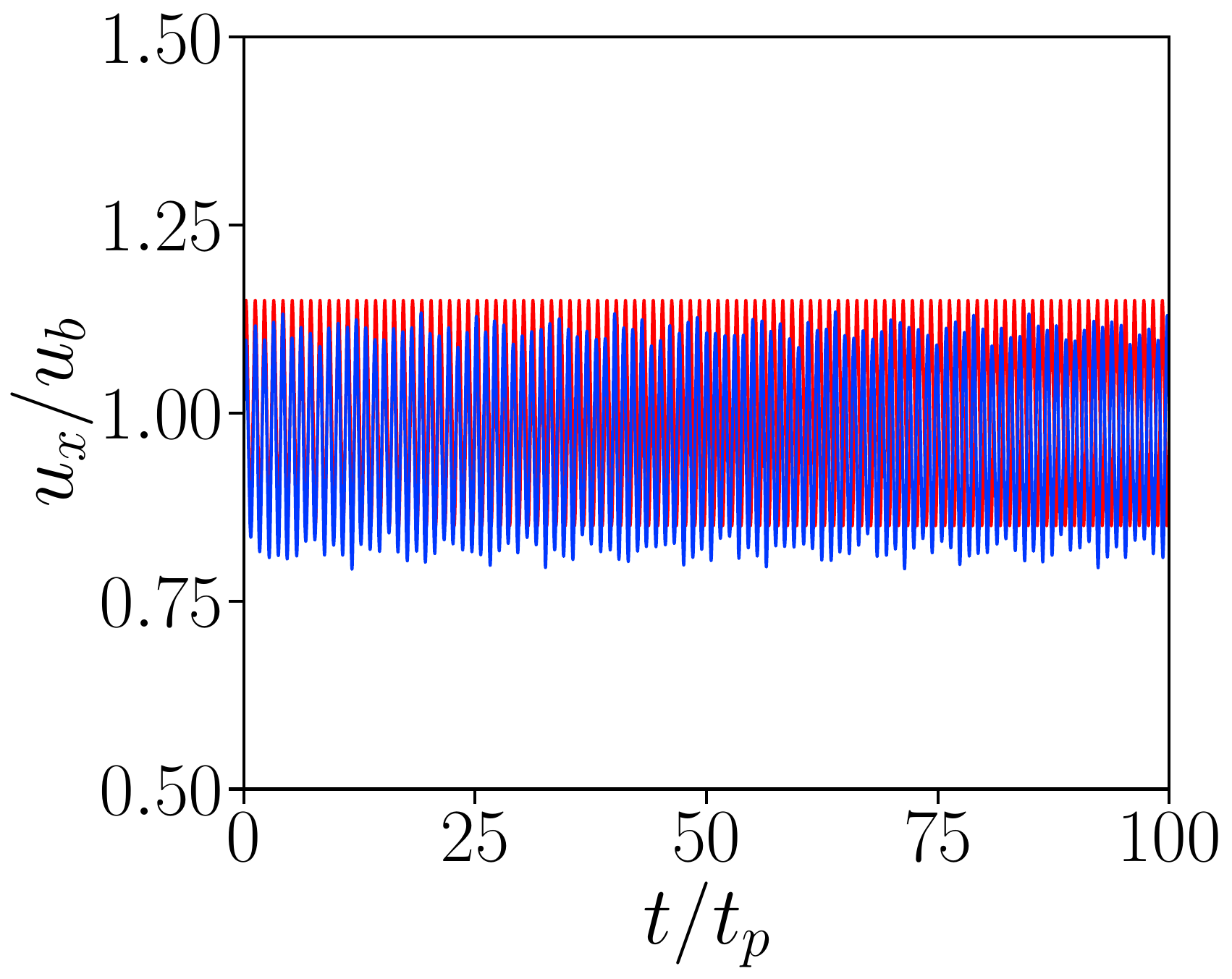}
        \caption{}
        \label{fig:prior3_015}
    \end{subfigure}
    \caption{Time evolution of the streamwise velocity imposed at the inlet. The red curve represents the law imposed for the run LES-HF, the blue curve is obtained using the coefficients obtained with the DA optimization. Results are shown for (top line) the distribution  \textit{prior state} 1 and (bottom line) the distribution \textit{prior state} 2. The inlet's velocity obtained via DA is shown for the coefficients optimized at (left column) $t=6t_p$, (centre column) $9t_p$ and (right column) $12t_p$.}
    \label{fig:beats}
\end{figure}

The parametric description of DA run is now sampled and investigated for some specific analysis phases. To this purpose, the inlet parameters sampled at times $t = 6t_p$, $t = 9t_p$ and $t=12t_p$ are considered to be frozen and they are used to visualize the resulting inlet behaviour over a longer time window. The plot of these inlet functions is shown in Fig. \ref{fig:beats}. The first row shows the inlet evolution calculated with the parameters optimized for \textit{prior state } 1 while the second row includes the same plot for \textit{prior state} 2. In both cases, the plot is performed for 100 cycles, enabling to observe the time evolution of the signal. These results confirm previous discussion about the optimization of the coefficients. For \textit{prior state } 1, the  inaccurate estimation of the amplitude coefficients requires time-dependent variations of the other free parameters to maintain a low discrepancy in time with the LES-HF simulation observations.In contrast, for \textit{prior state} 2, where the initial amplitude guess is more accurate, the optimization rapidly converges to a parametric configuration that effectively reproduces the inlet characteristic of the reference simulation.

In summary, the application of the HLEnKF algorithm successfully achieves a robust calibration of the inlet conditions for the DA ensemble members running on coarse grids, using instantaneous velocity samples from the high-fidelity LES-HF simulation as observations. However, this efficient calibration is obtained via a precise analysis of optimized time-windows between consecutive analysis phases, including knowledge of the characteristic times of the flow. Considering that such information is rarely available, the present analysis show the potential to perform DA optimization for turbulent flows as well as the challenges and difficulties that have to be considered in this process. Among the latter, it is demonstrated that the choice of different prior states strongly affects the results of the optimization. On the one hand, the results of the DA optimization starting from \textit{prior state} 2 rapidly converges towards the expected coefficients. One could also see that the resulting set of inlet parameter take into account the numerical and modelling errors developing in the intake pipe, in order to minimize the discrepancy between model runs and observation. On the other hand,  The algorithm fails to converge to the expected parameter values when using \textit{prior state} 1. To address this issue, the DA algorithm adjusts the parameters at each analysis phase to enhance the quality of the prediction. Despite the differing outcomes in parameter optimization for the two DA runs, both successfully capture a convincing representation of the inlet characteristics. From this perspective, the HLEnKF proves effective in achieving robust parametric optimization for both prior configurations used.  

\section{Flow synchronization using Data Assimilation}
\label{sec:localSynchronization}
\subsection{Setup details}
The results presented in Sec. \ref{sec:inletCalibration} showed that the HLEnKF performs an adequate calibration of the free parameters of an unsteady inlet condition when using observation from sensors. Therefore, the DA optimization acts so that the flow dynamics obtained for the ensemble members are similar to the ones of the simulation from which the observation was sampled. The attention is now focused on the capability of the DA algorithm to synchronize the instantaneous behaviour of the flow field predicted by the numerical models with the observation. This point is challenging for turbulent flows, considering that two identical configurations have the same statistical behaviour but may have very different instantaneous flow fields. It is important to stress that efficient synchronization may unlock advancement in multiple research fields dealing with engineering.    
In such applications, the synchronization of complete numerical models with localized data measured at sensors (such as for example in a digital twin architecture) provides tool for the analysis and potential anticipation of certain extreme conditions and events that could lead to a decrease in the efficiency of the physical system or harm its integrity. The objectives targeted in this section are two. The first one deals with the assessment of the synchronization capabilities of the HLEnKF when sparse observation is provided. The second one is to investigate how the DA state update performing the synchronization affects the statistical moments of the flow, and in particular its modal energy distribution. This second point is essential to validate the capabilities of the DA algorithm to improve the physical accuracy of the model when limited measurements are provided.

\begin{figure}
    \centering
    \includegraphics[width=0.6\linewidth]{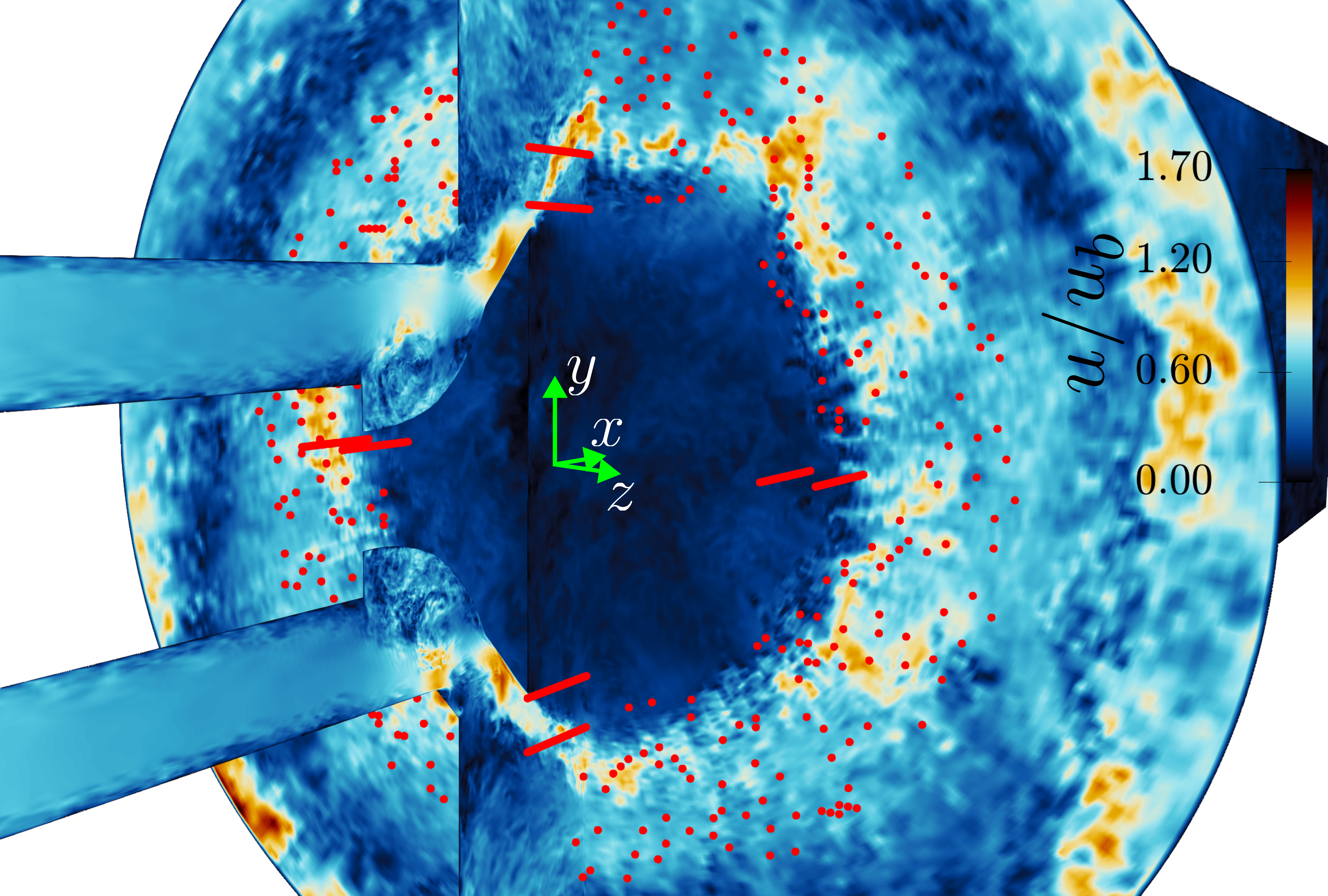}
    \caption{Qualitative representation of the sensors' distribution for the set SE-set. Sensors are represented with red dots.}
    \label{fig:SE-Obs-Set1}
\end{figure}

The starting point of this analysis is the DA calibration procedure performed with the \textit{prior state} 2. The inferred configuration obtain via the calibration of the inlet parameters for the latter will be referred to as LES-PE2 from now on. It is reminded that this DA run was performed for a total time of $t = 12t_p$. The 35 ensemble members show very similar but not identical parametric description of the inlet due to the inflation applied to the system. Therefore, their instantaneous physical fields for $t=12t_p$ are not identical. The analysis using DA state estimation to evaluate the synchronization properties is now performed starting from this set up and it is run for one complete cycle, thus ending at $t=13t_p$. 
Considering that the DA analysis period is only one $t_p$, the limitations in terms of time window between successive DA analyses are not any more relevant, and a maximum of 1250 analyses can be performed over the period of investigation, which corresponds to the sampling rate of simulation LES-HF. 
It is important to remind that observation from the LES-HF run was taken from cycles seven to ten and that the observation was reused once the set was over. Considering that the DA analysis started at the end of the second cycle for the run LES-PE2, this means that data from the ninth cycle of the run LES-HF is used as observation in the time window $t \in[12t_p, \, 13t_p]$. The LES-HF velocity field for the time $t=9t_p$ is shown in Fig. \ref{fig:velocityField_ref}. The sensors used in these investigations are now described. The set previously used for DA calibration in Sec. \ref{sec:inletCalibration} is now expanded, including additional sparsely distributed sensors on the plane $x=\SI{0.124}{\m}$ (see Fig. \ref{fig:positionSensorsAD}). A total of $582$ new sensors are included. This new sensor set, called SE-set (\textit{state estimation set}) is shown in Fig. \ref{fig:SE-Obs-Set1}. 

The size of the state matrix $\boldsymbol{\mathcal{U}}$ is now discussed. As previously stated in Sec. \ref{sec:inletCalibration} its size is equal to $N = 3 \, n_{cells}$ if no parametric optimization is performed ($N_\gamma=0$). However, the number of elements $n_{cells}$ changes for each local EnKF performed and it is a function of the volume of the region as well as of the local grid refinement. The volume is here considered to be spherical and the diameter is selected with respect to the covariance localization length used for the DA analysis. For this test case which exhibits turbulent features, the localization length is related to the integral length scale $\mathcal{L}$. This choice appears to be the most logical, as the correlation of the velocity field in turbulent flows rapidly decays for distances that are larger than $\mathcal{L}$. Therefore, it is an appropriate length scale for covariance localization. The integral length scale calculated using the velocity field from simulation LES-HF exhibits an average value of $\mathcal{L} \approx \SI{1.7d-3}{\metre}$. To take into account the distance between sensors and in order to reduce superpositions of the zones where different realizations of the HLEnKF are performed, the diameter of the localized regions for the HLEnKF is selected to be $\SI{2.5d-3}{\metre}$. Moreover, in order to obtain a DA state update of $\approx 1\%$ in the proximity of the external surface of each volume for the HLEnKF, the parameter $l$ in Eq. \ref{eqn:loc_matrix_v2}  
is set to $l=\SI{4.1d-4}{}$. This leads to $ \approx 63 \, 000$ grid elements or $N \approx 189 \, 000$ degrees of freedom for each of the $134$ realizations of the HLEnKF. The number of regions does not match the total number of sensors because some of them were clustered due to physical proximity. The number of elements in every region may exhibit variations depending on the local grid refinement. It is however important to stress that performing $134$ DA procedures for the selected volumes around sensors is significantly less computationally expensive than performing one single EnKF using data from eight million grid elements.

Tab. \ref{tab:SEconfigs} summarizes the features of the runs performed to evaluate the accuracy of DA's state estimation and synchronization. For all simulations, the time window between successive analyses is constant. However, the sensitivity to the length of such time window and its impact over the predictive capabilities of the DA algorithm is investigated. To this purpose, three runs are performed using the set SE-set. The first one used the shortest possible time window, corresponding to the sampling period of the simulation LES-HF. This time window is equal to $t_{DA} = 10\Delta_t = \SI{1e-6}{\s}$. 
The frequency at which DA is performed is therefore equal to $\SI{1000}{\kilo\hertz}$ in this case. The second configuration uses a time window equal to $t_{DA} = 50\Delta_t$ and for the third one  $t_{DA} = 250\Delta_t$. The corresponding assimilation frequencies are then $\SI{200}{\kilo\hertz}$ and $\SI{40}{\kilo\hertz}$. These runs are referred to as LES-TW10, LES-TW50 and LES-TW250, respectively. We stress that, when compared to the characteristic frequency scale of engine flows, the number of time sample per period used here could be typically obtained by using optical techniques such as high speed PIV \citep{leite2024_lxlaser, Voisine2011_ef} used currently in optical engines. Acquisition frequencies for PIV measurements are typically included in the range $\SI{1}{\kilo\hertz}$ - $\SI{10}{\kilo\hertz}$ \citep{galmiche2014turbulence}. However, the frequency of the velocity time signal imposed at the inlet for the run LES-HF is also $\approx 6$ times higher than the characteristic frequencies observed in internal combustion engines. Therefore, one can see that the second configuration investigated falls into a range of number of time samples per physical oscillation of the flow which is observed in experiments. These three runs are compared with the results obtained from the run LES-PE2, which is also continued for one cycle using the same analysis time windows used for calibration ($2500 \Delta_t$).

\begin{table}
    \centering
    \begin{tabular}{ccccc}
         \hline
         Name of the DA run & DA optimization & Set of sensors & DA time window \\
         \hline
         LES-TW10 & state & SE-set & 10 $\Delta_t$ \\
         \hline
         LES-TW50 & state & SE-set & 50 $\Delta_t$ \\
         \hline
         LES-TW250 & state & SE-set & 250 $\Delta_t$ \\
         \hline
         LES-PE2 & Parameters & calibration & 2500 $\Delta_t$ \\
         \hline
    \end{tabular}
    \caption{Summary of the runs performed to investigate DA state estimation for field synchronization, compared with the run LES-PE2.}
    \label{tab:SEconfigs}
\end{table}

\begin{figure}
    \centering
    \includegraphics[width=0.5\linewidth]{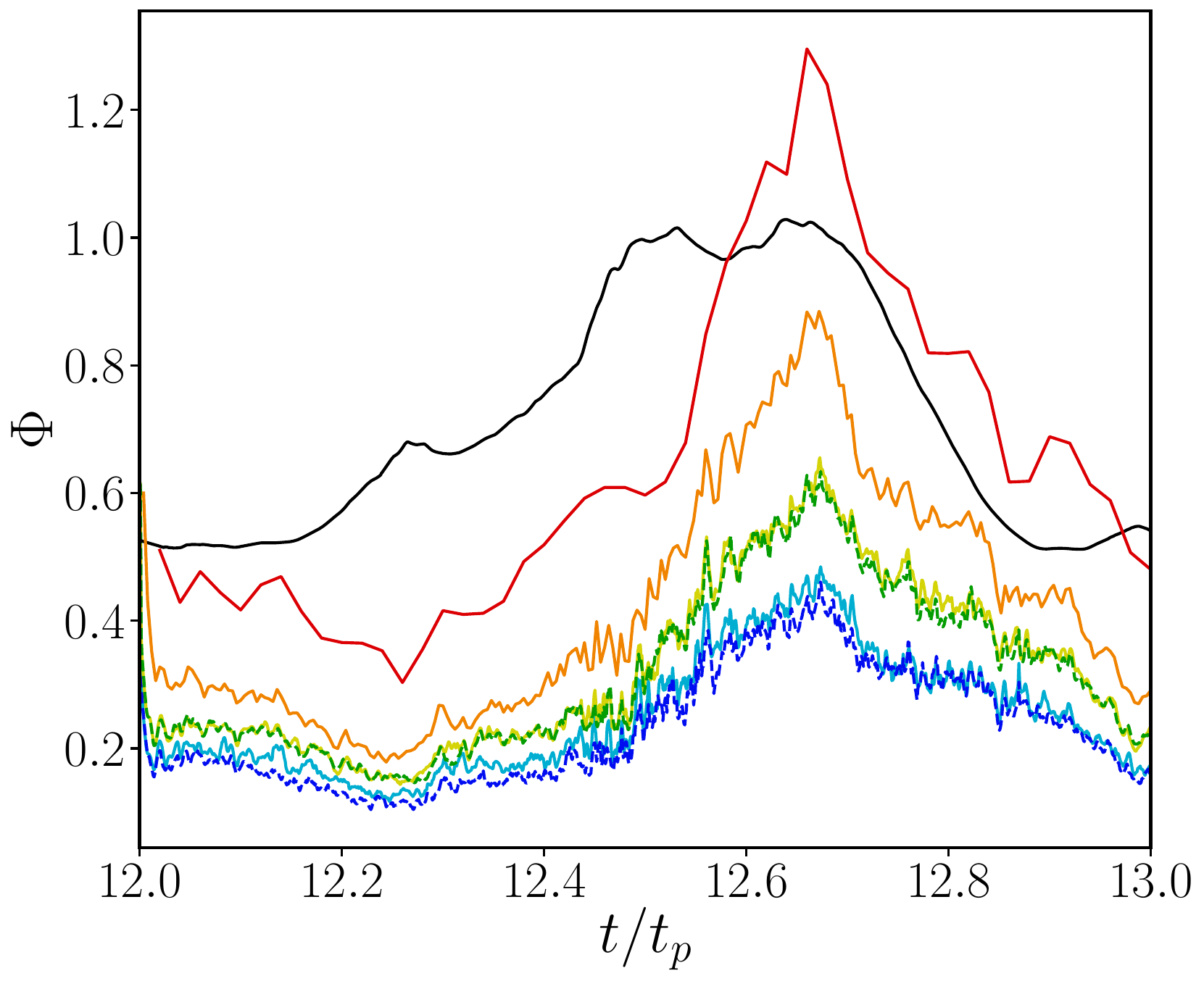}
    \caption{Evolution of $\Phi$ over the time of investigation. Black curve: run LES-PE2. Red curve: run LES-TW250 with $5\%$ confidence in the observation and without inflation. Orange curve: run LES-TW50 with $5\%$ confidence in the observation and without inflation. Green curves: run LES-TW10 with $20\%$ confidence in the observation and inflation equal to (light green) $0\%$ and (dark green) $5\%$. Blue curves: run LES-TW10 with $5\%$ confidence in the observation and inflation equal to (light blue) $0\%$ and (dark blue) $5\%$.}
    \label{fig:RMSE}
\end{figure}

The DA capabilities to perform efficient synchronization are evaluated via the calculation of a root-mean-square error $\Phi$ evaluated at each analysis phase $k$:

\begin{equation}
    \Phi(k) = \sqrt{\sum_{j=1}^{N_o}{(\langle \mathbf{s}_{j,k} \rangle-\boldsymbol{\alpha}_{j,k})^2}}\Biggl/N_o
    \label{eq:RMSE}
\end{equation}

where $\langle \mathbf{s}_{j,k} \rangle$ is the mean of all the velocity values calculated by the ensemble members at the coordinates of the sensors,  $\boldsymbol{\alpha}_{j,k}$ are the sampled observations and $N_o$ is the number of sensors. This indicator allows evaluating the global discrepancy between the assimilated velocity field and the high-fidelity reference at the sensors. The calculation of $\Phi(k)$ is carried out considering the 582 sensors for the three configurations LES-TW10, LES-TW50 and LES-TW250. For the first one, two levels of confidence in the observations are considered (uncertainty level of $5\%$ and $20\%$) as well as two levels for the inflation ($0\%$ and $5\%$). The evolution of $\Phi$ for these calculations is shown in Fig. \ref{fig:RMSE}. This indicator varies in time for all the runs, exhibiting values included in the range $[0.2, \, 1.2]$. The black curve represents the evolution of the error for the configuration without any state estimation LES-PE2. It is representative of the variability of the 35 simulations of the ensemble. The red and orange curves correspond to the runs LES-TW250 and LES-TW50, which are performed using a confidence level for the observations equal to $\sigma_m = 5\%$. For the former, the values of $\Phi$ are extremely similar to those observed for the case LES-PE2, which suggests that the DA time window is too large to obtain a permanent signature on the instantaneous structural organization of the flow. On the other hand, results for the run LES-TW50 show that the discrepancy is globally reduced at each time step over the cycle investigated.  
A similar behaviour is observed for $\Phi$ for the four realizations of the LES-TW10 run. The green curves (which superpose one with the other) are obtained with a confidence level of $20\%$ in the observations, while  the blue ones are obtained with a confidence level of $5\%$. One can see that variations for the values selected for the inflation do not strongly affect the results. 
Generally speaking, the best runs for the configuration LES-TW10 exhibit an error reduction of at least $60\%$ at all times when compared with the LES-PE2 run. This result shows the sensitivity of the noise affecting the observations in the DA procedure. While this parameter is governed by the quality of the acquisition system, these results indicate how this parameter can be adapted to improve the global synchronization of the flow. 
On the other hand, the low sensitivity to inflation can be justified by two factors. First, a good degree of initial variability of the flow field is granted by the selected \textit{prior} distribution. 
Second, the problem of variance collapse observed for classical EnKF algorithms is not as strong for the HLEnKF, due to the physical localization affecting the state update.

\subsection{Synchronization capabilities}
\label{sec:SE-results}

\begin{figure}
    \centering
    \includegraphics[width=0.35\linewidth]{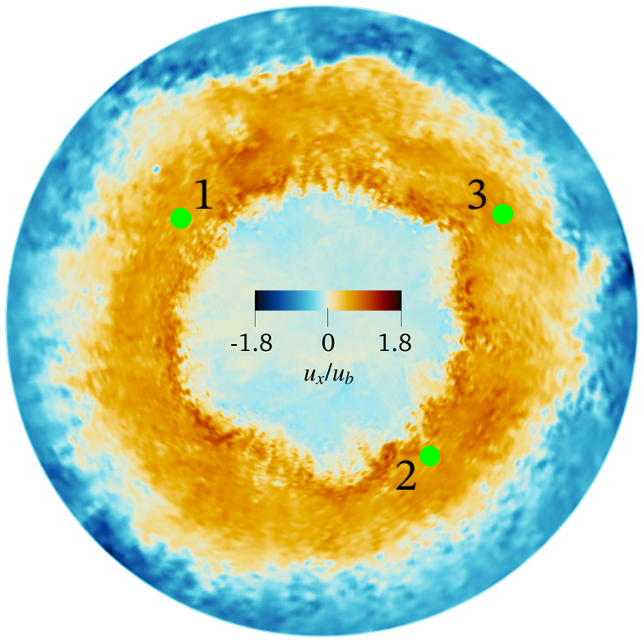}
    \qquad
    \begin{tabular}[b]{cccc}
        \hline
        Sensor & $x$ (m) & $y$ (m) & $z$ (m) \\
        \hline
        1 & 0.124 & -0.0267 & 0.0197 \\
        \hline
        2 & 0.124 & 0.0205 & -0.0254 \\
        \hline
        3 & 0.124 & 0.0342 & 0.0205 \\ 
        \hline
    \end{tabular}
    \captionlistentry[table]{Sensors used to check the synchronization of inferred velocities with the observation}
    \label{tab:monitoringProbes}
    \captionsetup{labelformat=andtable}
    \caption{Sensors used to assess the accuracy of the synchronization between the DA prediction and the observation.}
\end{figure}

The study of the root-mean-square error $\Phi$ using different hyperparameters showed that a confidence level in the observations of $5\%$ and zero inflation were providing the most accurate results. However, this evaluation is performed using a global indicator. The local synchronization capabilities for the runs performed are now investigated. To this purpose, three different sensors providing observation are monitored to study the evolution of the instantaneous velocities obtained via DA and compare them with the available high-fidelity data. These sensors, which are summarized in Tab. \ref{tab:monitoringProbes}, are located on the plane $x=\SI{0.124}{\m}$. Sensors 1 and 2 are located in the proximity of the valve outlet jet on opposite regions on the plane, while sensor 3 is slightly off-centre in a zone with higher shear.

\begin{figure}
    \centering
    
    \begin{subfigure}[b]{.313\textwidth}
        \centering
        \includegraphics[width=\textwidth]{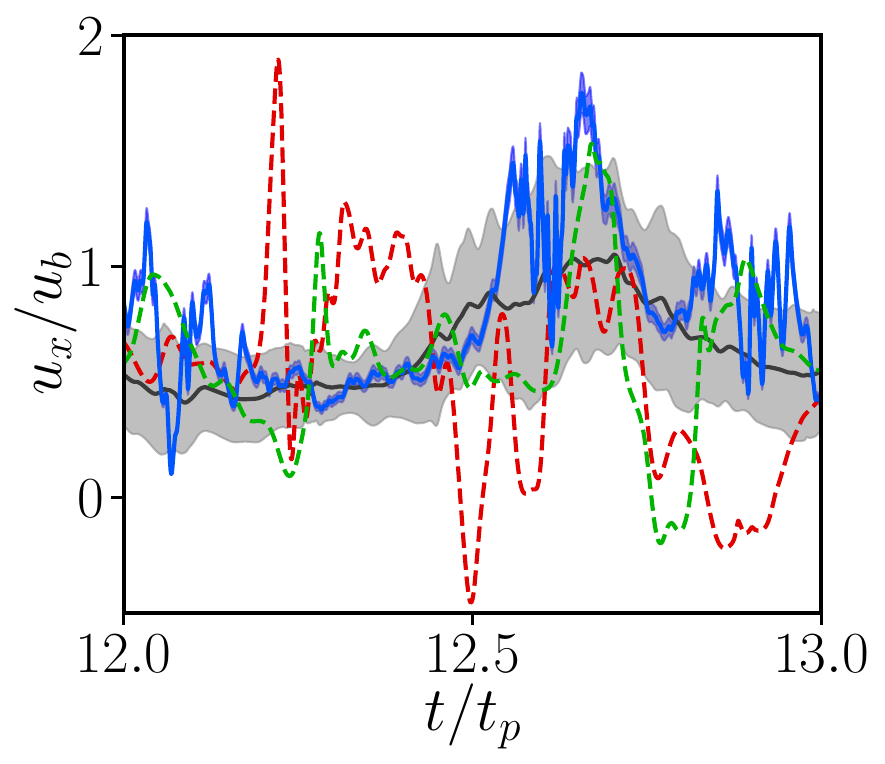}
        \caption{Sensor 1 - LES-PE2}
    \end{subfigure}%
    \hfill 
    \begin{subfigure}[b]{.343\textwidth}
        \centering
        \includegraphics[width=\textwidth]{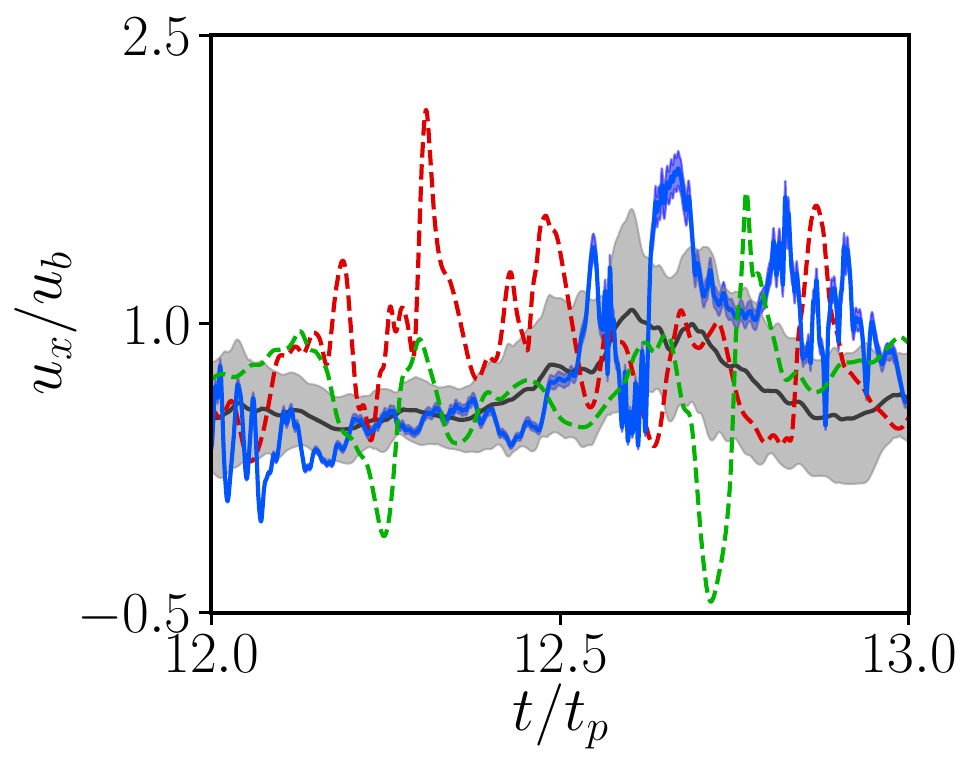}
        \caption{Sensor 2 - LES-PE2}
    \end{subfigure}%
    \hfill 
    \begin{subfigure}[b]{.343\textwidth}
        \centering
        \includegraphics[width=\textwidth]{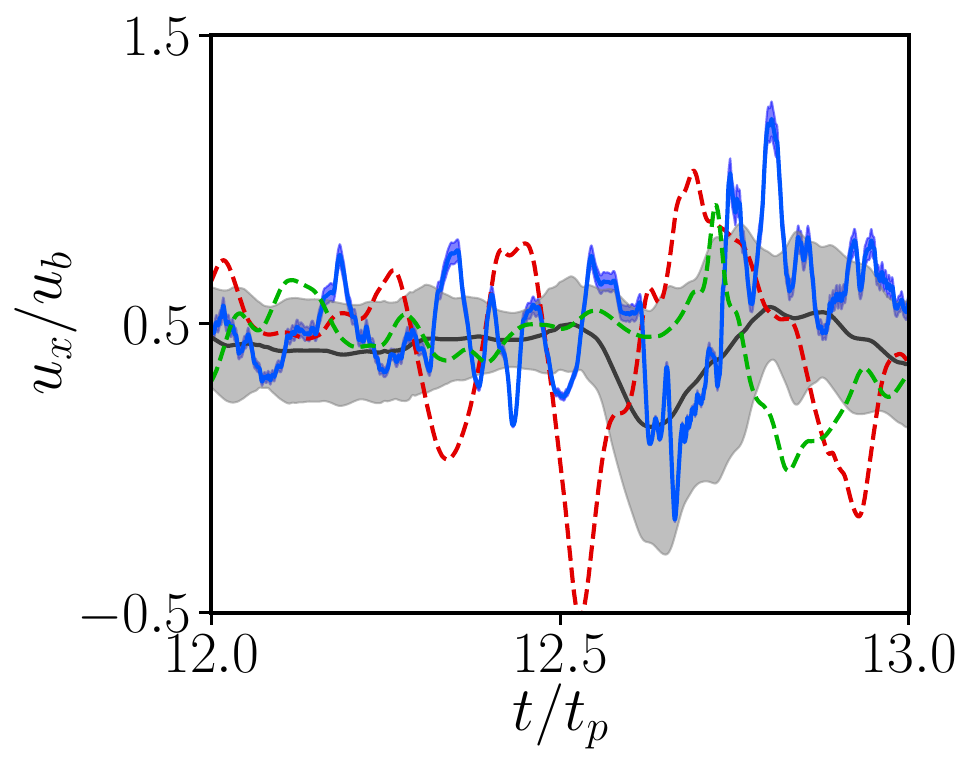}
        \caption{Sensor 3 - LES-PE2}
    \end{subfigure}

    \begin{subfigure}[b]{.313\textwidth}
        \centering
        \includegraphics[width=\textwidth]{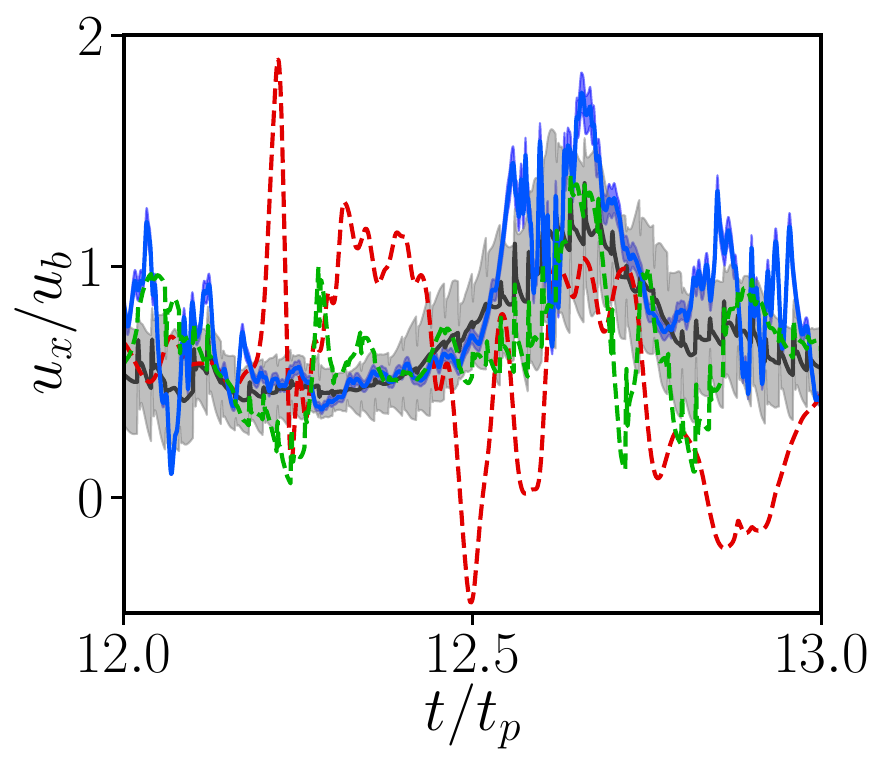}
        \caption{Sensor 1 - LES-TW250}
    \end{subfigure}%
    \hfill 
    \begin{subfigure}[b]{.343\textwidth}
        \centering
        \includegraphics[width=\textwidth]{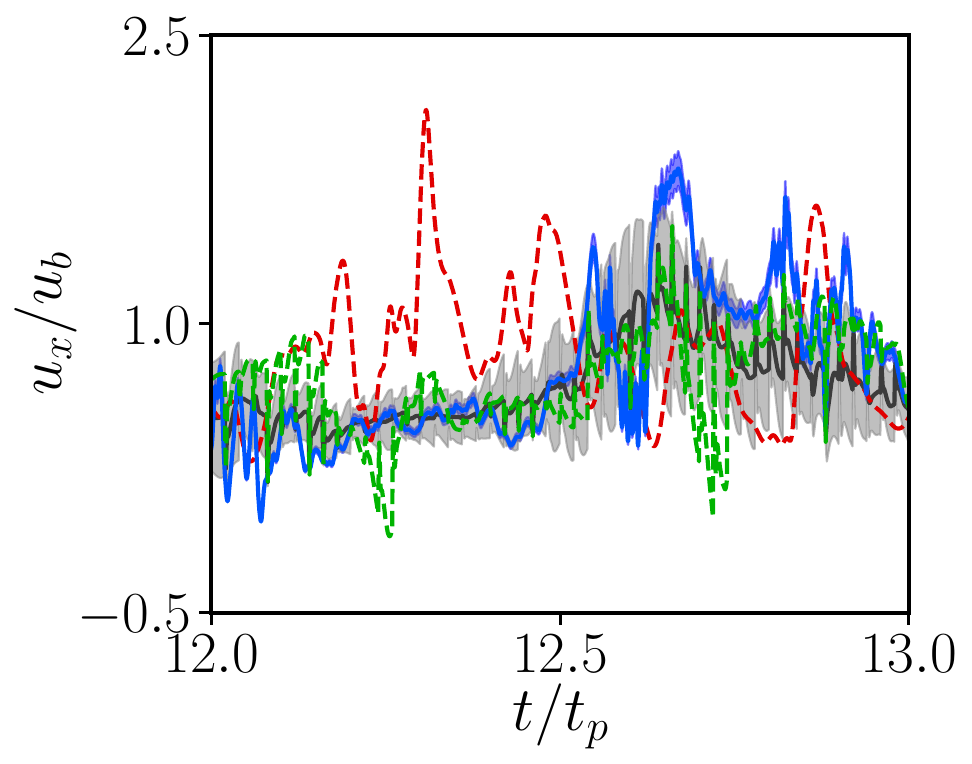}
        \caption{Sensor 2 - LES-TW250}
    \end{subfigure}%
    \hfill 
    \begin{subfigure}[b]{.343\textwidth}
        \centering
        \includegraphics[width=\textwidth]{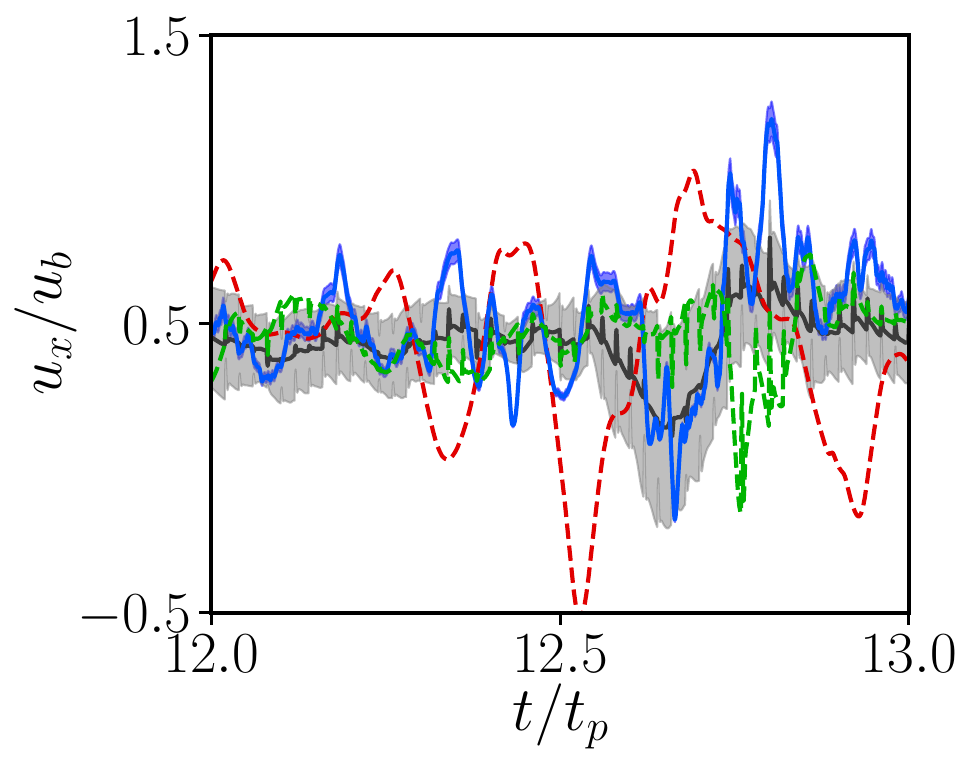}
        \caption{Sensor 3 - LES-TW250}
    \end{subfigure}

    \begin{subfigure}[b]{.313\textwidth}
        \centering
        \includegraphics[width=\textwidth]{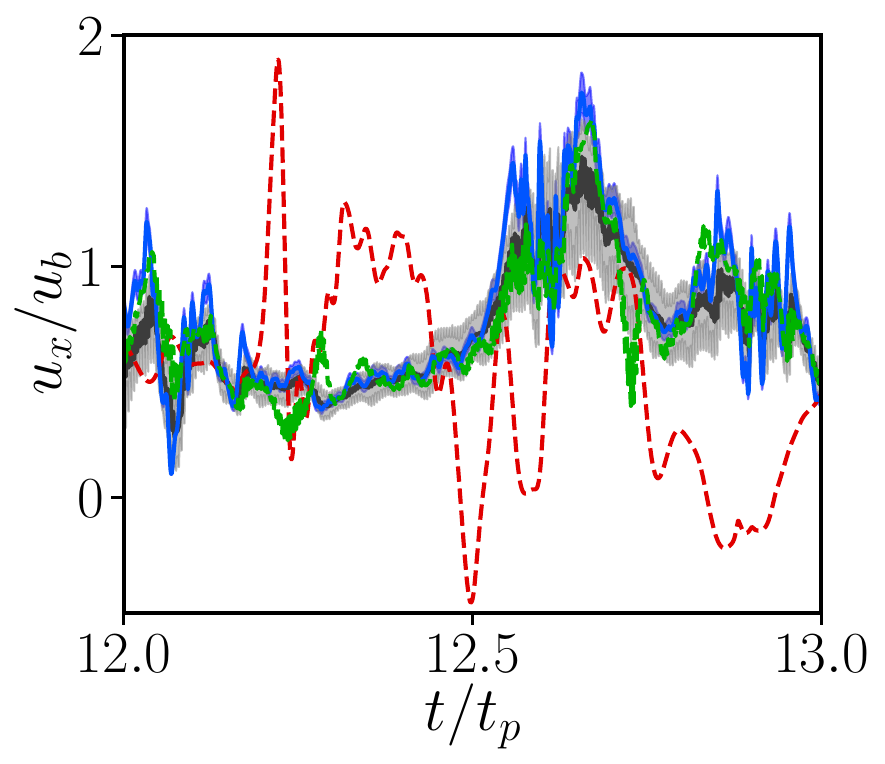}
        \caption{Sensor 1 - LES-TW50}
    \end{subfigure}%
    \hfill 
    \begin{subfigure}[b]{.343\textwidth}
        \centering
        \includegraphics[width=\textwidth]{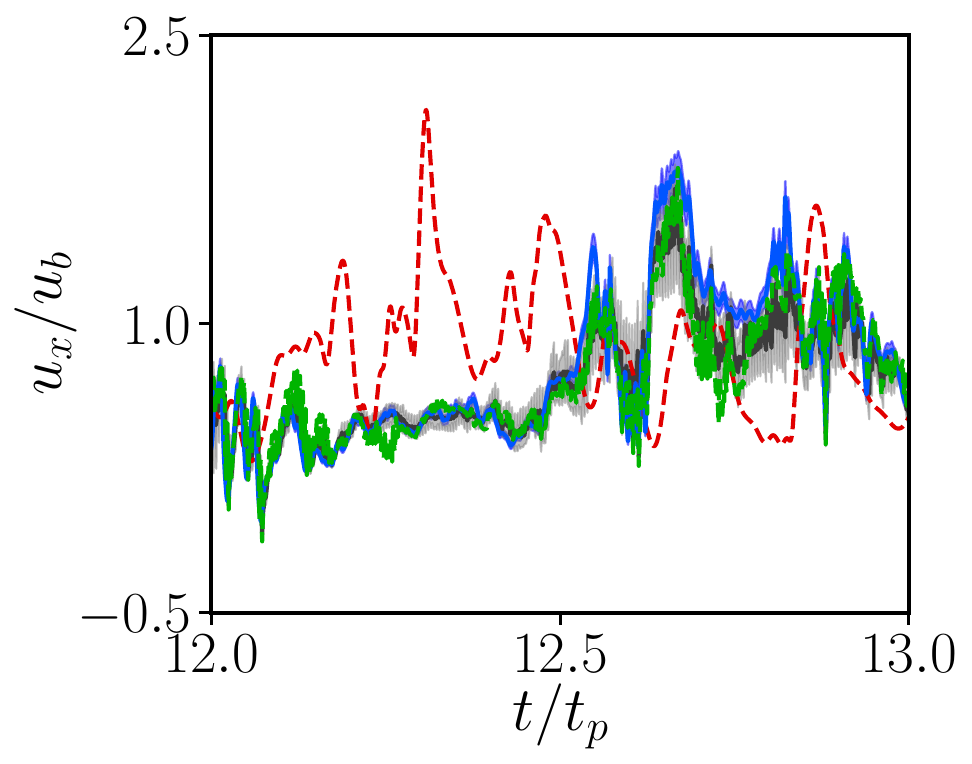}
        \caption{Sensor 2 - LES-TW50}
    \end{subfigure}%
    \hfill 
    \begin{subfigure}[b]{.343\textwidth}
        \centering
        \includegraphics[width=\textwidth]{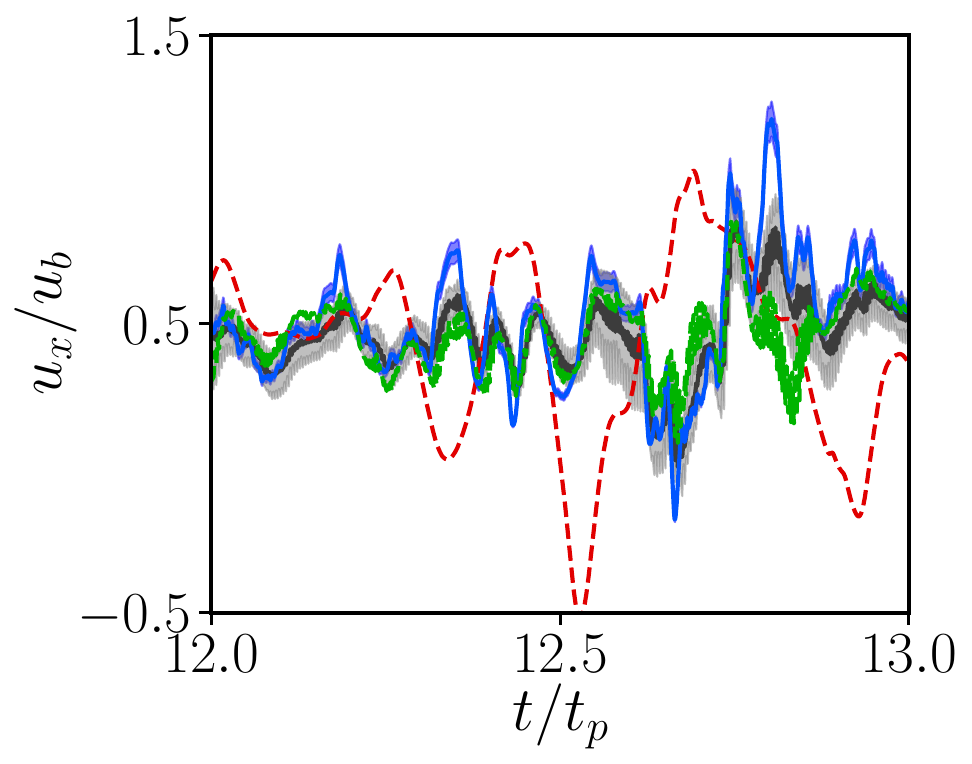}
        \caption{Sensor 3 - LES-TW50}
    \end{subfigure}
    
    \begin{subfigure}[b]{.313\textwidth}
        \centering
        \includegraphics[width=\textwidth]{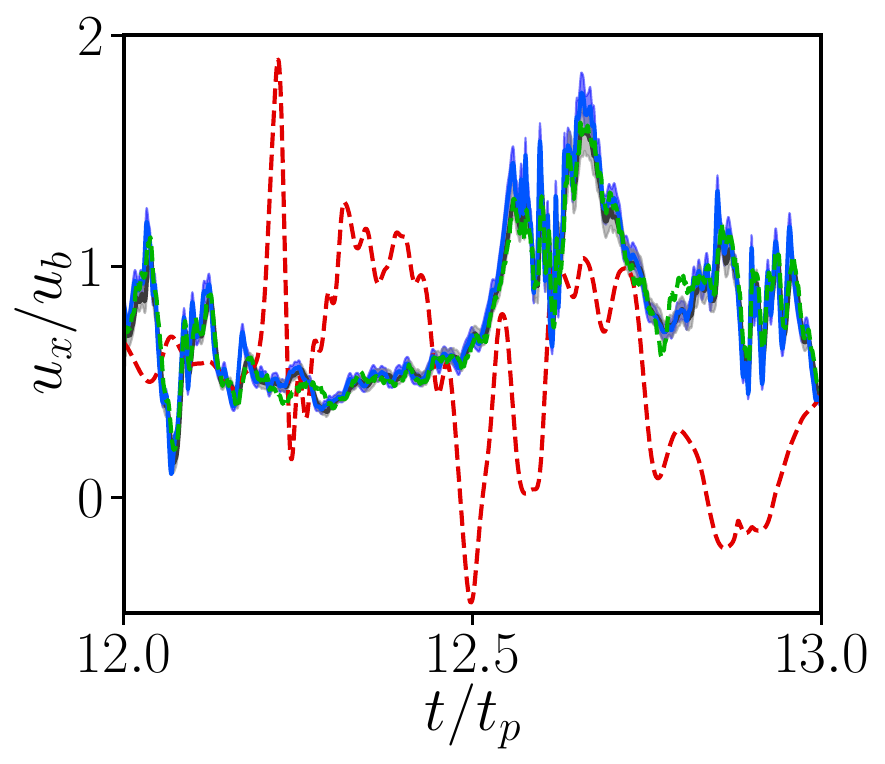}
        \caption{Sensor 1 - LES-TW10}
    \end{subfigure}%
    \hfill 
    \begin{subfigure}[b]{.343\textwidth}
        \centering
        \includegraphics[width=\textwidth]{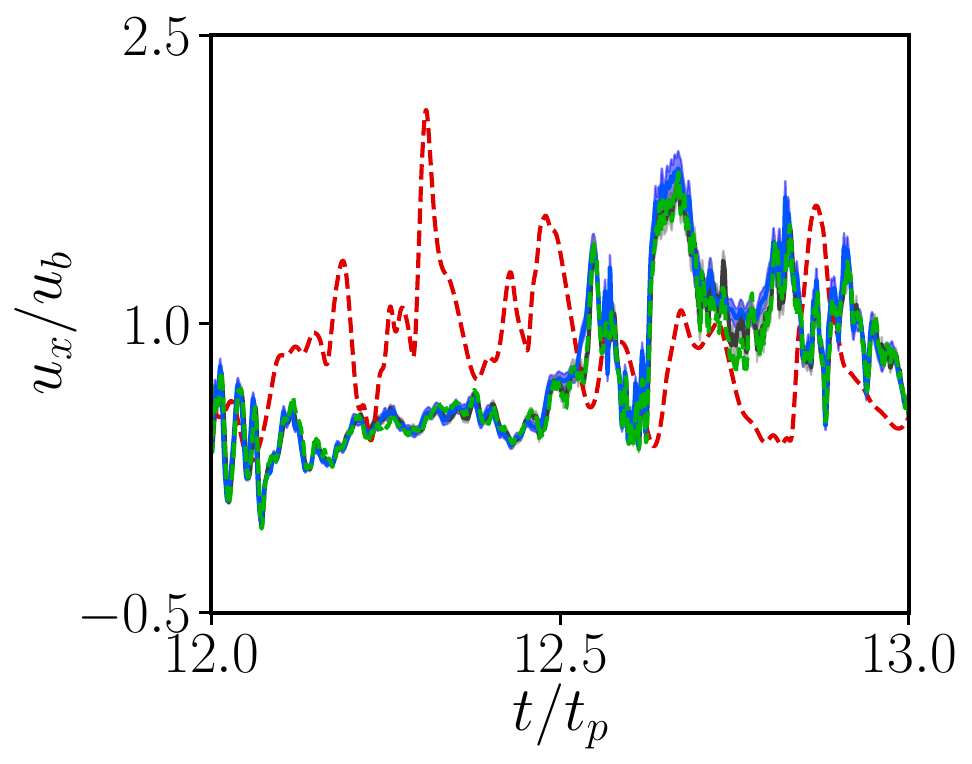}
        \caption{Sensor 2 - LES-TW10}
    \end{subfigure}%
    \hfill 
    \begin{subfigure}[b]{.343\textwidth}
        \centering
        \includegraphics[width=\textwidth]{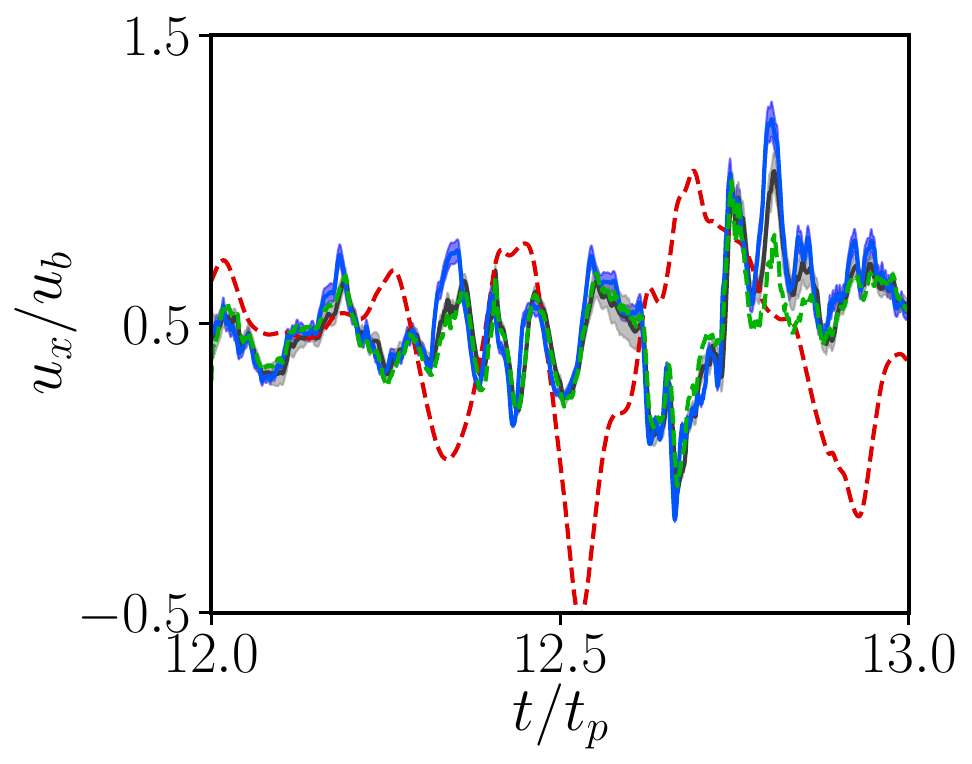}
        \caption{Sensor 3 - LES-TW10}
    \end{subfigure}
    
    \caption{Synchronization of the axial component of the velocity $u_x$. Results are shown for (rows) different DA runs and (column) at different sensors. The blue curves represent the observation. The black lines show the average prediction for the DA ensemble. The red lines provide results for member 27 of the ensemble without performing DA. The green lines  provide results for member 27 of the ensemble when using DA.}
    \label{fig:synchro_HLEnKF}
\end{figure}

Fig. \ref{fig:synchro_HLEnKF} shows the axial component of the velocity field for the runs performed at the three sensors investigated. The evolution of the instantaneous velocity component $u_x$ (normalized by the reference bulk velocity $u_b$) is shown in blue for the simulation LES-HF i.e. the one from which the observation is sampled. The results from the ensemble members of the DA procedure are shown in black (ensemble mean). Shaded areas around the blue and black curves correspond to the standard deviation of the measured velocity field. The red and green curves show the velocity evolution for member 27 of the database. The red curve corresponds to a run where the simulation is forecast in time without any DA correction, while the green line corresponds to the actual velocity profile when state estimation is performed. Results are now commented for each of the DA runs performed. The first row corresponds to the run LES-PE2, which does not include state estimation but was calibrated via the inlet inference in Sec. \ref{sec:inletCalibration}. As previously observed, the level of variance of the ensemble indicated by the shaded gray areas is relatively high. This is due to the variability imposed for the ensemble velocity fields via inflation. The analysis of the green lines show high instantaneous variations that are similar to the case without any data assimilation illustrated by the red curve. Neither of the two velocity profiles appears to synchronize with the blue curve representing the observation for any of the three sensors. Therefore, no synchronization is observed for LES-PE2 case. Rows 2 to 4 of Fig. \ref{fig:synchro_HLEnKF} present results for runs LES-TW250, LES-TW50 and LES-TW10. Here, a synchronization of the flow field is progressively observed. Improved flow reconstruction is obtained for shorter time windows between DA analyses. 
One can also see that a more accurate synchronization comes here with a significant reduction of the variance of the ensemble prediction. This point is clear comparing results from the second and the fourth row of Fig. \ref{fig:synchro_HLEnKF}.
On the other hand, when analyses are too far away in time, state updates tend to be too important due to the large discrepancy between model prediction and observation, which is here due to the cumulative numerical errors in time. Because of the strongly non-linear features of the flow, these corrections may lead to non-physical oscillations of the flow. This trend is visible for results from the run LES-TW250 in the second row.
The run LES-TW50 shows satisfactory results with an adequate synchronization capability. 
The best synchronization features are achieved by the LES-TW10 run. For all the three sensors considered, the ensemble mean  almost perfectly coincides with the observation. The use of the HLEnKF shows good efficacy in synchronizing the velocity field of the ensemble members with the high-fidelity observation. This result open perspectives of local flow synchronization while respecting global flow features, which is granted by the resolution of the Navier--Stokes equations performed in the ensemble runs. This point is now investigated further via the analysis of the statistical behaviour of the flow on the plane $x=\SI{0.124}{\m}$.

\begin{figure}
    \centering
    \begin{subfigure}{.495\textwidth}
    \centering
        \includegraphics[width=1\textwidth]{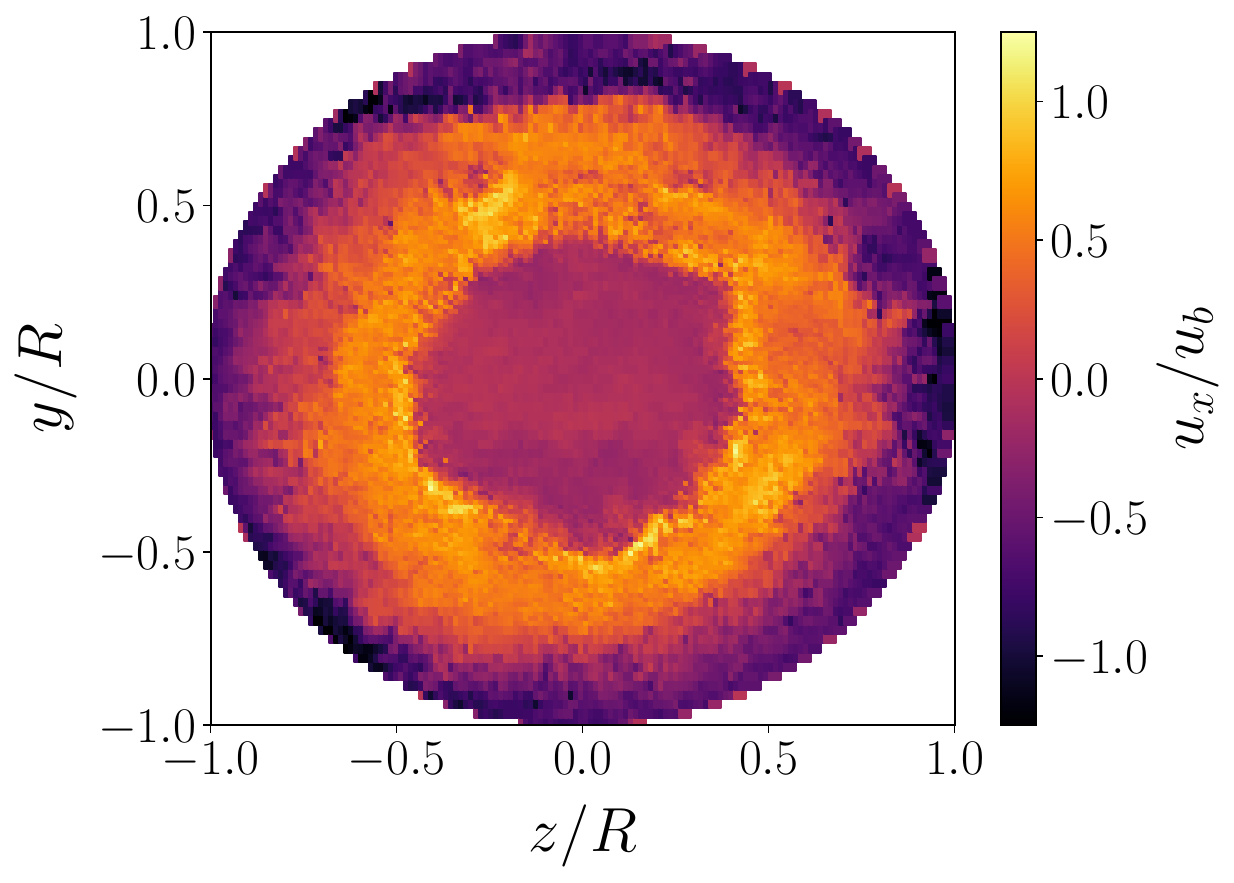}
        \label{fig:Ux_ref}
    \end{subfigure}
    \hfill
    \begin{subfigure}{.48\textwidth}
    \centering
        \includegraphics[width=1\textwidth]{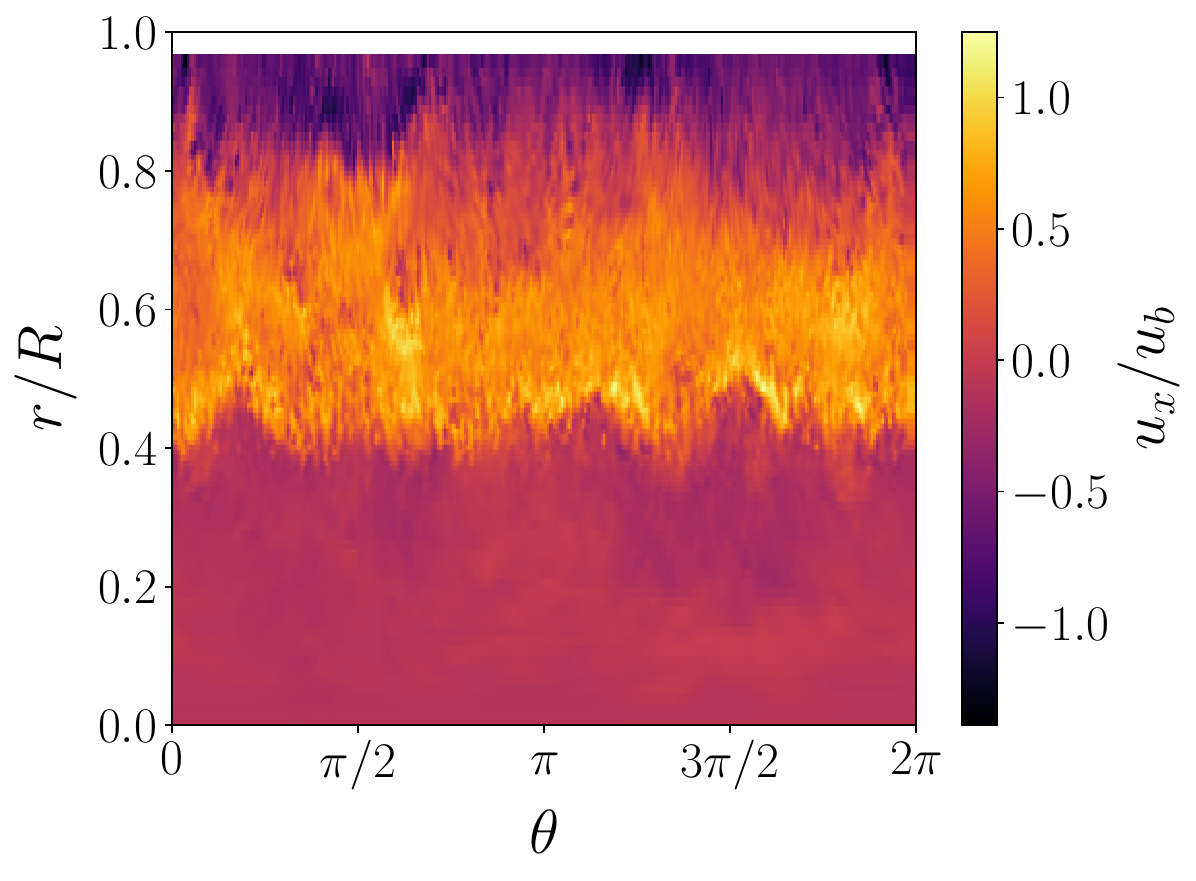}
        \label{fig:Ux-polar_ref}
    \end{subfigure}
    \caption{Normalized axial velocity for the run LES-HF. Data sampled on the plane for $x=0.124$. Left figure: measurements at sensors. Right figure: transformation in polar coordinates.}
    \label{fig:Ux_interpolation}
\end{figure}

The jet moving outside the valve directly crosses this plane, then the flow is redirected by the cylinder wall to create a  recirculation region, which forces the flow to cross again the investigated plane. The investigation of the modal energy distribution of the flow on this plane can therefore provide relevant information about the global accuracy of the HLEnKF. This investigation is complementary to the study of the local synchronization features in correspondence of the DA sensors.

The instantaneous axial velocity field $u_x$ sampled from the LES-HF reference simulation for the time $t=9t_p$ is shown in Fig. \ref{fig:Ux_interpolation}.
A Fourier transform of this field in the azimuthal direction is performed to investigate the characteristic modes representing the distribution of the kinetic energy. To do so, available data is first cast in the same polar coordinates for all the configurations (see  Fig. \ref{fig:Ux_interpolation}. (b)). The discrete Fourier transform of the velocity for each radius available can be expressed as :

\begin{equation}
    \hat{u}_m(r) = \sum^{N_{\theta}-1}_{n_\theta=0}u_{n_{\theta}}(r) e^{-2\pi i m n_\theta/N_{\theta}}
\end{equation}

with $n_\theta = 0,1,2, ...,N_{\theta-1}$ corresponding to the azimuthal discretization.
The modal energy distribution obtained for the DA runs is now compared with results obtained for the LES-HF reference. It is obtained calculating the power spectrum of each mode $P^m_s(r)=|\hat{u}_m(r)|^2$. 

\begin{figure}[h]
    \centering
    \begin{subfigure}{.32\textwidth}
        \centering
        \includegraphics[width=1\textwidth]{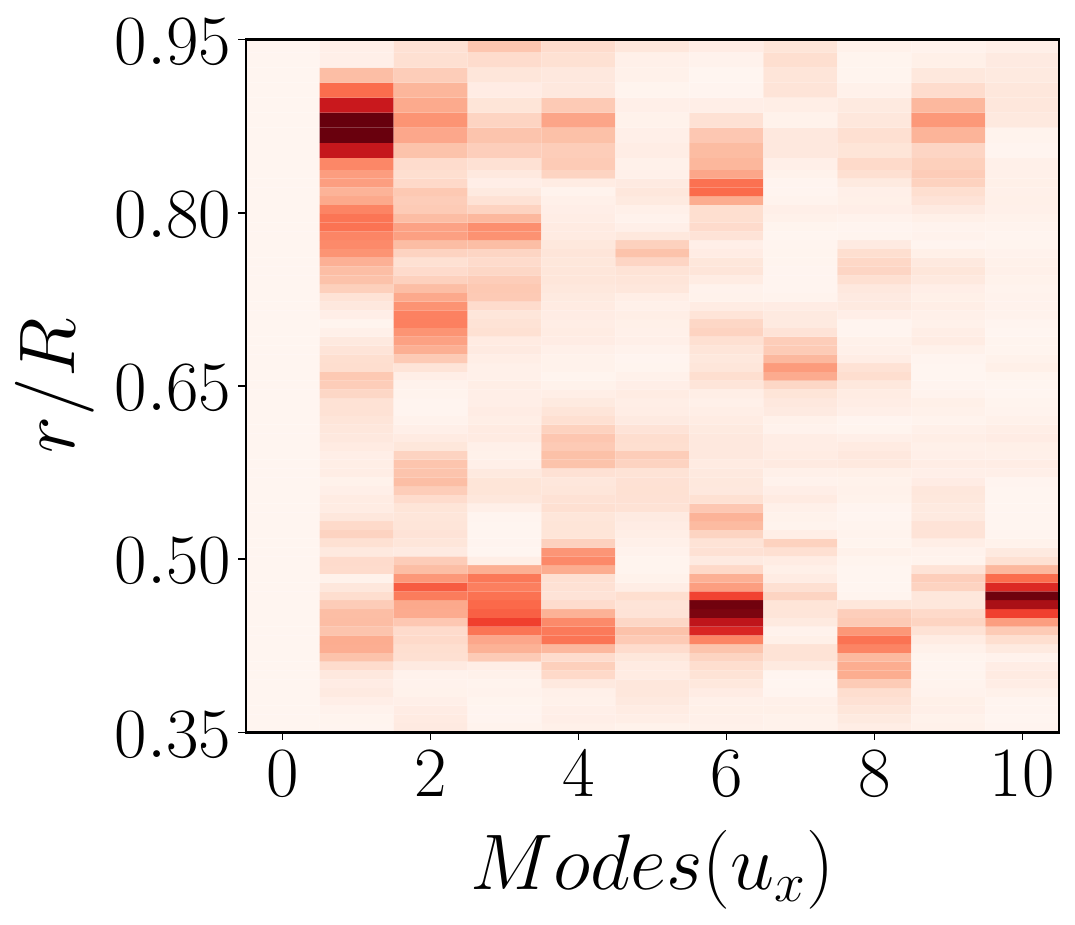}
        \caption{LES-HF}
        \label{fig:modes-map_x_ref}
    \end{subfigure}
    \hspace{1mm}
    \begin{subfigure}{.32\textwidth}
        \centering
        \includegraphics[width=1\textwidth]{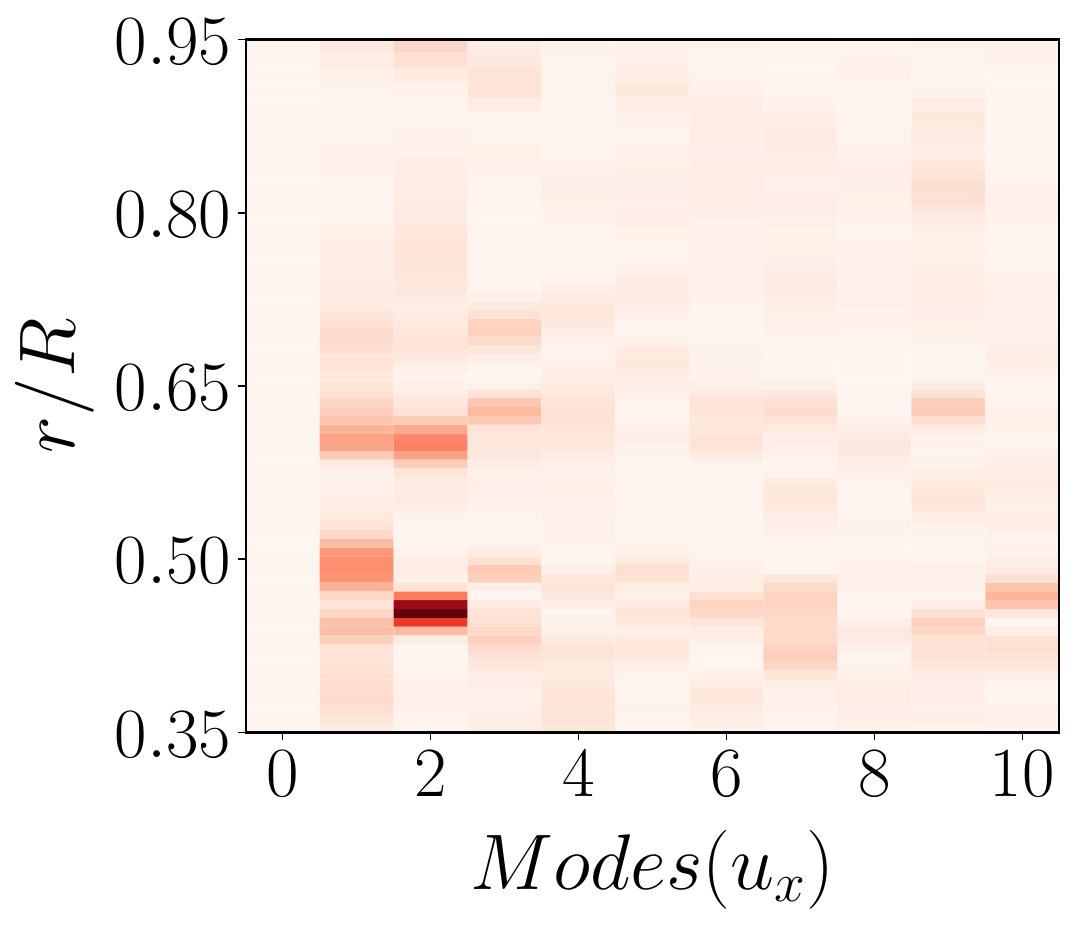}
        \caption{Member 27 \textit{prior} 2}
        \label{fig:modes-map_x_OFR3_member27_set6v3_noDA}
    \end{subfigure}
    \hspace{1mm}
    \begin{subfigure}{.32\textwidth}
        \centering
        \includegraphics[width=1\textwidth]{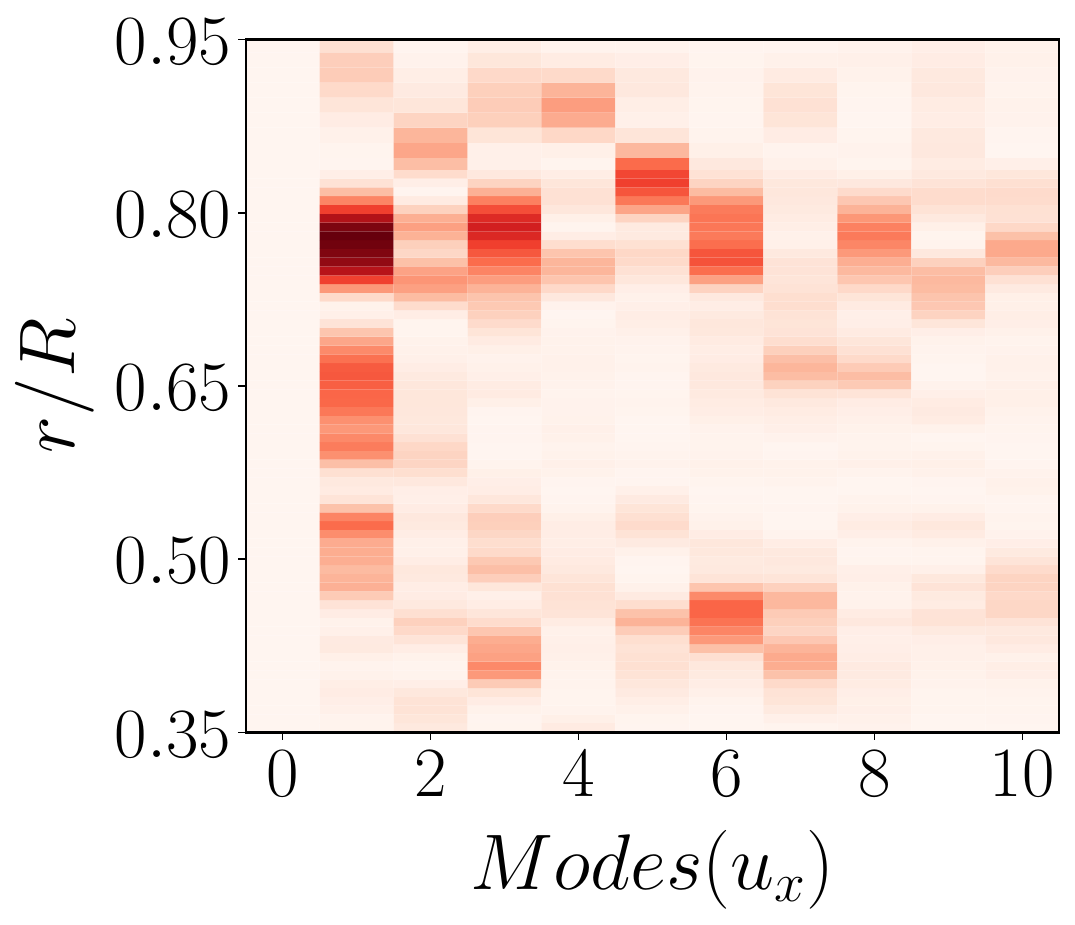}
        \caption{Member 27 LES-PE2}
        \label{fig:modes-map_x_OFR3_member27_set6v3_noState}
    \end{subfigure}
    \begin{subfigure}{.32\textwidth}
        \centering
        \includegraphics[width=1\textwidth]{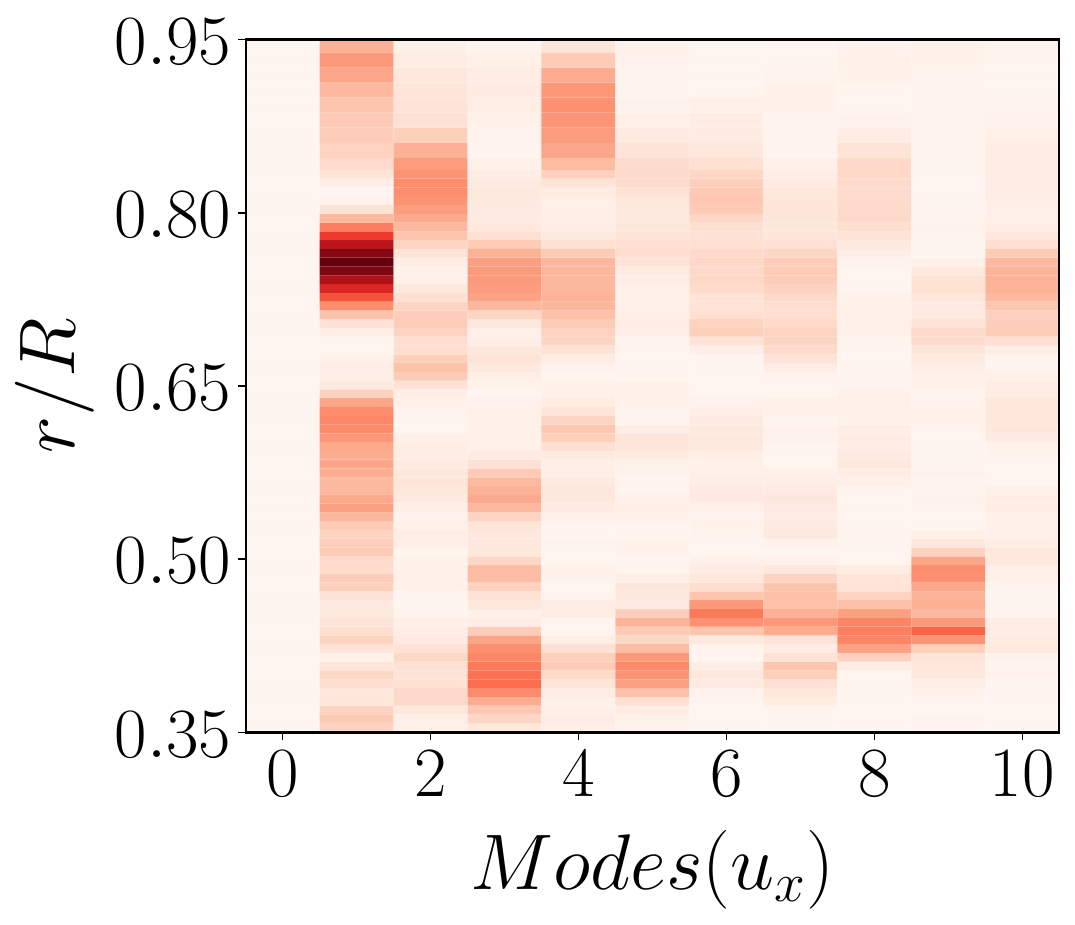}
        \caption{Member 27 LES-TW50}
        \label{fig:modes-map_x_OFR3_member27_set6v3_OW50_ON05}
    \end{subfigure}
    \hspace{1mm}
    \begin{subfigure}{.32\textwidth}
        \centering
        \includegraphics[width=1\textwidth]{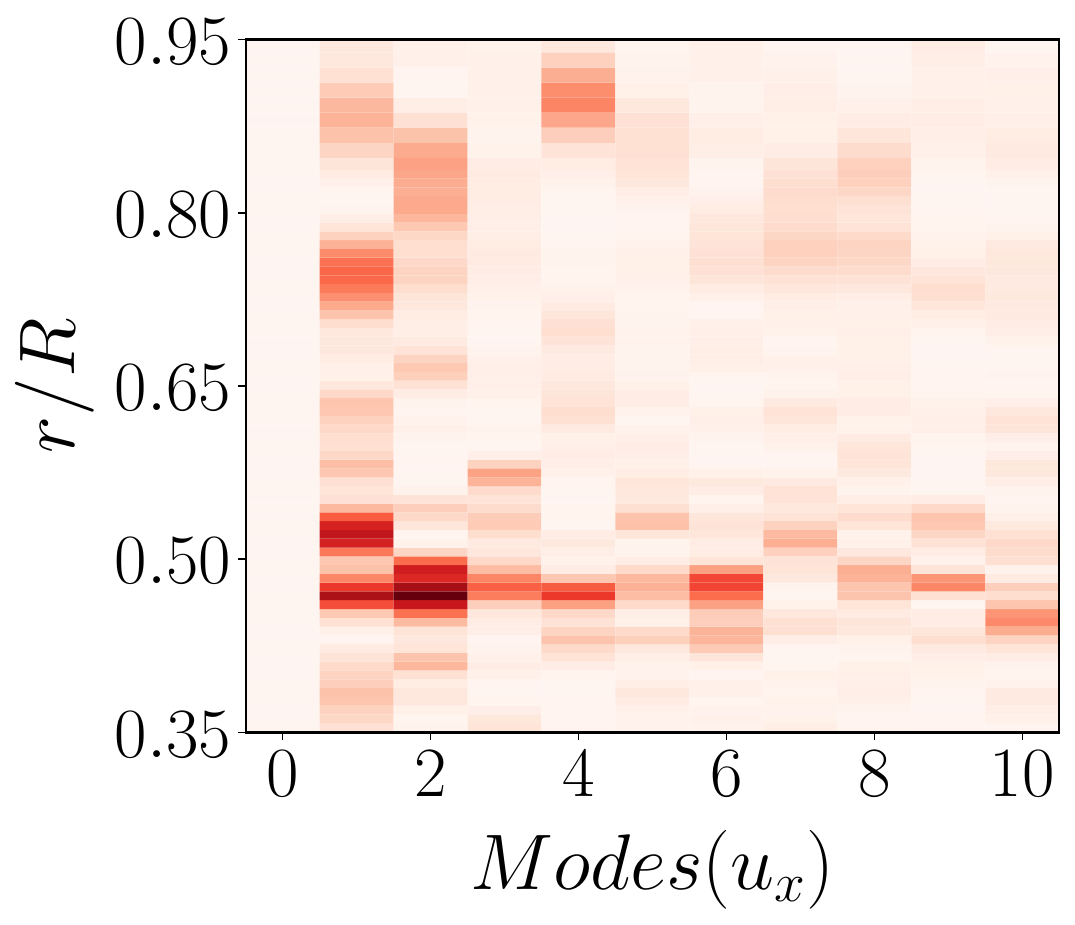}
        \caption{Member 27 LES-TW10}
        \label{fig:modes-map_x_OFR3_member27_set6v3_OW10_ON05_CorrEqnTimed}
    \end{subfigure}
    \hfill
    \begin{subfigure}{.07\textwidth}
        \centering
        \captionsetup{width=1.5\linewidth}
        \includegraphics[width=1\textwidth]{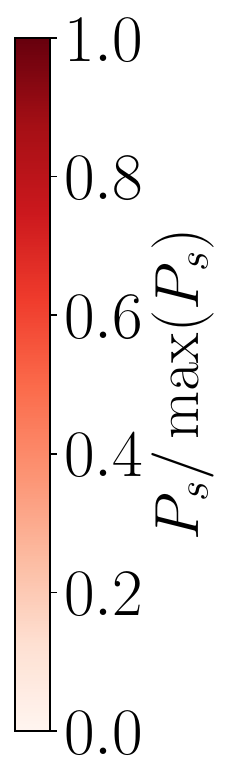}
        \vspace{0.01pt}
        \caption{Scale}
    \end{subfigure}
    \hspace{28mm}
    \caption{Modes of the azimuthal Fourier transform of the fluctuation of the axial velocity $u_x-\overline{u_x}$ at time $t=13t_p$. Results are shown for $r/R \in [0.35, \, 0.95]$, which represents the jet and recirculation region.}
    \label{fig:FFT_x_noMean}
\end{figure}

Fig. \ref{fig:FFT_x_noMean} shows the distribution of the $11$ first modes of the Fourier transform of the axial velocity. The energy carried by the remaining modes is not significant. The azimuthal mean (mode 0) $u_x-\overline{u_x}$ is here excluded to investigate the features of the fluctuating field. Results are shown  for the LES-HF reference simulation and for the simulation 27 of the ensemble for each DA run. In particular, the results for the \textit{prior state} 2, for the LES-PE2, LES-TW50 and LES-TW10 configurations are reported. The figures are coloured according to the spectral power $P_s$ normalized by the maximum power value $\max(P_s)$ for each case. These power spectra provide the kinetic energy (per radians) carried by each Fourier mode. The distribution for the simulation LES-HF presents three main modes: mode 1 contains a large amount of energy in the recirculation region of the flow ($r/R \approx 0.9$), while mode 6 and mode 10 present large amounts of energy in the jet shear region ($r/R \approx 0.5$). Modes 2 and 3 also provide important contributions in these two regions. The \textit{prior state} 2 simulation shows a very high energy level in the shear region ($r/R \approx 0.45)$ for mode 2 and partially mode 1, but the energy content for other modes is weak. In particular, the energy distribution observed in the recirculation region is very low when compared with the results from the LES-HF run. The calibration of the inlet condition for the LES-PE2 run strongly affect the energy distribution. A very high level of energy is now observed in the recirculation region for several modes. Results for the DA runs LES-TW50 and TW10, which are shown in Fig. \ref{fig:FFT_x_noMean} (d) and (e), exhibit the closest match to the predicted modal distribution of the reference simulation. The energy distribution for both DA run exhibits a clear multi-modal behaviour in the shear region, while modes 1, 2 and 4 are dominant in the recirculation region. 
This last result is promising, as DA sensors are not close to the recirculation region. Therefore, one can see that the usage of a CFD model based on Navier-Stokes equations is able to correctly propagate a local state update to obtain an improved global prediction. These findings are inline with previous results by \cite{Meldi2017_jcp}, who observed similar results for DA applications using scale-resolved CFD.

\begin{figure}[h]
    \centering
    \begin{subfigure}{.32\textwidth}
        \centering
        \includegraphics[width=1\textwidth]{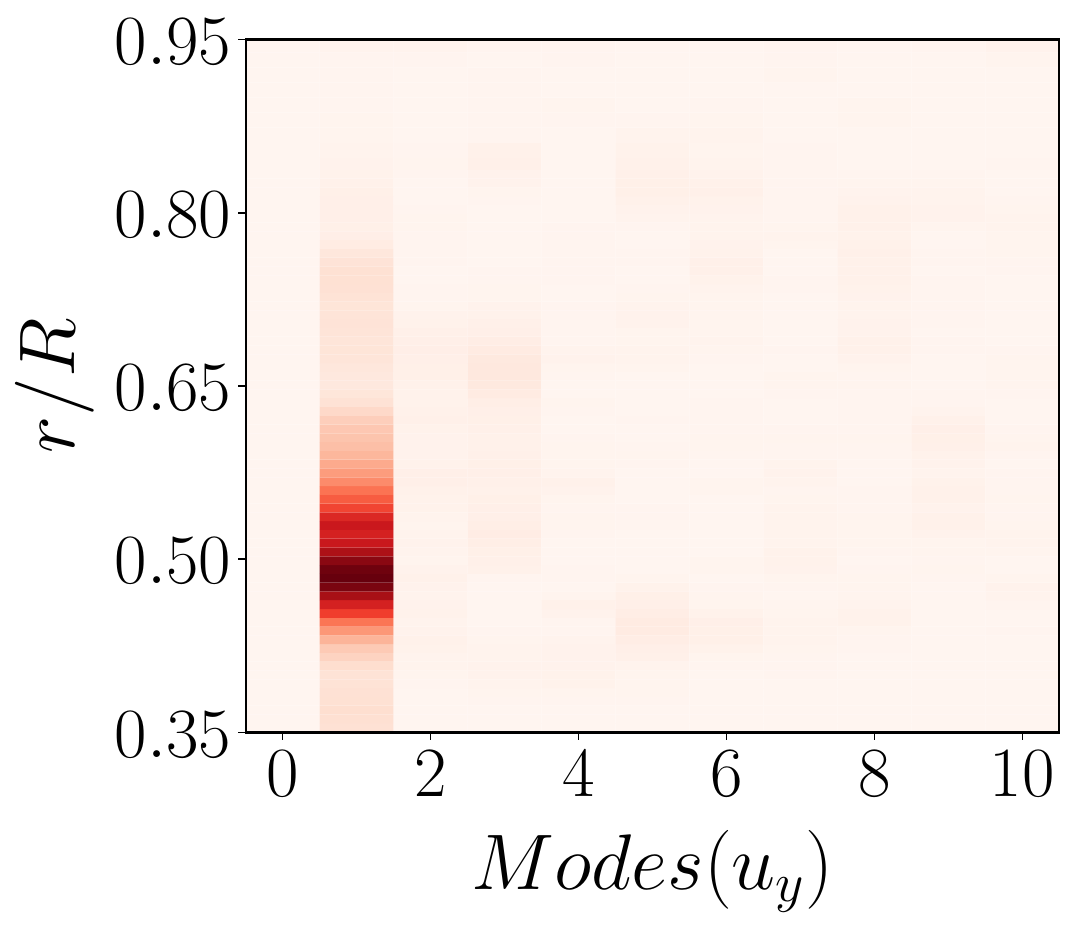}
        \caption{LES-HF}
        \label{fig:modes-map_y_ref}
    \end{subfigure}
    \hspace{1mm}
    \begin{subfigure}{.32\textwidth}
        \centering
        \includegraphics[width=1\textwidth]{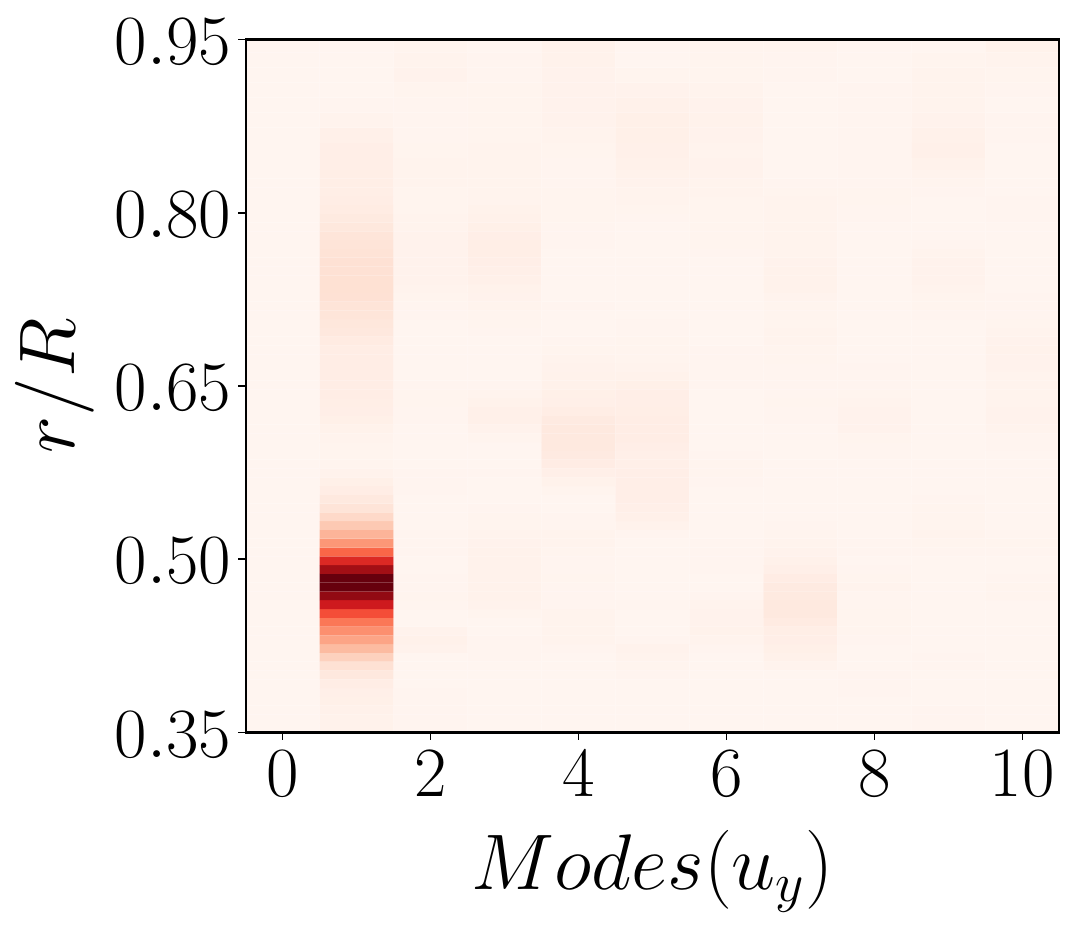}
        \caption{Member 27 \textit{priorstate} 2}
        \label{fig:modes-map_y_OFR3_member27_noDA}
    \end{subfigure}
    \hspace{1mm}
    \begin{subfigure}{.32\textwidth}
        \centering
        \includegraphics[width=1\textwidth]{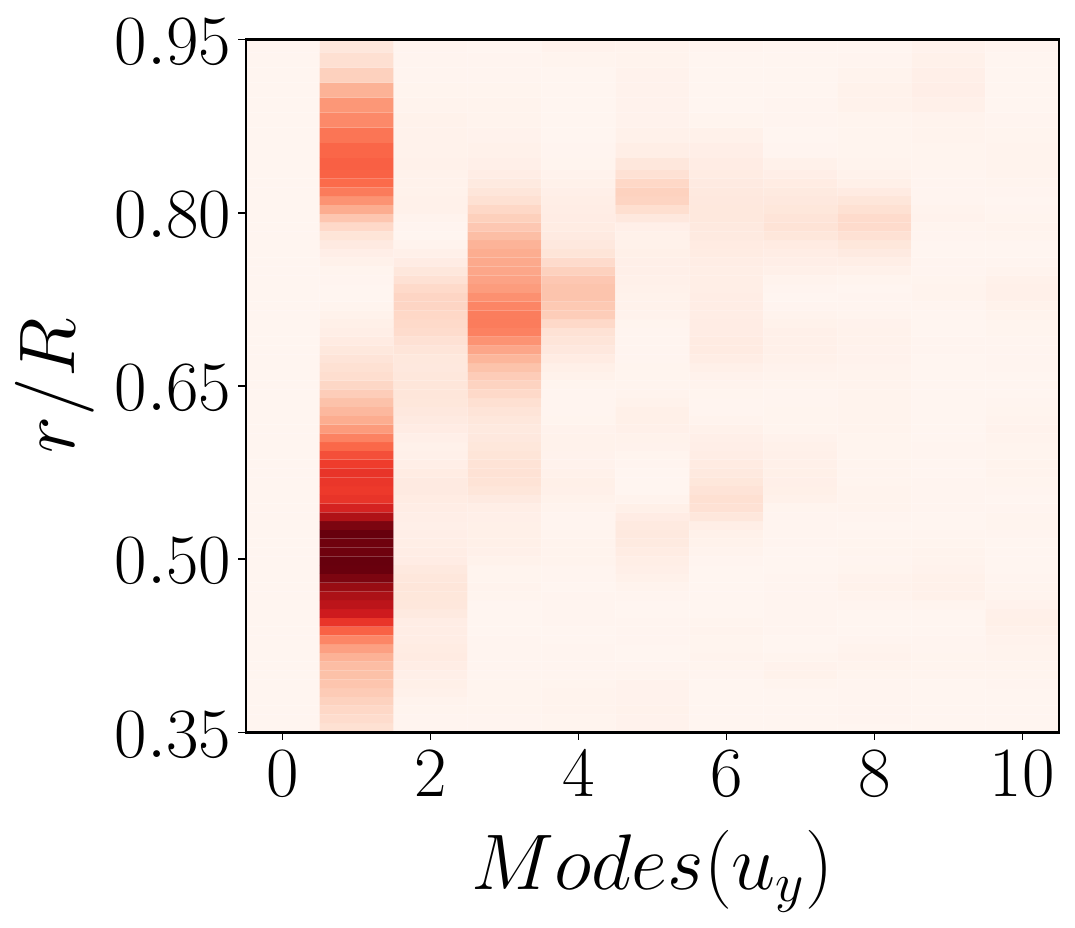}
        \caption{Member 27 LES-PE2}
        \label{fig:modes-map_y_OFR3_member27_noState}
    \end{subfigure}
    \begin{subfigure}{.32\textwidth}
        \centering
        \includegraphics[width=1\textwidth]{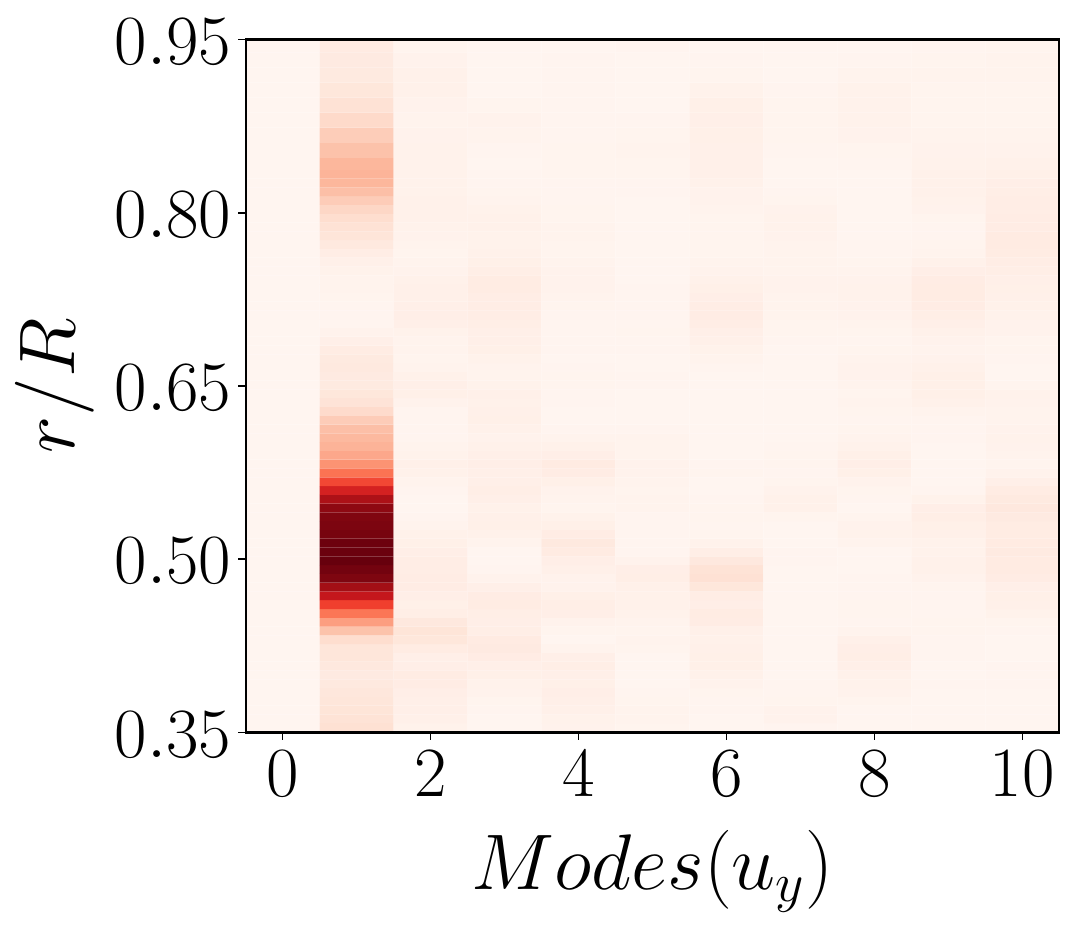}
        \caption{Member 27 LES-TW50}
        \label{fig:modes-map_y_OFR3_member27_set6v3_OW50_ON05}
    \end{subfigure}
    \hspace{1mm}
    \begin{subfigure}{.32\textwidth}
        \centering
        \includegraphics[width=1\textwidth]{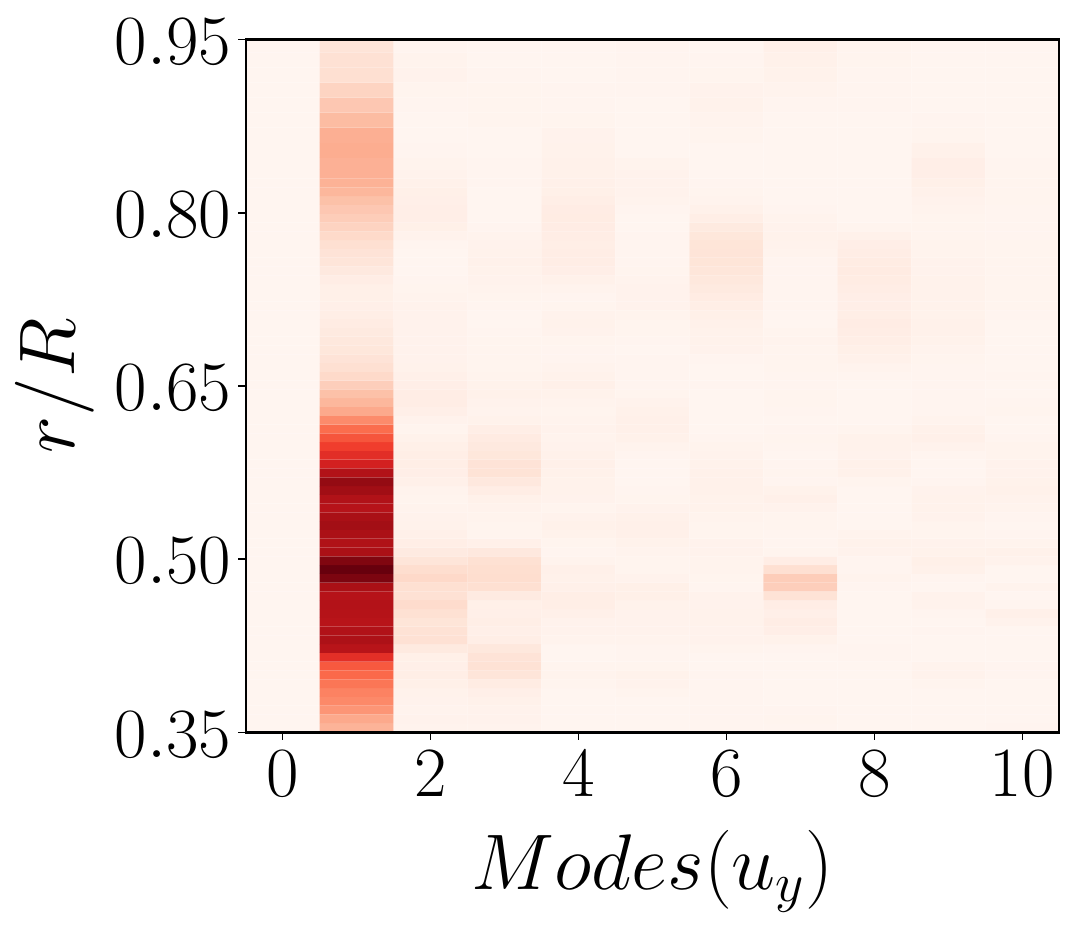}
        \caption{Member 27 LES-TW10}
        \label{fig:modes-map_y_OFR3_member27_set6v3_OW10_ON05_CorrEqnTimed}
    \end{subfigure}
    \hfill
    \begin{subfigure}{.07\textwidth}
        \centering
        \captionsetup{width=1.5\linewidth}
        \includegraphics[width=1\textwidth]{estimation_etat/FFT/modes-map/modes-colorBar.pdf}
        \vspace{0.01pt}
        \caption{Scale}
    \end{subfigure}
    \hspace{28mm}
    \caption{Modes of the azimuthal Fourier transform of the velocity $u_y$ at time $t=13t_p$. Results are shown for $r/R \in [0.35, \, 0.95]$, which represents the jet and recirculation region.}
    \label{fig:FFT_y}
\end{figure}

At last, the component $u_y$ of the velocity field is studied. Considering the symmetric behaviour of the test case, $u_z$ is not investigated, as the results are statistically the same. Fig. \ref{fig:FFT_y} presents the distribution of the first $11$ modes for the Fourier transform of $u_y$. In this case, mode 1 is predominant over all the other modes. The energy is concentrated in the shear region, where the jet has not yet impacted the cylinder wall. This strong energy concentration is observed from $r/R = 0.45$ to $r/R = 0.65$ for the reference simulation LES-HF. For the \textit{prior state} 2 the distribution is smaller and high density of energy is observed for $r/R = 0.45$ and $r/R = 0.55$. The results for the run LES-PE2 show a significant  over-prediction of the energy content for modes 1 and 3 in the proximity of the recirculation region, showing again a significant discrepancy. The runs LES-TW50 and LES-TW10 also show a high energy concentration of energy for mode 1 in the recirculation region, but the results are in adequate agreement with the reference simulation.

\begin{figure}[h]
    \centering
    \begin{subfigure}{.45\textwidth}
        \centering
        \includegraphics[width=1\textwidth]{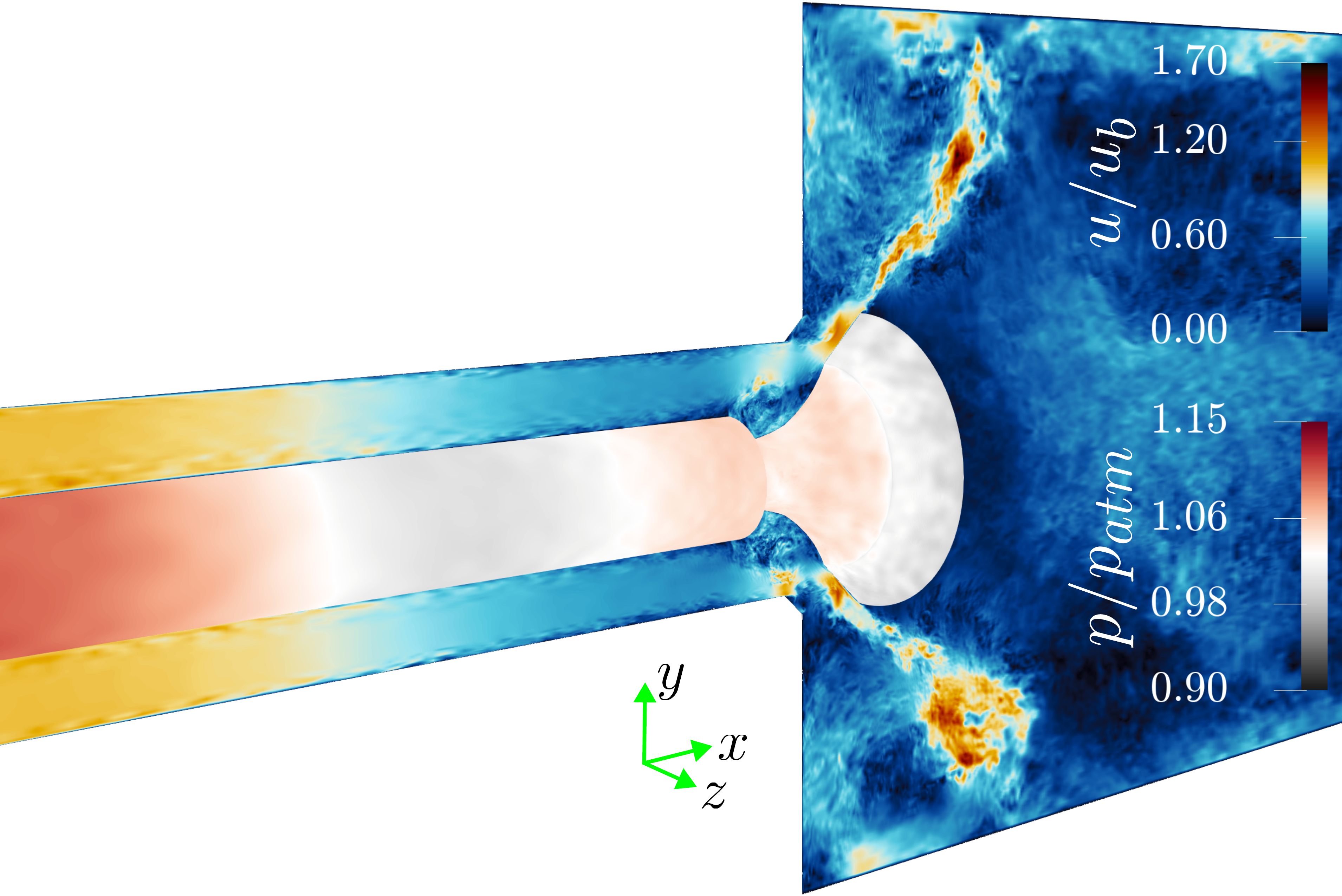}
        \caption{LES-HF}
        \label{fig:velocityField_ref_comp}
    \end{subfigure}
    \begin{subfigure}{.45\textwidth}
        \centering
        \includegraphics[width=1\textwidth]{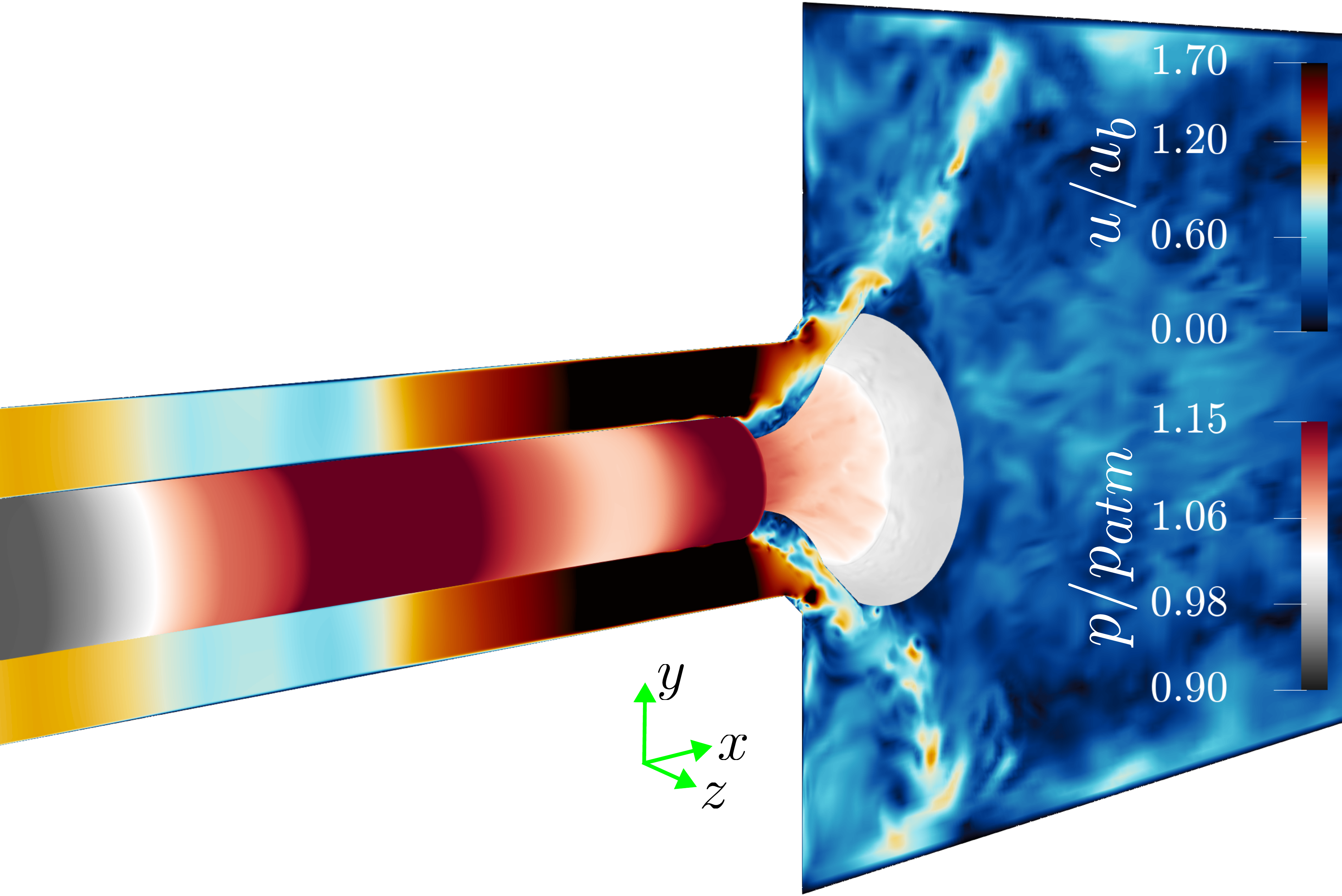}
        \caption{Member 27 \textit{prior} 2}
        \label{fig:member27_noDA_comp}
    \end{subfigure}
    \begin{subfigure}{.45\textwidth}
        \centering
        \includegraphics[width=1\textwidth]{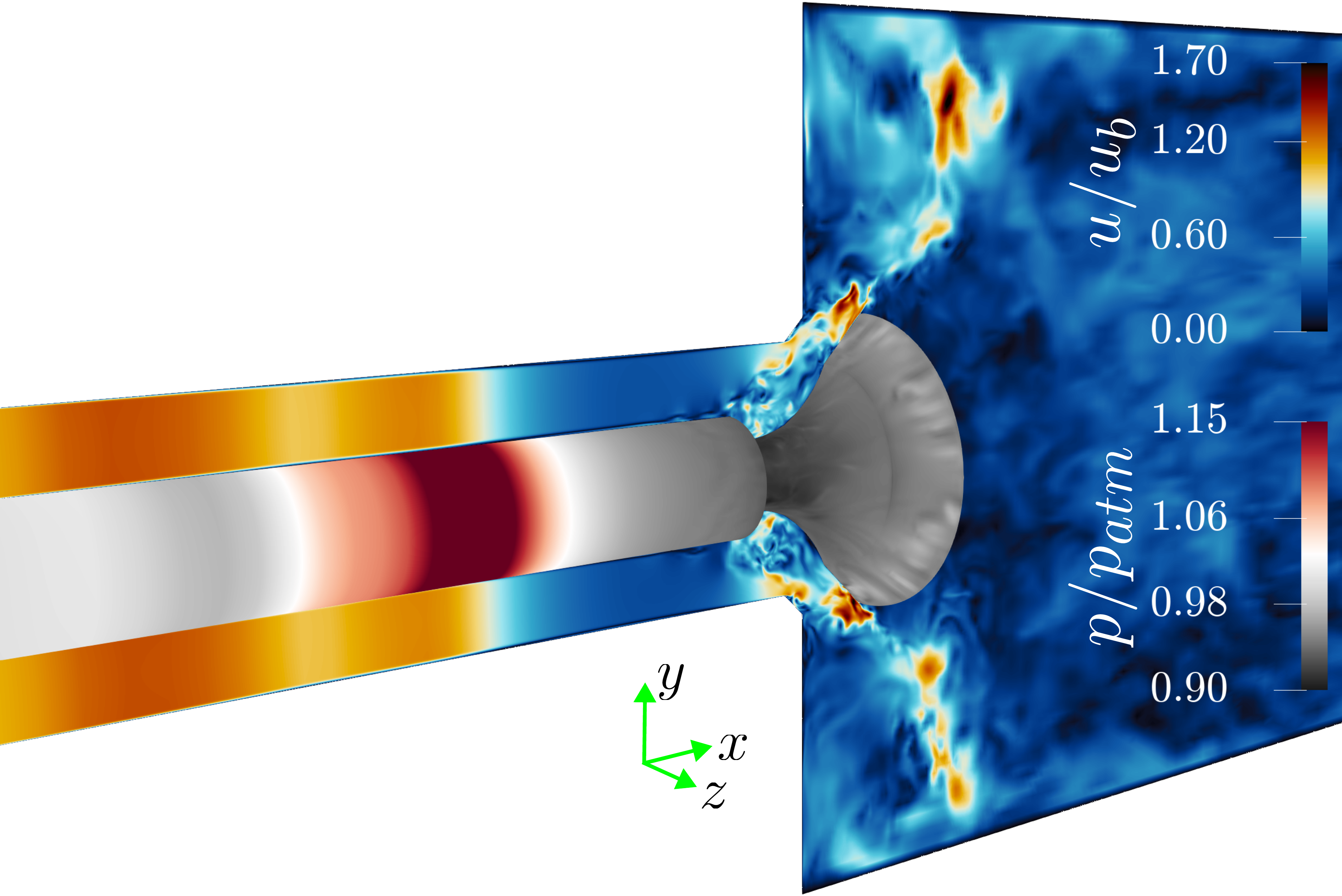}
        \caption{Member 27 LES-PE2}
        \label{fig:member27_DA_noState_comp}
    \end{subfigure}
    \begin{subfigure}{.45\textwidth}
        \centering
        \includegraphics[width=1\textwidth]{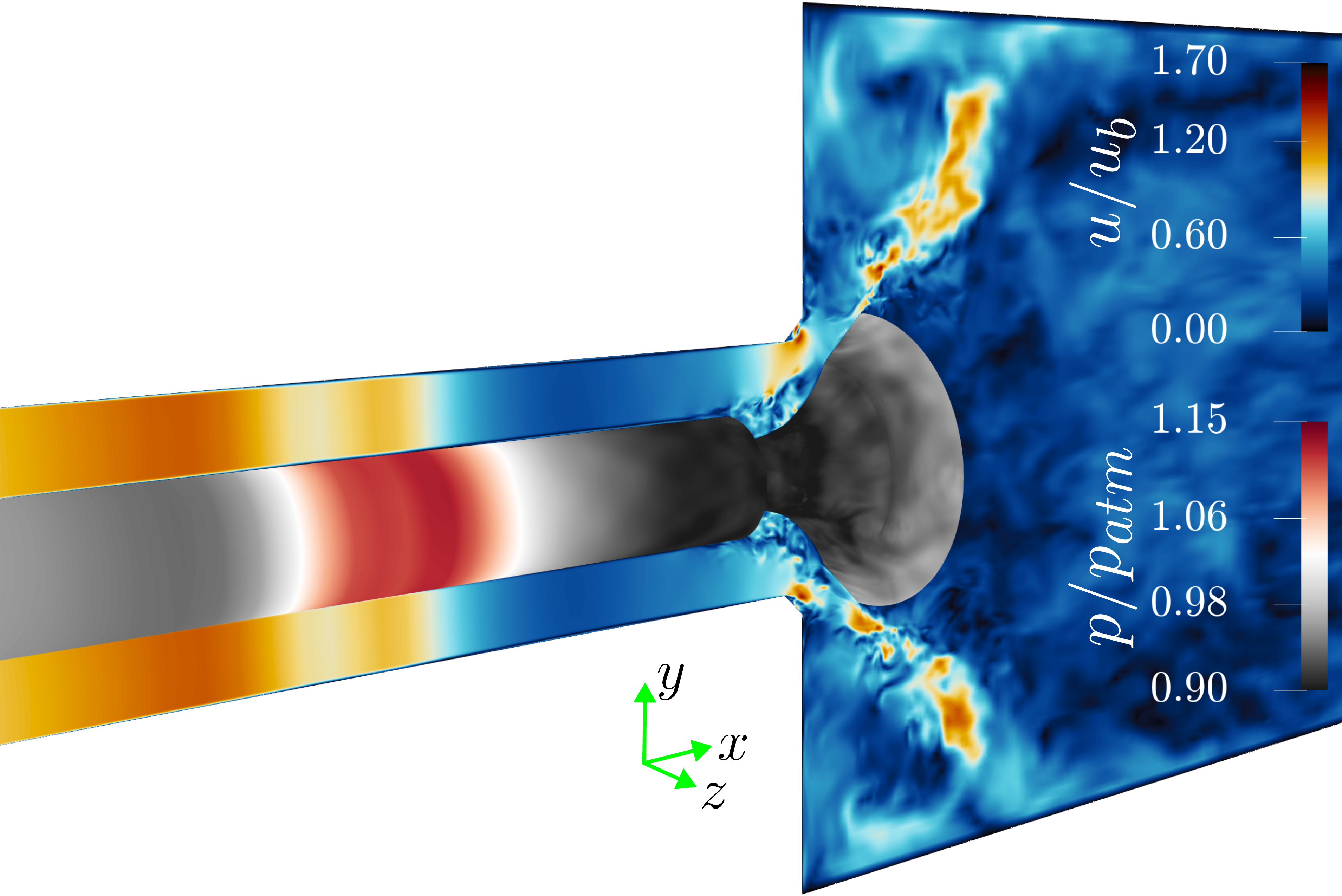}
        \caption{Member 27 LES-TW10}
        \label{fig:member27_DA_set6_TW10_ON05_comp}
    \end{subfigure}
    
    \caption{Instantaneous velocity fields for the different runs. Results are shown on the plane for $z=0$. Velocity magnitudes are normalized by the reference flow velocity $u_b$.}
    \label{fig:qualitativeComp}
\end{figure}

In summary, the investigation of the energy distribution using a discrete Fourier transform shows that the HLEnKF is able to provide an accurate prediction of the statistics of the flow field. Despite the local behaviour of the state update, the distribution of the kinetic energy exhibits an accurate behaviour even far from the assimilation regions. The energy distribution of the non-axial component of the velocity, however, shows differences between all the studied configurations despite the excellent synchronization of all the velocity components. A qualitative comparison of the velocity fields for the plane $z=0$ is shown in Fig.~\ref{fig:qualitativeComp}. It illustrates in particular the phase shift of the \textit{prior state} 2 compared to the calibrated LES-PE2 configuration. Furthermore, the LES-TW10 presents topological improvements of the velocity structures without however recreating the characteristic “clamp” shape showed on the LES-HF reference simulation. These topological improvements, the excellent synchronization of the velocities in the inferred zones as well as the improved prediction of the energy distribution of the flow highlight the potential of highly localized sequential data assimilation in the digital twin paradigm.

\section{Conclusion}
\label{sec:conclusion}

A Data Assimilation tool based on the Ensemble Kalman filter has been used to augment scale resolving numerical simulation. The analysis, which combined ensemble numerical runs using LES and high-fidelity data available on local sensors, has been performed for the investigation of the flow rig test case. 
The application of DA techniques to such a complex case, in terms of multiscale interactions and of degrees of freedom investigated, was possible thanks to the development of the HLEnKF technique and its implementation in the online library CONES. The investigation performed for this test case targeted two main objectives, which are i) the calibration of the free parameters driving an unsteady inlet condition and ii) the efficient synchronization of the flow field predicted by the LES model with the observation.

The DA calibration was performed twice, changing the initial values and variability of different parameters for the two prior states selected for investigation. In both cases, the HLEnKF shows a satisfactory calibration for scale-resolving turbulent conditions. However, the usage of different prior states creates different attractors for the parametric configuration, highlighting the difficulties in the optimization of such complex, non-linear systems.

The HLEnKF state estimation was able to provide a local synchronization of the inferred velocity fields with the observed data. However, it was observed that the frequency of DA analysis is a critical parameter here and that if the time period between successive analyses is not of the same order of magnitude of the physical turnover time, synchronization is not reached.
The study of the Fourier modes of the velocity fluctuation provides additional information on the quality of the reconstruction of the velocity field. Although unchanged for the $u_y$ and $u_z$ components of the velocity field, the modal distribution for the axial component $u_x$ is improved by the DA algorithm. A lower discrepancy with the reference data is also visible in the recirculation zone. No direct DA state update is performed in this region, but the advection of the flow from zones where the HLEnKF is performed has a positive effect on the accuracy of the solver.

This investigation permitted to highlight the potential of such DA techniques for the investigation of turbulent flows using scale resolving techniques, as well as to identify the challenging aspects to be faced in the near future to obtain robust predictive tools. Among the positive points, the potential for the integration of sequential DA in the digital twin paradigm is the most exciting. Future advancement of these tools targets applications to test cases of industrial interest. In particular, the usage of observation from experiments for complex cases will allow to assess more clearly the potential and limitations of these techniques for realistic applications.

\subsection*{Acknowledgements}{The author L.V. acknowledges the support of CERFACS during the editing and proofreading of this manuscript.}

\subsection*{Funding}{This work was supported by the French National Research Agency (ANR) to the ANR-20-CE05-0007 ALEKCIA project (https://www.ifpenergiesnouvelles.fr/alekcia).
This work was also granted access to the HPC resources of TGCC under allocation no. A0162B10763 from the GENCI (Grand Equipement National de Calcul Intensif) eDARI program.}

\subsection*{Declaration of interests}{The authors report no conflict of interest.}

\bibliographystyle{abbrvnat}
\bibliography{bibliography}

\end{document}